\documentstyle[sprocl,epsfig]{article} 
\def\beq{\begin{equation}}
\def\eeq{\end{equation}}

\def\beqn{\begin{eqnarray}}
\def\eeqn{\end{eqnarray}}

\catcode`@=11
\def\citenum#1{\csname b@#1\endcsname}
\catcode`@=12
\begin{document}
\title{THE MSSM AND WHY IT WORKS}
\author{S. Dawson}
\address{Physics Department\\
Brookhaven National Laboratory\\
Upton, NY, 11973}
\maketitle\abstracts{
An introduction to
the minimal supersymmetric standard model is presented.
We emphasize phenomenological motivations for this
model, along with examples of experimental tests.
Particular attention is paid to the Higgs sector of
the theory.
} 
\section{INTRODUCTION}

The Standard Model of particle physics is in stupendous agreement
with experimental measurements; in some cases it has been tested to a 
precision of greater than $.1\%$.  Why then expand our model?  The
reason, of course, is that the Standard Model contains 
a variety of  nagging 
theoretical problems which cannot be solved without the introduction
of some new physics.
We have no understanding of masses or why there are three
generations of quarks and leptons.  The origin of electroweak
symmetry breaking is a complete mystery.  The source of
$CP$ violation is not known.   The list
goes on and on...... 
  
  Supersymmetry is, at present, many theorists' 
favorite  candidate for  new physics beyond the Standard
Model.  Unfortunately, merely constructing a supersymmetric
version of the Standard Model does not answer many of the open
questions. String theories may answer some of these questions,
although a phenomenologically viable string theory has yet to
be constructed.  In these lectures, we will try to be very
explicit about the benefits and drawbacks of various SUSY models.

The most important  aspect of the Standard Model which has not
yet been
verified experimentally is the Higgs sector.  
The Standard Model without the Higgs boson is
incomplete
 since  it predicts massless fermions and gauge
bosons.  Furthermore, the electroweak radiative corrections
to observables such as the $W$ and $Z$ boson masses 
would be  infinite
if there were no Higgs boson.
  The simplest means of curing these defects
is to introduce  a single $SU(2)_L$ doublet of Higgs bosons.
When the neutral component of the Higgs boson gets a 
vacuum expectation value, (VEV),  the $SU(2)_L\times U(1)_Y$ gauge
symmetry is broken, giving the $W$ and $Z$ gauge bosons
their masses.  The chiral symmetry forbidding fermion masses
is broken at the same time, allowing the fermions to become
massive. The coupling of the Higgs
boson to gauge bosons is just that required to cancel
the  infinities
in electroweak radiative corrections.

The electroweak symmetry breaking of the Standard Model has
the special feature that we know the energy scale at which it
must occur.  The argument follows from the scattering of
longitudinally polarized gauge bosons.  At high energy,
$\sqrt{s}>>M_W$, the amplitude for 
this process  is, \cite{lqt}
\beq
{\cal A}(W_L^+W_L^- \rightarrow W_L^+W_L^-)=-
{G_F M_h^2\over 8 \sqrt{2}\pi}\biggl\{
2+{M_h^2\over s-M_h^2}-{M_h^2\over s}\log
\biggl(1+{s\over M_h^2}\biggr)\biggr\}
\quad .
\eeq
For a light Higgs boson, we have the
limit,
\beq
{\cal A}(W_L^+W_L^- \rightarrow W_L^+W_L^-)
\longrightarrow_{s>> M_h^2} -{G_F M_h^2\over 4 \pi\sqrt{2}}
\quad .
\label{wwwwamp}
\eeq
Applying the unitarity condition
to  the $I=J=0$ partial wave for this
process, $\mid a_0^0 \mid < {1\over 2}$,
 gives the restriction,
\beq
M_h < 860~GeV
\quad .
\eeq
We thus are reasonably confident that a weakly interacting Higgs
boson, if it exists, will appear below the $TeV$ scale.

Given the nice features of the Standard Model with a single Higgs boson,
  what then is the
problem with this simple and economical picture?
The argument against the simplest  version of the Standard Model
 is purely theoretical and arises
when radiative corrections to the Higgs boson mass are computed.
 At one loop, the quartic self- interactions
of the Higgs boson  generate a quadratically
divergent contribution
 to the Higgs boson mass which must be cancelled
by a  mass  counterterm.  This counterterm must be fine tuned
at each order in perturbation theory.  
    The quadratic growth
of the Higgs  boson mass beyond  tree level in
perturbation theory  is one of
 the driving motivations  behind
the introduction of supersymmetry, which we will see cures this
problem.  
The cancellation of quadratic divergences will be discussed
extensively in the next section.  

In these lectures, we  discuss the theoretical motivation for supersymmetric
theories and introduce the minimal low energy effective supersymmetric
theory, (MSSM).
We consider  the MSSM and its simplest grand unified extension,
along with models where the supersymmetry is broken by the gauge
interactions, (GMSB).  
  The particles and their interactions are examined 
with  particular attention paid to the Higgs sector of SUSY
models.  
Finally, we discuss indirect limits on the SUSY partners
of ordinary matter coming
from precision measurements at LEP and 
direct production searches at the Tevatron and LEPII.
Search strategies for SUSY
at both future  $e^+e^-$ and hadron colliders are briefly 
touched on.
  There exist numerous excellent
reviews of both the more formal aspects of supersymmetric model
building \cite{hkrep,bagtasi,martin}
 and the phenomenology of these
models \cite{martin,xerxes,peskin} and the reader is referred to
these for more details.

\section{Quadratic Divergences}
  
The vanishing of quadratic divergences in a supersymmetric
theory is typically advertised as one of the primary motivations
for introducing  supersymmetry.  As such, it
is important to examine the question of quadratic divergences in
detail.  A nice discussion is given in the lecture notes by
 Drees.\cite{drees}   
 We begin by considering  a theory with a single  fermion, $\psi$,
coupled to a massive  Higgs scalar,
\beq
{\cal L}_\phi=
{\overline \psi}(i \gamma^\mu\partial_\mu)\psi
+\mid \partial_\mu \phi\mid^2-m_S^2 \mid \phi\mid^2
-\biggl( {\lambda_F\over 2}
{\overline \psi}\psi \phi +{\rm h.c.} \biggr)  
\quad .
\label{ephilag}
\eeq
We will assume that this Lagrangian leads to spontaneous symmetry 
breaking and so take $\phi=(h+v)/\sqrt{2}$, with $h$ the physical Higgs
boson.  (The $\sqrt{2}$ is arbitrary at this point and is put in only
because it is conventional.)  After spontaneous symmetry breaking,
the fermion acquires a mass, $m_F=\lambda_F v/\sqrt{2}$. 
First, let us consider the fermion self-energy  arising from
the  scalar loop corresponding to Fig. 1.
\begin{figure}[tb]
\vspace*{-4.05in} 
\centerline{\epsfig{file=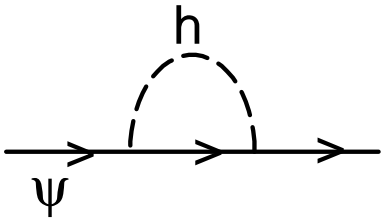,height=6.in}}
\vspace*{-.5in} 
\caption{Fermion mass renormalization from a Higgs loop.}
\end{figure}  
\beq
-i\Sigma_F(p)=\biggl({-i\lambda_F\over \sqrt{2}}\biggr)^2
(i)^2
\int{d^4k\over (2\pi)^4}{(k+m_F)\over [k^2-m_F^2][(k-p)^2-m_S^2]}
\quad .
\eeq
The renormalized fermion mass is $m_F^r=m_F+\delta m_F$, with
\beqn
\delta m_F &=& \Sigma_F(p)\mid_{ p=m_F}
\nonumber \\
&=& i {\lambda_F^2\over 32\pi^4}\int_0^1 dx \int  d^4 k^\prime
{m_F (1+x)\over [k^{\prime 2} -m_F^2 x^2-m_S^2(1-x)]^2}
\quad .
\label{meren}  
\eeqn

Since many of you probably only know how to calculate loop diagrams
using dimensional regularization, we will take a brief aside to
discuss the calculation of Eq. \ref{meren} using a momentum space
cutoff.
(This discussion directly parallels that of the renormalization
of the electron self-energy in Bjorken and Drell.\cite{bd})
  The integral can be performed in Euclidean space, which
amounts to making the following transformations,
\beqn
k_0  &\rightarrow & i k_4
\nonumber \\
d^4k^\prime & \rightarrow & i d^4 k_E
\nonumber \\
k^{\prime 2} & \rightarrow & - k_E^2
\quad .
\eeqn
Since the integral of Eq. \ref{meren} depends only on
$k_E^2$, it can easily be performed using the
result (valid for symmetric integrands),
\beq \int d^4 k_E f(k_E^2)=\pi^2
\int^{\Lambda^2}_0 y dy f(y)
\quad .
\label{intfacs}  
\eeq 
In Eq. \ref{intfacs}, $\Lambda$ is a high energy cut-off, presumably of the
order of the Planck scale or a GUT scale.
The renormalization of the fermion   mass is then, 
\beqn
\delta m_F&=& -{\lambda_F^2 m_F\over 32\pi^2}\int_0^1~dx (1+x)
\int^{\Lambda^2} _0 {y dy\over [y+m_F^2x^2+m_S^2(1-x)]^2}
\nonumber \\
&=& -{3 \lambda_F^2 m_F\over 64\pi^2} \log\biggl({\Lambda^2
\over m_F^2}\biggr) + ....
\eeqn
where the $....$  indicates terms independent of the cutoff
or which vanish when $\Lambda\rightarrow\infty$.
This correction clearly corresponds to a well-defined expansion for $m_F$.
Fermion masses are said to be {\it natural}.  In the limit
in which the fermion mass vanishes,
Eq. \ref{ephilag} is invariant under the chiral transformations,
\beqn
\psi_L & \rightarrow  & e^{i\theta_L}\psi_L
\nonumber \\
\psi_R &\rightarrow & e^{i\theta_R}\psi_R,
\eeqn
and so we
see that setting the fermion mass to zero increases the symmetry
of the theory.  
Since the Yukawa coupling (proportional to the fermion
 mass term) breaks
this symmetry, the corrections to the mass must be proportional to $m_F$. 

\begin{figure}[tb]
\vspace*{-1.2in} 
{\centerline{\epsfig{file=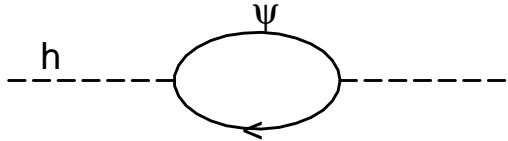,height=4.in}}}
\vspace*{-1.5in} 
\caption{Higgs mass renormalization from a fermion loop.}
\end{figure}  
The situation is quite different, however, when we consider the
renormalization of the scalar  mass  from
a fermion loop (Fig. 2) using the same Lagrangian (Eq. \ref{ephilag}),
\beq
-i\Sigma_S(p^2)=\biggl({-i\lambda_F\over\sqrt{2}}\biggr)^2
(i)^2 (-1)\int {d^4 k \over (2 \pi)^4}
{Tr[(k+m_F)((k-p)+m_F)]\over
(k^2-m_F^2)[(k-p)^2-m_F^2]}
\quad .
\eeq 
The minus sign is the consequence of Fermi statistics
and will be quite important later.  Integrating with a momentum
space cutoff as above we
 find the contribution to the Higgs mass,
$(\delta M_h^2)_a\equiv\Sigma_S(m_S^2)$, 
\beqn
(\delta M_h^2)_a&
=&-{\lambda_F^2\over 8\pi^2}\biggl[
\Lambda^2+(m_S^2-6m_F^2)\log\biggl({\Lambda\over m_F}\biggr)
\nonumber \\
&&
+
(2m_F^2-{m_S^2\over 2})
\biggl(1+I_1\biggl({m_S^2\over m_F^2}\biggr)\biggr)\biggr]~
+{\cal O}\biggl({1\over \Lambda^2}\biggr),
\label{quads}
\eeqn
where $I_1(a)\equiv \int^1_0 dx \log(1-ax(1-x))$.
The Higgs boson mass diverges {\bf{\it quadratically}}! 
The Higgs boson thus does not obey the decoupling theorem
and  this quadratic divergence appears independent of
the mass of the Higgs boson. 
Note that the correction is {\it not} proportional to $M_h$.  This 
is because setting the Higgs mass equal to zero does not 
increase the symmetry of the Lagrangian.  There is nothing
that protects the Higgs mass from these large corrections
and, in fact, the Higgs mass wants to be close to the largest mass
scale in the theory. 

Since we know that in the Standard Model,
the physical Higgs boson
 mass, $M_h$, must be less than around
$1~TeV$ (in order to keep the $WW$ scattering cross section from
violating unitarity), we have the unpleasant result,
\beq
M_h^2=M_{h,0}^2+\delta M_h^2+{\hbox{counterterm}}
,
\eeq
where the counterterm must be adjusted to a precision of
roughly $1$ part
in $10^{15}$ in order to cancel the 
quadratically divergent contributions to $\delta M_h^2$.
This adjustment must be made at each order in perturbation
theory.
This is known as the ``{\bf {\it hierarchy problem}}''.  

Of course, the quadratic divergence can be renormalized
 away in exactly the same manner  
as is done for logarithmic divergences by adjusting the
cut-off. There is nothing formally wrong with this fine tuning.  
 Most theorists, however,
regard this solution as unattractive.
On the other hand, 
 if the calculation is performed in
dimensional regularization, one obtains only $1/\epsilon$
singularities which are absorbed into the definitions
of the counterterms.    Hence, the problem of quadratic
divergences does not become apparent when using dimensional regularization.
It arises  only when one attempts to import a physical significance
to the cut-off $\Lambda$.   
  
The effect of scalar particles on the Higgs mass renormalization is
quite different from that of fermions.
We
introduce  two  complex scalar fields, $\phi_1$ and $\phi_2$,
interacting with the Standard Model Higgs boson, $h$,
(the reason for introducing $2$ scalars is that with foresight we
know that a 
supersymmetric theory associates $2$ complex scalars with
each fermion -- we could just as easily make the argument given
below with one additional scalar and slightly different couplings), 
\beqn
{\cal L}&=&
\mid \partial_\mu \phi_1\mid^2+
\mid \partial_\mu \phi_2\mid^2
-m_{s_1}^2\mid \phi_1\mid^2-m_{s_2}^2 \mid\phi_2\mid^2
\nonumber \\
&&
+\lambda_S\mid \phi\mid^2
\biggr(\mid \phi_1\mid^2 +\mid \phi_1\mid^2\biggr)
+{\cal L}_\phi  \quad .
\label{lag2h}
\eeqn
\begin{figure}[tb]
\vspace{-1.4in}
\centerline{\epsfig{file=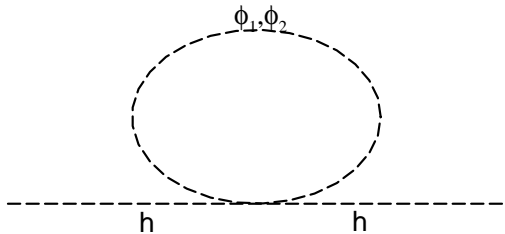,height=3in}}
\caption{Higgs mass renormalization from scalar loops.}
\end{figure} 
\noindent From the diagram of Fig. 3,
 we find the contribution to the Higgs mass
renormalization,
\beqn
(\delta M_h^2)_b& =& -\lambda_S
\int {d^4k\over (2\pi)^4}\biggl[
{i\over k^2-m_{s_1}^2}
+{i\over k^2-m_{s_2}^2}\biggr]
\nonumber \\
&=& {\lambda_S\over 16\pi^2}\biggl\{
2 \Lambda^2 -2m_{s_1}^2\log\biggl({\Lambda\over m_{s_1}}\biggr)
-2m_{s_2}^2\log\biggl({\Lambda\over m_{s_2}}\biggr) \biggr\}
\nonumber \\  &&
+{\cal O}\biggl({1\over \Lambda^2}\biggr).
\label{quad2}
\eeqn       
From Eqs. \ref{quads} and \ref{quad2}, we see that if 
\beq
\lambda_S=\lambda_F^2
,
\label{susyreq}
\eeq
the quadratic divergences coming from these two terms cancel
each other.  Notice that the cancellation occurs independent
of the masses, $m_F$ and $m_{s_i}$, and of the magnitude of
the couplings $\lambda_S$ and $\lambda_F$.

In the Standard Model, one could attempt to cancel the quadratic
divergences in the Higgs boson mass by balancing the contribution from
the Standard Model
Higgs quartic coupling with that from the top quark loop
in exactly the same manner as above.  This approach
would give a prediction
 for the Higgs boson mass in terms of the top quark mass. 
However, since there is no symmetry to enforce this relationship,
 this attempt to cancel quadratic 
divergences must fail at $2-$ loops. 

\begin{figure}[tb]
\vspace*{-2.75in} 
\centerline{\epsfig{file=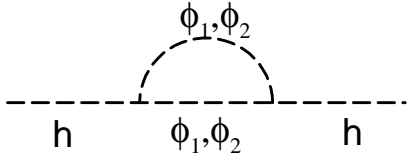,height=4in}}
\caption{Higgs mass renormalization from scalar loops.}
\end{figure} 
After the spontaneous symmetry breaking, Eq. \ref{lag2h} also
leads to a cubic interaction shown in Fig. 4.  These  also graphs
give a contribution to the Higgs mass renormalization,
although they are not quadratically divergent.
\beq
(\delta M_h^2)_c={\lambda_S^2 v^2\over 16\pi^2}\biggl\{ -1+2 \log\biggl(
{\Lambda\over m_{s_1}}\biggr)-I_1\biggl({M_h^2\over m_{s_1}^2}
\biggr)\biggr\} +(m_{s_1}\rightarrow m_{s_2})
+{\cal O}\biggl({1\over \Lambda^2}\biggr).
\eeq
Combining the three contributions to the Higgs mass and assuming
$\Lambda_S=\Lambda_F^2$ and $m_{s_1}=m_{s_2}$, we find
no quadratic divergences in the Higgs mass renormalization,
\beq
(\delta M_h^2)_{tot}={\lambda_F^2\over 4\pi^2}
\biggl\{ m_{s_1}^2\log\biggl({\Lambda\over m_{s_1}}\biggr)
-m_F^2\log\biggl({\Lambda\over m_F}\biggr)
\biggr\}+{\cal O}\biggl({1\over\Lambda^2}\biggr).
\eeq
If the mass splitting between the fermion and scalar is small,
$\delta m^2\equiv m_F^2-m_{s_1}^2$, then we have
the approximate result,
\beq
(\delta M_h^2)_{tot}={\lambda_F^2\over 4\pi^2}\delta^2
\quad .
\label{splitting}
\eeq
Therefore, as long as the mass splitting between scalars and fermions is
``small'', no unnatural cancellations will be required and
the theory can be considered ``natural''.  
In this manner, a theory with nearly degenerate fermions and scalars
and carefully adjusted couplings solves the hierarchy problem.  

\section{WHAT IS SUPERSYMMETRY? }

In the previous section we saw that if the fermion
and scalar couplings of a theory are carefully adjusted,
it is possible to cancel the quadratically divergent contributions
to the Higgs boson mass.  
Of course, in order for this cancellation
to persist to all orders in perturbation
theory it must be the result of a symmetry.
This symmetry is ${\it{\bf supersymmetry}}$.

Supersymmetry is a symmetry which relates
the masses and couplings of  particles of differing
spin, (in the above example,
fermions and scalars).
  The particles are combined into a ${\it superfield}$,
which contains fields differing by one-half
 unit of spin.\cite{wess,west}  The
simplest example, the scalar (or chiral)
 superfield, contains a complex
scalar, $S$, and a two- component Majorana fermion, $\zeta$.
(A Majorana fermion, $\zeta$, is one which
is equal to its  charge conjugate,
$\zeta^c=\zeta$. A familiar example is a  Majorana neutrino.)
The supersymmetry completely specifies the allowed interactions.
In this simple case, the Lagrangian is
\beqn 
{\cal L}&=& -\partial_\mu S^*\partial^\mu S - i
{\overline \zeta}{\overline \sigma}^\mu \partial_\mu\zeta
-{1\over 2} m(\zeta\zeta+{\overline \zeta}
{\overline\zeta})
\nonumber \\
&&-c S\zeta\zeta-c^* S^*{\overline \zeta}{\overline \zeta}
-\mid m S+c S^2\mid ^2
 ,  
\label{suslag}
\eeqn 
(where ${\overline \sigma}^\mu\equiv (1, -{\vec \sigma})$,
${\vec \sigma}$ are the Pauli matrices, 
and $c$ is an arbitrary coupling constant.)
This Lagrangian is invariant (up to a total derivative)
under supersymmetry transformations which take the scalar into the
fermion and ${\it vice~versa}$. 
The scalar self interactions of Eq. \ref{suslag},
\beq
V(S,S^*)=\mid m S+c S^2\mid^2
\eeq
clearly yield a potential which is positive definite, which is a general
feature of an unbroken supersymmetric theory.  
An unbroken supersymmetric theory has its minimum at $\langle V
\rangle =0$.

Since  the scalar and fermion interactions have the
same coupling, the cancellation of quadratic divergences
occurs automatically, as in Eq. \ref{susyreq}.  
One thing that is immediately obvious is that this Lagrangian
contains both a scalar and a fermion ${\it of~equal ~mass}$.
Supersymmetry connects particles of differing spin, but
with all other
characteristics the same.  That is, they have
the same quantum numbers and the same mass.
\begin{itemize}
\item
Particles in a superfield have the same masses and
quantum numbers and differ by $1/2$ unit of spin in 
a theory with unbroken supersymmetry.
\end{itemize}

It is clear, then, that {\bf
supersymmetry must be a broken symmetry} if it is to be a theory
of low energy interactions.
There is no scalar particle, for example, with the mass and
quantum numbers of the electron.  In fact, there are no candidate
supersymmetric scalar partners for any of the fermions in the
experimentally observed 
spectrum.  We will take a non-zero
 mass splitting between the particles of
a superfield as a signal for supersymmetry breaking.

Supersymmetric theories are easily constructed according to the rules
of  supersymmetry.  I present here a cookbook approach to
constructing the minimal supersymmetric version
of the Standard Model.  The first step is to pick the particles in
superfields.\cite{gates} 
 There are two types of superfields relevant for
our purposes: 
\begin{enumerate}
\item  ${\it Chiral~(Scalar) ~Superfields}$:  These consist of 
a complex scalar field, $S$, and a $2$-component Majorana fermion
field, ${\zeta}$.
\item ${\it Massless Vector~Superfields}$:  These consist of a massless
gauge field with field strength $F_{\mu\nu}^A$
 and a $2$-component Majorana fermion field, $\lambda_A$, termed
a ${\it gaugino} $. 
The index $A$ is the gauge index.  
\end{enumerate} 
 A chiral superfield, $\Phi$,
 can be written in terms of anti-commuting
Grassman variables, $\theta$ as,
\beq
\Phi(x)=S(x)+\sqrt{2}\theta \zeta(x)+
\theta\theta F(x)
\label{thetadef}
\eeq
and all of the interactions can be written in terms of $\Phi$.
This form makes it clear that all the components of a superfield
must have the same mass and gauge interactions.
In the chiral superfield we see that the number of scalar degrees
of freedom is equal to the number of fermion degrees of freedom.
Similarly, a massless vector boson and a Majorana gaugino also
have 2 degrees of freedom.
The superfields also contain ``auxiliary fields'', $F$,  which
are fields with no kinetic energy terms in
the Lagrangian.\cite{wess}  These fields are not important
for our purposes, although they are important for constructing
the interactions.

\subsection{The Particles of the MSSM}

The MSSM respects the same $SU(3)\times
SU(2)_L\times U(1)_Y$ gauge symmetries as does the
Standard Model.  
The particles necessary to construct the minimal supersymmetric 
version of the Standard Model are shown in Tables 1 and 2
in terms of the superfields, (which are denoted by the
superscript ``hat'').  
Since there are no candidates for supersymmetric partners of
the observed particles, we must double the entire spectrum,
placing the observed particles in superfields with new
postulated superpartners.  
 There are, of course, quark and
lepton superfields for all $3$ generations and we have
listed  in Table 1
only the members of the first generation to simplify the
notation.  
The superfield ${\hat Q}$ thus consists of an $SU(2)_L$
doublet of quarks:
\beq
Q= 
\left( \begin{array}{c} u \\
d\end{array}\right)_L
\eeq
and their scalar partners which are also in
an $SU(2)_L$ doublet,
\beq
{\tilde Q}=
\left(\begin{array}{c}  {\tilde u}_L\\ 
{\tilde d}_L\end{array}\right)
\quad .  
\eeq   Similarly, the
superfield ${\hat U}^c$ (${\hat D}^c$)
 contains the right-handed
up  (down) anti-quark, ${\overline u}_R$ (${\overline d}_R$), 
 and its scalar partner, ${\tilde u}_R^*$ (${\tilde d}_R^*$).
The scalar partners of the quarks are called
squarks.  
We see that each quark has $2$ scalar partners, one corresponding
to each quark chirality.
The leptons are contained in the $SU(2)_L$ doublet superfield
${\hat L}$ which contains the left-handed fermions,
\beq
L=\left(\begin{array}{c} \nu \\
e\end{array}\right)_L
\eeq
and their scalar partners,
\beq
{\tilde L}=\left(\begin{array}{c}
{\tilde \nu}_L\\
{\tilde e}_L\end{array}\right)
\quad .
\eeq
Finally, the right-handed anti-electron, ${\overline e}_R$, is contained
in the superfield ${\hat E}^c$ and has a scalar partner
${\tilde e}_R^*$.  The scalar partners of the leptons
are termed sleptons.  

  The $SU(3)\times SU(2)_L\times U(1)_Y$  gauge fields all obtain
Majorana fermion partners in a SUSY model.
  The ${\hat G}^a$ superfield contains
the gluons, $G^{a\mu}$, and their partners the gluinos, ${\tilde g}^a$;
${\hat W}_i$ contains the $SU(2)_L$ gauge bosons, $W_i^\mu$ and
their fermion partners, ${\tilde \omega}_i$ (winos);
and ${\hat B}$ contains the $U(1)_Y$ gauge field, $B^\mu$,
and its fermion partner, ${\tilde b}$ (bino).  The usual
notation is to denote the supersymmetric partner of a fermion
or gauge field with the same letter, but with a tilde over it.   
\begin{table}[htb]
\begin{center}
{Table 1: Chiral Superfields of the MSSM}
\vskip6pt
\renewcommand\arraystretch{1.2}
\begin{tabular}{|lccrc|}
\hline
\multicolumn{1}{|c}{Superfield}& SU(3)& $SU(2)_L$& $U(1)_Y$
& Particle Content 
\\
\hline
${\hat Q}$   &    $3$          & $2$&  $~{1\over 6}$
& ($u_L,d_L$), (${\tilde u}_L,{\tilde d}_L$)\\
${\hat U}^c$ & ${\overline 3}$ & $1$& $-{2\over 3}$
&${\overline u}_R$, ${\tilde u}_R^*$\\
${\hat D}^c$ & ${\overline 3}$ & $1$&  $~{1\over 3}$
&${\overline d}_R$, ${\tilde d}_R^*$\\
${\hat L}$   & $1$             & $2$& $~-{1\over 2}$
& $(\nu_L,e_L)$, (${\tilde \nu}_L, {\tilde e}_L$)\\
${\hat E}^c$ & $1$             & $1$& $~1$ 
& ${\overline e}_R$, ${\tilde e}_R^*$\\
${\hat H_1}$ & $1$             & $2$& $-{1\over 2}$ 
&($H_1, {\tilde h}_1$)\\
${\hat H_2}$ & $1$             & $2$& $~{1\over 2}$
& $(H_2, {\tilde h}_2)$ \\ 
\hline
\end{tabular}
\end{center}
\end{table}
 
\begin{table}[htb]
\begin{center}
{Table 2: Vector Superfields of the MSSM}
\vskip6pt
\renewcommand\arraystretch{1.2}
\begin{tabular}{|lcccc|}
\hline
\multicolumn{1}{|c}{Superfield}&SU(3)&$SU(2)_L$&$U(1)_Y$
& Particle Content\\
\hline
${\hat G^a}$  &  $8$  &  $1$  &  $0$ 
&$G^\mu$, ${\tilde g}$ \\
${\hat W^i}$  &  $1$  &  $3$  &  $0$ 
& $W_i^\mu$, ${\tilde \omega}_i$  \\
${\hat B}$  &  $1$  &  $1$  &  $0$ 
& $B^\mu$, ${\tilde b}$  \\
\hline
\end{tabular}
\end{center}
\end{table}

One feature of Table 1 requires explanation.  The Standard Model
contains a single $SU(2)_L$ doublet of scalar particles, dubbed the
``Higgs doublet".  In the supersymmetric extension of
the Standard Model, this scalar doublet acquires a SUSY
partner which is an $SU(2)_L$ doublet of
Majorana  fermion fields, 
 ${\tilde h}_1$  (the Higgsinos), which   
 contribute to the triangle $SU(2)_L$ and $U(1)_Y$ gauge anomalies.
Since the fermions of the Standard Model have exactly the
right quantum numbers to cancel these anomalies, it follows
that the contribution from the fermionic partner of the
Higgs doublet remains uncancelled.\cite{anoms}    
Since gauge theories cannot have anomalies, these contributions
must be cancelled somehow if the SUSY theory is to be sensible.
The simplest way is to add a second Higgs doublet with 
precisely the opposite $U(1)_Y$ quantum numbers as the
first Higgs doublet.  In a SUSY model,
 this second
Higgs doublet will also have fermionic partners, ${\tilde h}_2$,
 and the contributions of the fermion partners of the
two Higgs doublets to gauge anomalies
will precisely cancel each other, leaving
an anomaly free theory.  
It is easy to check that the fermions of Table 1 satisfy the
conditions for anomaly cancellation:
\beq
Tr(Y^3)=Tr(T_{3L}^2Y)=0\quad .
\eeq 
We will see later that $2$ Higgs doublets are also required in order to 
give both the up and down quarks masses in a SUSY theory.  
The requirement that there be at least $2$  $SU(2)_L$
Higgs doublets is a feature
of all models with weak scale supersymmetry.  
\begin{itemize}
\item  
In  general,  supersymmetric extensions of the
Standard Model  have extended Higgs sectors leading to a rich
phenomenology of Higgs scalars.  
\end{itemize} 
  
It is instructive to consider the cancellation of quadratic 
divergences in the Higgs boson mass renormalization from the complete
set of particles contained in the MSSM.\cite{quaddiv}
  Now gauge bosons, gauginos,
Higgs self-interactions, Higgsinos, fermions, and sfermions all
contribute.  The cancellation of quadratic divergences in this case
uses in a fundamental manner the fact that the trace of the hypercharge
generator over the particle spectrum vanishes.\cite{drees}
In order not to have large contributions of the form of Eq. 
\ref{splitting}, the argument is usually made that the SUSY particles
must have masses below around $1~TeV$.

\subsection{Aside on $2-$ Component Notation}

Supersymmetry is most naturally formulated using
$2-$ component Majorana spinors.  However, for practical
purposes, it is usually more convenient to obtain Feynman rules for 
the fermion
 interactions in terms of $4-$ component spinors.  It is necessary,
then, to develop techniques for going back- and- forth between the
two formulations.\cite{wess}

A $4-$ component fermion, $\psi$, can be written in terms of
$2-$ component spinors, $\zeta$ and ${\overline \eta}$, as,
\beq
\psi=\left(\begin{array}{c}
\zeta\\
{\overline \eta}
\end{array}
\right)  ~.
\eeq	
For a Majorana spinor, $\psi_M$, we have $\psi_M=\psi_M^{c}$, where
$\psi_M^c$ is the charge conjugated spinor. This requires
that a Majorana fermion have the form,
\beq  
\psi_M=\left(\begin{array}{c}
\zeta\\
{\overline \zeta}
\end{array}
\right)~.
\label{majdef}
\eeq
It is most straightforward to work with fermions of definite helicity,
$\psi_{R,L}=P_\pm\psi$ with $P_\pm={1\over 2}(1\pm\gamma_5)$, 
\beq 
P_+=\left(\begin{array}{cc}
1 & 0\\
0 & 0
\end{array}\right)
, \quad 
P_-=\left(
\begin{array}{cc}
0 & 0 \\
0 & 1
\end{array}\right)                    ~~.
\eeq
We then have the following useful results for translating between
$2-$ and $4-$ component notation,
\beq
\begin{array}{ll}
{\overline \psi}_a P_- \psi_b&=   \eta_a \zeta_b \\
{\overline \psi}_a P_+\psi_b&=  {\overline \eta}_b {\overline \zeta}_a \\
{\overline \psi}_a \gamma^\mu P_- \psi_b&= 
{\overline \zeta}_a {\overline \sigma}^\mu \zeta_b \\
{\overline \psi}_a \gamma^\mu P_+\psi_b&= 
- {\overline \eta}_b {\overline \sigma}^\mu \eta_a
\quad .
\end{array} 
\label{twofour}
\eeq  
Many more results of this type can be found in Refs.~
 \citenum{hkrep} and \citenum{wess}.
The results of Eq. \ref{twofour} contain  suppressed 
$\epsilon$ tensors,
\beq
\zeta\eta\equiv \zeta^\alpha\eta_\alpha=
\zeta^\alpha \epsilon_{\alpha\beta}\eta^\beta
=-\eta^\beta\epsilon_{\alpha \beta}\zeta_\alpha=
\eta\zeta
\eeq
with $\epsilon_{12}=-\epsilon^{21}=\epsilon^{12}=-\epsilon_{21}=1$,
$\epsilon_{11}=\epsilon_{22}=\epsilon^{11}=\epsilon^{22}=0$.  We also
have,
\beq
{\overline \zeta}{\overline \sigma}^\mu \eta=- \eta \sigma^\mu
{\overline \zeta}   \quad .
\label{switch}
\eeq
The minus sign of Eq. \ref{switch} is vital for deriving the correct
Feynman rules for Majorana particles.

As an example, consider the Lagrangian for a $4-$ component Dirac
fermion,
\beq
{\cal L}= \overline \psi\biggl(
i\partial_\mu\gamma^\mu -m\biggr)  \psi 
,
\eeq
where we work in the basis,
\beq
\gamma^\mu=
\left(
\begin{array}{ll}
0&\sigma^\mu\\
{\overline \sigma}^\mu & 0
\end{array}
\right)
\quad .
\eeq
Using Eq. \ref{twofour}, the Dirac Lagrangian therefore
becomes in $2-$ component notation,
\beq
{\cal L}=
-i {\overline \zeta} {\overline \sigma}^\mu\partial_\mu\zeta
-i {\overline \eta} {\overline \sigma}^\mu\partial_\mu\eta
-m (\eta\zeta +{\overline \eta}{\overline\zeta}
)                                              
  \quad .
\eeq 

Another interesting application of Eq. \ref{twofour}
 is to consider the coupling
of two gluinos to a gluon.  From gauge invariance, 
the coupling must have the form
\beq
{\cal L}=f_{ijk}{\overline{\tilde g}}^i\gamma^\mu (a+b\gamma_5){\tilde g}^j
G_\mu^k,
\label{glucoups}
\eeq  
where ${\tilde g}^i,~i=1,.,8$, is the 
color octet Majorana gluino (in 4-component notation), 
\beq
{\tilde g}^i=\left(
\begin{array}{cc}
\zeta^i
\\ 
{\overline \zeta}^i
\end{array}
\right)
.
\eeq 
and $f_{ijk}$
is the anti-symmetric $SU(3)$ tensor.               
Consider the axial vector piece of Eq. \ref{glucoups}:
\beq
\begin{array}
{ll} 
f_{ijk}{\overline{\tilde g}}^i \gamma^\mu\gamma_5
{\tilde g}^j G_\mu^k  &=f_{ijk}\biggl[
{\overline{\tilde g}}^i \gamma^\mu P_+ {\tilde g}^j
-{\tilde g}^i \gamma^\mu P_- {\tilde g}^j G_\mu^k\biggr]  \\
 &=f_{ijk} \biggl[ - {\overline \zeta}^j {\overline \sigma}^\mu
\zeta^i - {\overline \zeta}^i {\overline \sigma}^\mu \zeta^j
\biggl] G_\mu^k
\\
 &=0,
\end{array}
\eeq  
where the last line follows from the anti-symmetry of $f_{ijk}$.
Hence the fact that the gluino is a Majorana particle has a physical
consequence:  it can only have a vector coupling to the gluon.

\subsection{The Interactions of the MSSM} 
Having specified the superfields of the theory, the next
step is to construct the supersymmetric Lagrangian.\cite{early}  There
is very little freedom in the allowed interactions between
the ordinary particles and their supersymmetric partners.
It is this feature of a SUSY model which gives it predictive
power (and makes it attractive to theorists!). 
It is important to note here, however,
 that there is nothing to stop us from
adding more superfields to those shown in Tables 1 and 2 as
long as we are careful to add them in such a way that 
any  additional contributions to gauge anomalies cancel
among themselves.  The MSSM which we concentrate on,
 however, contains only those fields given in the tables.    

The supersymmetry  associates  each $2$-component Majorana fermion
with a complex scalar.  The massive
fermions of the Standard Model are, however,
Dirac fermions
and we will use the more familiar $4$- component
notation when writing the fermion interactions.\cite{hkrep}   
The fields of the MSSM all have canonical kinetic energies:
\begin{eqnarray} 
{\cal L}_{KE}&=&\sum_i\biggl\{ (D_\mu S_i^*)(D^\mu S_i)
+i{\overline \psi}_i D \psi_i\biggr\}\nonumber \\
&&+\sum_A\biggl\{ -{1\over 4} F_{\mu\nu}^A F^{\mu\nu A}
+{i\over 2} {\overline \lambda_A} D \lambda_A
\biggr\}
, 
\label{lkin}
\eeqn 
where $D$ is the $SU(3)\times SU(2)_L\times U(1)_Y $
gauge invariant derivative.  
  The $\sum_i$ is over all the fermion fields of the Standard
Model, $\psi_i$, and their scalar partners, $S_i$,    
 and also over the $2$ Higgs
doublets with their fermion partners as given in Table 1.
  The $\sum_A$  is over
the $SU(3)$, $SU(2)_L$ and $U(1)_Y$ gauge fields with their
fermion partners, the gauginos, $\lambda_A$.  
  
The interactions between the chiral superfields of Table 1 
  and the
gauginos  and the gauge fields
 of Table 2
are completely specified by the gauge symmetries
and by the supersymmetry, as are the quartic interactions of
the scalars,                
\beq
{\cal L}_{int}=-\sqrt{2}\sum_{i,A}
 g_A\biggl[S_i^* T^A {\overline \psi}_{iL}
\lambda_A +{\rm h.c.}\biggr] -{1\over 2} 
\sum_A \biggl( \sum_i g_A S_i^* T^A S_i\biggr)^2
\quad ,  
\label{scalints}  
\eeq       
where 
$g_A$ is the relevant gauge
coupling constant. We see that the interaction strengths
are given in terms of the gauge couplings.
{\bf There are no adjustable parameters.}
 For example, the
interaction between a quark, its scalar partner, the squark,  
 and the gluino is governed
 by the strong coupling
constant, $g_s$. A  complete set of Feynman rules for the minimal
SUSY model described here is given in the review by Haber
and Kane.\cite{hkrep}  A good rule of thumb is to take an interaction
involving Standard Model particles
 and replace two of the particles
by their SUSY partners to get an approximate strength for
the  interaction.

The only freedom in constructing the supersymmetric Lagrangian
(once the superfields and the gauge symmetries are chosen)
is contained in a function called the ${\it {\bf superpotential}}$,
$W$.
The superpotential is a function of the chiral superfields
of Table 1 only
(it is not allowed to contain their complex conjugates) and it
contains terms with $2$ and $3$ chiral superfields.  Terms in
the superpotential with more than $3$ chiral superfields would
yield non-renormalizable interactions in the Lagrangian.
The superpotential  is an analytic function of the superfields
and thus is not allowed to contain derivative interactions.
  From the superpotential can be found both the scalar potential and
the Yukawa interactions of the fermions with the scalars:
\beq 
{\cal L}_{W}=-\sum_i \mid {\partial W\over
\partial z_i}\mid ^2 -{1\over 2}\sum_{ij}
\biggl[ {\overline \psi}_{iL} {\partial^2 W
\over \partial z_i \partial z_j}\psi_j+{\rm
h.c.}\biggr]
, 
\label{lagw} 
\eeq
where $z_i$ is a chiral superfield.
To obtain the interactions, we take the derivatives of $W$ with
respect to the superfields, $z_i$, and then evaluate the result in 
terms of the scalar components of $z_i$, $\phi_i$.   
This form of the Lagrangian is dictated by the supersymmetry
and by the requirement that it be renormalizable.  An explicit
derivation of Eq. \ref{lagw} can be found in Ref.~\citenum{wess}.
  
The usual approach is to write the most general
$SU(3)\times SU(2)_L\times U(1)_Y$ invariant 
superpotential with arbitrary coefficients for the interactions,  
\begin{eqnarray}
W &=& -\epsilon_{ij} \mu {\hat H}_1 ^i {\hat H}_2^j 
+\epsilon_{ij}
\biggl[ \lambda_L  {\hat H}_{1 }^i {\hat L}^{cj}{\hat E}^c +
\lambda_D  {\hat H}_1^i {\hat Q}^j {\hat D}^c
+\lambda_U  {\hat H}_2^j {\hat Q}^i {\hat U}^c\biggr]
\nonumber \\
&& + \epsilon_{ij}\biggl[ {\lambda_1} {\hat L}^i 
{\hat L}^j {\hat E}^c +
\lambda_2 {\hat L}^i {\hat Q}^j {\hat D}^c
\biggr] 
+\lambda_3 {\hat U}^c {\hat D}^c {\hat D}^c
, 
\label{superpot} 
\eeqn  
(where  $i,j$ are $SU(2)$ indices which are typically not written
explicitely).
In principle, a bi-linear term $\epsilon_{ij}
{\hat L}^i{\hat H_2}^j$ could also be included in the
superpotential since ${\hat L}$ and ${\hat H_1}$ have the
same gauge and Lorentz quantum numbers. 
 It is possible, however, to rotate the
lepton field, ${\hat L}$, such that this term vanishes.
It can be reintroduced through renormalization group effects
if the  rotation is performed at the GUT scale, but
these effects are small except for neutrino masses and so
we will ignore them.\cite{hz}
We have written the superpotential in terms of the fields of the first
generation.  In principle, the $\lambda_i$
 could all be  matrices which
mix the interactions of the $3$ generations.

The $\mu {\hat H}_1 {\hat H}_2$ term in the superpotential
gives mass terms for the
Higgs bosons when we apply $\mid \partial W/\partial z\mid^2$
 and so $\mu$ is often called the Higgs mass
parameter.   
We shall see later that the physics
is very sensitive to the sign of $\mu$.  The
terms in the square brackets proportional
to $\lambda_L$, $\lambda_D$, and $\lambda_U$  give the usual
Yukawa interactions of the fermions with the
Higgs bosons from the term
${\overline \psi}_i (\partial^2 W/\partial z_i 
\partial z_j) \psi_j$.
 Hence   these coefficients are
determined  in terms of the fermion masses and
the 
vacuum expectation values of the neutral
members of the scalar components of the Higgs doublets and are not
free parameters at all.

The Lagrangians of Eqs. \ref{lkin}, \ref{scalints}
and \ref{lagw} cannot, however, be the
whole story as all the particles (fermions, scalars,
gauge fields) are massless at this point.
 
\subsection{R Parity}  
The terms in the second line of Eq. \ref{superpot} (proportional
to $\lambda_1, \lambda_2$ and $\lambda_3$) are
a problem.  They contribute to  lepton and
baryon number violating interactions and
can mediate proton decay at tree level through the exchange
of the scalar partner of the down quark. 
If the SUSY partners of the Standard Model 
particles have masses on the TeV scale, then
these interactions are severely restricted by 
experimental measurements.\cite{proton}     
We write the R-parity violating contributions
to the superpotential as
\beq
W_{RP}=  \lambda_1^{\alpha\beta\gamma} {\hat L}^\alpha 
{\hat L}^\beta {\hat E}^{c\gamma} +
\lambda_2^{\alpha\beta\gamma}
 {\hat L}^\alpha {\hat Q}^\beta {\hat D}^{c\gamma}
+\lambda_3^{\alpha\beta\gamma}
 {\hat U}^{c\alpha} {\hat D}^{c\beta} {\hat D}^{c\gamma}
, 
\label{rplag}
\eeq
where the indices $\alpha,\beta,\gamma$
 label the $3$ generations of quarks
and leptons.

There are several possible approaches to the
problem of the lepton and baryon number violating interactions.
The first is simply to make the coefficients, $\lambda_1,
\lambda_2$, and $\lambda_3$ 
small enough to avoid experimental limits.\cite{sher}   This artificial
tuning of parameters is regarded as unacceptable  by 
many theorists, but is certainly
allowed experimentally.  Another tactic is to
make either  the lepton number violating
interactions, $\lambda_1$ and $\lambda_2$, or
the baryon number violating interaction,  $\lambda_3$, zero,
(while allowing the others to be non-zero) which
would  forbid proton decay.
The experimental limit on proton decay involves the couplings
to the first generation,\cite{goity}
\beq
\mid \lambda_2^{11\alpha}\lambda_3^{11\alpha}
 \mid < 10^{-27}\biggl(
{M_{\tilde d_\alpha}\over 100~GeV}\biggr)^2
.
\eeq  
Many other possible interactions involving
the  $\lambda_2^{\alpha\beta\gamma}$ are forbidden
by low energy interactions.  For example, $\nu$-less double
beta decay requires,\cite{dreiner}
\beq
\mid \lambda_2^{111}\mid < 7\times 10^{-3} \biggl({M_{\tilde q}
\over 200~GeV}\biggr)\biggl({M_{\tilde g}\over 1~TeV}\biggr)^{1/2}
\quad .
\eeq 
Other processes, such as $\mu\rightarrow e \gamma$ restrict 
different combinations
of R parity violating operators.  A review is given in 
Ref.~\citenum{dreiner}.
One could simply fine tune these couplings to be small. 
There is, however, not much theoretical support for this
approach since one of the motivations for introducing supersymmetry is
to avoid  the fine tuning of couplings.

  The usual strategy   is to require
that  all of these undesirable  lepton and baryon number
violating terms be forbidden by a symmetry.
(If they are forbidden by a symmetry, they will not
re-appear at higher orders of perturbation theory.)  
The symmetry which does the job is 
called ${\it\bf R~parity}$.\cite{hz,rp}
R parity can be defined as a discrete $Z_2$ symmetry under
which the anti-commuting variable
$\theta\rightarrow -\theta$.  Eq. \ref{thetadef}
makes it clear
that a scalar and its fermionic partner transform 
oppositely under R parity.  If we define the superfields
to transform under R parity such that
\beqn
({\hat Q}, {\hat U}^c, {\hat D}^c, {\hat L}, {\hat E}^c)&
 \longrightarrow
&
- ({\hat Q}, {\hat U}^c, {\hat D}^c, {\hat L}, {\hat E}^c)
\nonumber \\
({\hat H}_1, {\hat H}_2)&\longrightarrow & ({\hat H}_1, {\hat H}_2)
\eeqn
then it is clear that the terms of Eq. \ref{rplag} are
forbidden.    
R parity is thus a multiplicative quantum number
such that all particles of the Standard Model have R parity
+1, while their SUSY partners have R parity -1.
R parity can also be defined as,
\beq
R\equiv (-1)^{3(B-L)+s}
\quad ,
\eeq
for a particle of spin $s$.     
 It is worth
noting that in the Standard Model, this problem
of baryon and lepton number violating interactions  does not
arise, since such interactions are 
forbidden by the gauge symmetries  to contribute
to dimension- $4$ operators
and first arise in dimension- $6$ operators which are
suppressed by factors of some heavy mass scale.

The assumption of R parity conservation has profound
experimental consequences which go beyond the details
of a specific model.  Because R parity is a multiplicative
quantum number, it implies that the number of SUSY partners
in a given interaction is always conserved modulo 2.  
\begin{itemize}
\item
 SUSY
partners can only be pair produced from  Standard Model
particles.  
\end{itemize}
Furthermore, a SUSY particle will decay in a chain until
the lightest SUSY particle is produced (such a decay is
called a ${\it cascade~decay}$).  This lightest SUSY
particle, called the LSP, must be absolutely stable when R
parity is conserved.
\begin{itemize}
\item
A theory with $R$ parity conservation will have a 
lightest SUSY particle (LSP) which is stable.  
\end{itemize}
 The LSP must be neutral since there
are stringent cosmological bounds on  light
charged or colored
particles which are stable.\cite{pdg}
If $R$ parity is violated then it is possible
for some other particle (such as the gluino) to
be the LSP.\cite{farr}  Hence the LSP is
stable and  neutral and
is not seen in a detector (much like a neutrino)
since it interacts only by the exchange of a heavy virtual SUSY
particle.  
\begin{itemize}
\item  The LSP will interact very weakly with
	ordinary  matter.  
\item  
A generic signal for R parity conserving SUSY theories 
is missing transverse energy from the non-observed LSP.  
\end{itemize}  
In theories without $R$ parity conservation, there will
not be a stable  LSP, and the lightest SUSY
particle will decay into ordinary particles (possibly
within the detector).   Missing transverse energy 
will no 
longer be a robust signature
 for SUSY particle production.\cite{baer1,snow}  

The assumption of R-parity conservation in SUSY models has
been under attack because of some recent experimental results.
The HERA experiments see an excess of events at large
$Q^2$, ($Q^2> 1.5\times 10^4 GeV^2$), in $e^+p$ deep inelastic
scattering events, although the statistics are poor.\cite{hera}  This
excess is difficult to explain within the context of 
the Standard Model.  One possibility is that these events
are a signal for SUSY models with R-parity violating
interactions.  An $R$-parity
violating  superpotential could introduce interactions of the
form,
\beq
e^+ d\rightarrow {\tilde u}_L, {\tilde c}_L, {\tilde t}_L
\quad ,
\label{rsusy}
\eeq
along with interactions involving sea quarks.  
The excess events at HERA could be interpreted as
resonant squark production through the $\lambda_2$ interaction
of Eq. \ref{rplag}.  The kinematics and number of events would then
give restrictions of the mass of the exchanged squark and the
relevant $\lambda_2^{\alpha\beta\gamma}$ operator.\cite{dreiner}

To summarize, R parity violating theories have a number of features
which are different from those of the Standard Model:
\begin{itemize}
\item
Baryon and/or lepton number is violated in some interactions.
\item
The SUSY partners of ordinary particles can be singly produced.
\item
The LSP is not stable.  Hence the missing $E_T$ signature
for supersymmetry is degraded.
\item
The LSP need not be the neutralino, since the LSP is no longer stable. 
\end{itemize}

\subsection{Supersymmetry Breaking}  

The mechanism for supersymmetry breaking is not well understood.
At this point we have 
constructed a SUSY theory containing all of the
Standard Model particles, but the supersymmetry
remains unbroken and the particles and their SUSY partners 
are massless. This is clearly unacceptable.  
Suppose we try to break the supersymmetry spontaneously, by
giving some set of scalars VEVs.  The difficulty in doing
this arises from a fundamental problem.  Just as in
our simple example of Eq. \ref{suslag}, the scalar potential of
the MSSM is positive definite,
\beq
V(\phi_i,\phi_i^*)=\sum_{i}\mid {\partial W\over \partial z_i}
\mid^2
+{1\over 2} \sum_{A,i,j} g_A^2 \mid
\phi_i^* T^A_{ij} \phi_j\mid^2 \ge 0
\quad .
\label{vpot}
\eeq
Clearly the second term, ${1\over 2} g_A^2 \mid
\phi_i^* T^A_{ij} \phi_j\mid^2 $ is minimized
 if $\langle \phi_i\rangle=0$. 
In order to spontaneously break the supersymmetry, we need a non-
trivial contribution from the first term.  Such a contribution
will contribute to the mass matrices and a spontaneouly broken
SUSY theory satisfies a mass-squared sum rule:\cite{hkrep} 
\beq S Tr M^2 \equiv
3 Tr M_V^2 -2 Tr M_F^2 +Tr M_S^2 =0.
\eeq
This sum rule holds independent of the values of 
the scalar fields.  Satisfying it is problematic, since we 
want ${\it all}$ SUSY particles to be heavier than their
Standard Model partners.  This sum rule has caused most
theorists  to give up on models with spontaneous supersymmetry
breaking.

 At the
moment the usual approach is  to assume that the MSSM, which
is the theory at the electroweak scale, is an effective
low energy theory.\cite{wein} 
It is typically
assumed that the SUSY breaking occurs
 at a high scale, say $M_{pl}$, and perhaps
results from some complete theory encompassing gravity. 
The supersymmetry breaking is implemented by including  explicit
``soft''
 mass terms for the scalar members of the
chiral multiplets and for the gaugino members of the vector
supermultiplets in the
Lagrangian.
  These interactions are termed soft because
they do not re-introduce the quadratic divergences which motivated
the introduction of the supersymmetry in the first place.  
The dimension of soft operators in the Lagrangian must be
$3$ or less, which means that the possible
 soft operators are mass terms,
bi-linear mixing terms (``B'' terms), and 
tri-linear scalar mixing terms (`` A terms'').  
The origin of these supersymmetry breaking terms is left
unspecified.  
The complete set of soft SUSY breaking terms
(which respect R parity and the $SU(3)\times SU(2)_L\times U(1)_Y$
gauge symmetry)
for the first generation  
 is given by the Lagrangian:\cite{early,soft} 
\beqn
-{\cal L}_{soft}&=&
m_1^2 \mid H_1\mid^2 +m_2^2 \mid H_2\mid^2 - B \mu 
\epsilon_{ij} (H_1^i H_2^j + {\rm h.c.}) 
+M_{\tilde Q}^2 ({\tilde u_L}^* {\tilde u_L}
+{\tilde d_L}^* {\tilde d_L})
\nonumber \\  && +M_{\tilde u}^2 
{\tilde u_R}^* {\tilde u_R} +M_{\tilde d}^2
{\tilde d_R}^* {\tilde d_R} 
+
M_{\tilde L}^2({\tilde e_L}^*{\tilde e_L}
+{\tilde \nu_L}^*{\tilde \nu}_L)
+M_{\tilde e}^2 {\tilde e}_R^*{\tilde e_R}
\nonumber \\  &&
+{1\over 2}\biggl[ M_3 {\overline {\tilde g}} {\tilde g}
+M_2 {\overline {\tilde \omega_i}}{\tilde \omega_i}
+M_1{\overline {\tilde b}}{\tilde b}
\biggr] 
+{g\over \sqrt{2}M_W}\epsilon_{ij}
\biggl[ {M_d\over \cos\beta}A_dH_1^i
{\tilde Q}^j {\tilde d_R}^*
\nonumber \\  && +
{M_u\over \sin\beta}A_u H_2^j {\tilde Q}^i
{\tilde u_R}^*   
+{M_e\over \cos\beta}A_e H_1^i{\tilde L}^j {\tilde e}_R^*
+{\rm h.c.} \biggr]
\quad .
\label{lagsoft} 
\eeqn
This Lagrangian has arbitrary masses for the scalars and
gauginos and also arbitrary tri-linear and bi-linear
mixing terms.  
The scalar and gaugino mass terms 
have the desired effect of breaking the degeneracy 
between the particles and their SUSY partners.
The tri-linear A-terms have been defined with an
explicit factor of mass and we will see later that they
affect primarily the particles of the third generation.\footnote{
We have also included an angle $\beta$ in the normalization of the $A$
terms.  The factor
$\beta$ is related to the vacuum expectation values
of the neutral components of the Higgs fields and is defined in
the next section.  The normalization is, of course, arbitrary
and the normalization we have chosen reflects theoretical
prejudices.}    
When the $A_i$  terms are non-zero, the scalar partners of the
left- and right-handed fermions can mix
when the Higgs bosons get vacuum
expectation values  and so they  are no longer mass
eigenstates.  The $B$ term mixes the scalar components of the $2$  
Higgs doublets.   

The philosophy is to add all of the mass and mixing terms
which are allowed by the gauge symmetries.  
To further complicate matters, all of the mass  and interaction terms 
 of Eq. \ref{lagsoft} 
 may be  matrices involving all three generations.  
${\cal L}_{soft}$ has clearly broken the supersymmetry
since the SUSY partners of the ordinary particles have
been given arbitrary masses.  This has come at the
tremendous
expense, however, of introducing a large number of
unknown parameters (Haber \cite{hnew} has called this model
the MSSM-124 to emphasize the number of free parameters!). 
 It is one of the wonderful features
of supersymmetry that even with all these new parameters,
the theory is still able to make some definitive predictions.  
This is, of course, because the gauge interactions of the
SUSY particles are completely fixed.  

We have now constructed the Lagrangian describing
a softly broken supersymmetric theory which is
assumed to be the effective theory at the weak scale.
In the next
section we will examine how the electroweak symmetry
is broken in this model and study the mass spectrum and
interactions of the new particles.  

\subsection{The Higgs Sector and Electroweak Symmetry Breaking}
\subsubsection{The Higgs Potential}
The Higgs sector of the MSSM is very similar to that of
a general $2$ Higgs doublet model
and  can be constructed from Eq. \ref{vpot}. The first 
contribution to the Higgs potential is called
the ``D'' term, $V_D$,
\beqn
V_D=& \sum_A{1\over 2} D^A D^A\nonumber \\
D^A \equiv & -g_A \phi_i^*T^A_{ij}\phi_j \quad ,
\eeqn
is called the ``D''-term.  For $H_1 (H_2)$, we have
$Y=-{1\over 2} ({1\over 2})$ and
so the D-terms are given by,
\beqn
U(1)_Y:\quad D^1=& - {g^\prime\over 2} \biggl(
\mid H_2\mid^2-\mid H_1\mid^2\biggr)
\nonumber \\
SU(2)_L:\quad D^a=& - {g\over 2}\biggl(H_1^{i*} \tau^a_{ij}H_1^j
    +H_2^{i*} \tau^a_{ij}H_2^j\biggr),
\eeqn
(where $T^a={\tau^a\over 2}$).  The D terms then contribute
to the scalar potential:
\beq
V_D={g^{\prime 2} \over 8}\biggl(\mid H_2\mid^2 -\mid H_1\mid^2\biggr)^2
    +{g^2\over 8}\biggl( H_1^{i*}
\tau_{ij}^a H_1^j+ H_2^{i*}\tau^a_{ij}H_2^j\biggr)^2
\quad .
\eeq   
Using the $SU(2)$ identity,
\beq
\tau_{ij}^a\tau_{kl}^a=2 \delta_{il}\delta_{jk}-\delta_{ij}\delta_{kl}
\eeq
we find,
\beqn
V_D&=&{g^2\over 8}\biggl\{ 4\mid H_1^*\cdot H_2\mid^2
-2 (H_1^*\cdot H_2)(H_2^*\cdot H_2)+
\biggl(\mid H_1\mid^2+\mid H_2\mid^2\biggr)^2\biggr\}                                           
\nonumber \\
&& +{g^{\prime 2}\over 8}\biggl( \mid H_2\mid^2-\mid H_1\mid^2\biggr)^2
.
\eeqn
Adding the remaining contribution to the potential, the
``F''-term, 
\beq
V_F=\sum_i\mid {\partial W\over \partial z_i}\mid^2
,
\eeq
we  find the supersymmetric potential,
\beq
V=\mid \mu\mid^2\biggl(\mid H_1\mid^2+\mid H_2\mid^2\biggr)
+{g^2+g^{\prime 2}\over 8}\biggl(\mid H_2\mid^2
-\mid H_1\mid^2\biggr)^2+{g^2\over 2}\mid H_1^*\cdot H_2\mid^2
 .
\eeq
This potential has its minimum at $\langle H_1^0\rangle=
\langle H_2^0\rangle=0$, giving $\langle V\rangle=0$ and
so represents a model with no electroweak symmetry breaking
and no SUSY breaking.
Adding all possible soft SUSY breaking terms
as in Eq. \ref{lagsoft}, we find
  the scalar potential
involving the Higgs bosons,  
\beqn
V_H&=&
\biggl(\mid \mu\mid^2 +m_1^2\biggr)\mid H_1\mid^2
+\biggl(\mid \mu \mid^2+m_2^2\biggr)\mid H_2\mid^2
-\mu B \epsilon_{ij}\biggl(H_1^i H_2^j+{\rm h.c.} \biggr)
\nonumber \\
&&
+{g^2+g^{\prime 2}\over 8}\biggl(
\mid H_1\mid^2 - \mid H_2\mid^2\biggr)^2
+{1\over 2} g^2 \mid H_1^*H_2\mid^2
\quad .
\label{higgspot} 
\eeqn 
The Higgs potential of the SUSY model can be seen
to depend on $3$ independent combinations of parameters, 
\beqn 
&& \mid\mu\mid^2+m_1^2,
\nonumber \\
&& 
\mid \mu \mid^2+m_2^2,
\nonumber \\
&&~~~  \mu B~~, 
\label{higgspar}
\eeqn   
where $B$ is a new mass parameter.  
This is in 
contrast to the general $2$ Higgs doublet model where 
there are $6$ arbitrary coupling constants (and a phase)
 in the potential.  
From Eq. \ref{scalints}, it is clear that the quartic couplings
are fixed in terms of the gauge couplings and so they are
not free parameters.  
This leaves only the mass terms of Eq. \ref{higgspot} unspecified. 
Note that $V_H$ automatically conserves CP since
any complex phase in $\mu B$ can be absorbed into the
definitions of the Higgs fields.    
  
Clearly, if $\mu B=0$ then all the terms in the potential
are positive and the minimum  of the potential
occurs with $V=0$ and $\langle H_1^0\rangle=\langle H_2^0
\rangle=0$.
  Hence all $3$ parameters of Eq. \ref{higgspar}
must be non-zero in order for the electroweak symmetry to be
broken.  
\footnote{
We assume that the parameters are arranged in such
a way that the scalar partners of the quarks and leptons
do not obtain vacuum expectation values.  Such vacuum
expectation values would spontaneously break the $SU(3)$ 
color  gauge symmetry or lepton number.
This requirement that the proper vacuum be
chosen  gives a restriction on
\protect $ A_i/{\tilde M}$, where 
\protect 
${\tilde M}$ is a generic squark or slepton mass.}
                                                      
In order for the electroweak symmetry to be broken
and for the potential to be stable at large values
of the fields, the
parameters must satisfy the relations,
\beqn
(\mu B)^2 & >&\biggl(\mid \mu\mid^2+m_1^2\biggr)
\biggl(\mid \mu\mid^2+m_2^2\biggr)
\nonumber \\
\mid \mu\mid^2+{m_1^2+m_2^2\over 2}& >& \mid \mu B\mid
\quad .
\eeqn
We will assume that these conditions are  met.  
The symmetry is broken when the neutral components of the Higgs doublets
get vacuum expectation values,
\beqn
\langle H_1^0\rangle & \equiv & v_1
\nonumber \\
\langle H_2^0\rangle & \equiv & v_2
\quad . 
\eeqn  
By redefining the Higgs fields, we can always 
choose $v_1$ and $v_2$ positive.

\begin{figure}[tb]
\vspace*{.25in}  
\centerline{\epsfig{file=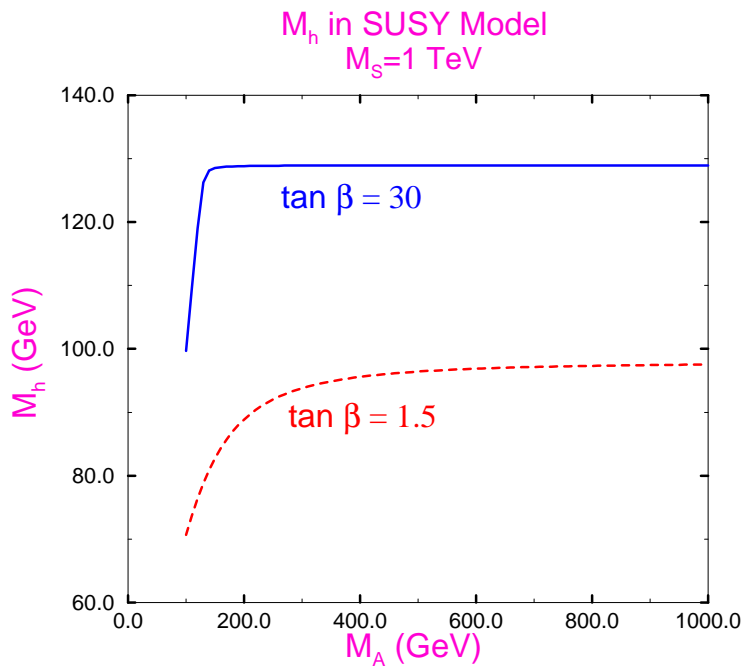,height=3.5in}}
\caption{Mass of the lightest neutral Higgs boson as
a function of the pseudoscalar mass, $M_A$,
and $\tan\beta$. This figure includes one-loop
radiative corrections
to the Higgs mass as in Eq. 70,  assumes a common scalar mass of
$1~TeV$, and neglects mixing effects, ($A_i=\mu=0$).}
\vspace*{.5in}
\label{mlight}  
\end{figure} 

In the MSSM, the Higgs mechanism works in the same manner
as in the Standard Model.  
When the electroweak symmetry is broken,
the $W$ gauge boson gets a mass which
is fixed by $v_1$ and $v_2$,
\beq
M_W^2={g^2\over 2}(v_1^2+v_2^2)\quad .\eeq
Before the symmetry was broken, the $2$ complex $SU(2)_L$ 
Higgs doublets had $8$ degrees of freedom.  Three of
these were  absorbed to give the $W$ and $Z$ gauge bosons
their masses, leaving $5$ physical degrees of freedom.
A charged Higgs boson, $H^\pm$, a CP -odd neutral
Higgs boson, $A$, and $2$ CP-even neutral Higgs bosons, $h$ and $H$
remain in the spectrum.
After fixing $v_1^2+v_2^2$ such that the $W$  boson gets the correct
mass, the Higgs sector is  then described by $2$ additional
parameters which can be chosen however you like.  The
usual choice is
\beq
\tan\beta\equiv {v_2\over v_1}\eeq
and $M_A$, the mass of the pseudoscalar Higgs boson.
Once these two parameters are given, then the  masses of
the remaining Higgs bosons can be calculated in terms
of $M_A$ and $\tan\beta$.
Note that we can chose $0 \le \beta\le {\pi\over 2}$ since we have
chosen $v_1, v_2 > 0$.

\begin{figure}[tb]
\vspace*{.25in} 
\centerline{\epsfig{file=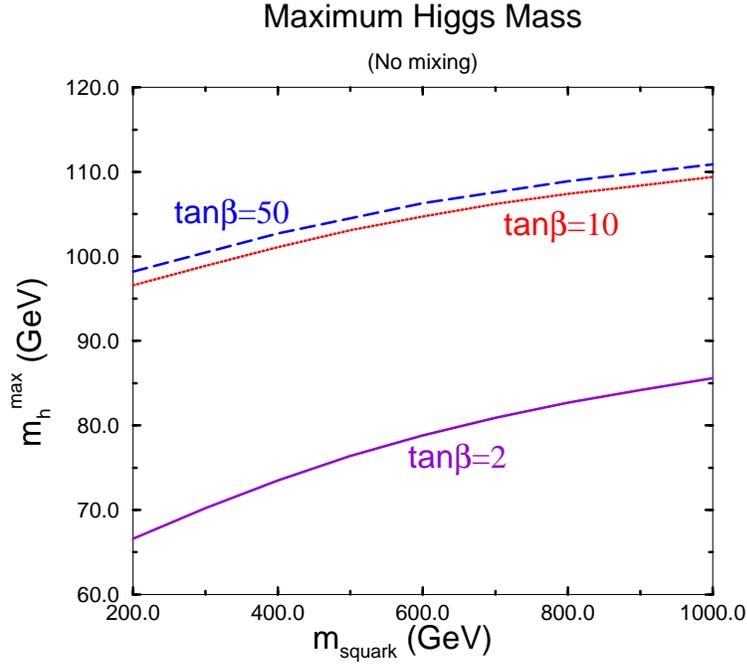,height=3.5in}}
\caption{Maximum value of the lightest Higgs boson  mass as a
function of the  squark mass.
This figure includes 2-loop radiative corrections and renormalization
group improvements.
(We have assumed degenerate squarks and set  the
mixing parameters $A_i=\mu=0$.)}  
\vspace*{.5in}
\label{mhmaxfig}
\end{figure}

\subsubsection{Higgs Boson Masses}
It is straightforward to find the physical Higgs bosons 
and their masses in terms of the parameters of 
Eq. \ref{higgspot}.  Details can be found in Ref.~\citenum{hhg}.  
The neutral Higgs masses are found by diagonalizing
the $2\times 2$ Higgs mass matrix and    
by convention, $h$ is taken to be the lighter of the neutral
Higgs bosons.   At tree level, the masses of the neutral Higgs
bosons are given by,
\beq
M_{h,H}^2={1\over 2}\biggl\{
M_A^2 +M_Z^2 \mp \biggl( (M_A^2+M_Z^2)^2-4 M_Z^2 M_A^2\cos^2
2 \beta \biggr)^{1/2}\biggr\}
.
\eeq
The pseudoscalar mass is given by,
\beq 
M_A^2={2 \mid \mu B \mid \over \sin 2 \beta},
\eeq
and the charged scalar mass is,
\beq
M_{H^\pm}^2=M_W^2+M_A^2
\quad .  
\eeq
We see that   at tree level \cite{hhg},  
 Eq. \ref{higgspot}  gives important predictions about
the relative masses of  the Higgs bosons,
\beqn
M_{H^+} &>& M_W \nonumber \\
M_H &>& M_Z \nonumber \\
M_h &<& M_A\nonumber \\
M_h &<& M_Z\mid \cos 2 \beta\mid 
 \quad .  
\label{higgmass}
\eeqn
These relations yield  the desirable
prediction that the lightest neutral
Higgs boson is lighter than the $Z$  boson and
so must be observable at LEPII.  However,
it was realized 
several years ago that loop corrections to the
relations of Eq. \ref{higgmass} are large.  
In fact the corrections  to $M_h^2$ grow like $G_F M_T^4$
and receive contributions   from loops with both top
quarks 
and squarks.  In a model with unbroken supersymmetry,
these contributions would cancel. Since the supersymmetry
has been broken by splitting the masses of the
fermions and their scalar partners, 
 the neutral Higgs boson masses become
at one- loop,\cite{massloop} 
\beqn
M_{h,H}^2&=&{1\over 2}\biggl\{ M_A^2+M_Z^2+{\epsilon_h\over \sin^2
\beta}\pm\biggl[
\biggl(M_A^2-M_Z^2)\cos 2 \beta +
{\epsilon_h\over \sin^2\beta}\biggr)^2
\nonumber \\  &&
+\biggl(M_A^2+M_Z^2\biggr)^2\sin^2 2 \beta\biggr]^{1/2}\biggr\}
\eeqn 
where  $\epsilon_h$ is the contribution
of the one-loop  corrections,
\beq
\epsilon_h\equiv {3 G_F\over \sqrt{2}\pi^2}M_T^4
\log \biggl( {{\tilde M}^2\over M_T^2}\biggr)
\quad .
\eeq 
We have assumed that all of  the squarks
have equal masses,  ${\tilde M}$, and have 
 neglected the smaller effects from the mixing parameters,
$A_i$ and $\mu$.  In Fig. \ref{mlight},
 we show the lightest Higgs boson
mass
as a function of the pseudoscalar mass, $M_A$,
 and for two 
values of $\tan\beta$.  
For $\tan\beta > 1$, the mass eigenvalues increase monotonically
with increasing $M_A$ and give an upper bound to the
mass of the lightest Higgs boson,
\beq
M_h^2 < M_Z^2 \cos^2 2 \beta +\epsilon_h
\quad .
\label{mhlimcor}
\eeq  
The corrections from $\epsilon_h$ are always positive and
increase the mass of the lightest neutral Higgs boson with
increasing top quark mass.  
From Fig. \ref{mlight},
 we see that $M_h$ always  obtains its maximal value for
rather modest values of the pseudoscalar mass, $M_A > 300~GeV$.  
The radiative corrections to the charged Higgs mass-squared
are proportional to $M_T^2$ and so are much smaller than
the corrections to the neutral masses.

\begin{figure}[tb]
\vspace*{.25in}
\centerline{\epsfig{file=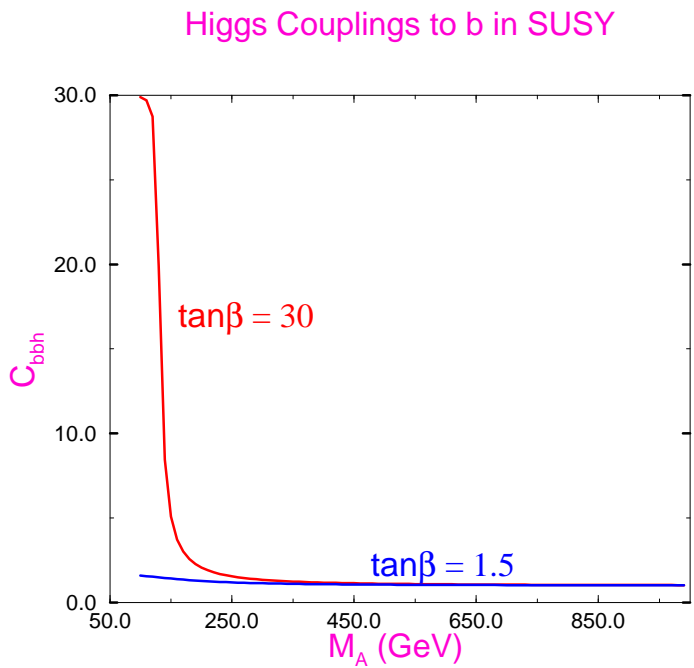,height=3.5in}}
\caption{Coupling of the lightest Higgs boson
to  charge $-1/3$  quarks including radiative
corrections [34] in terms of the couplings
defined in Eq. 77. The value  
$C_{bbh}=1$ corresponds to the Standard Model coupling
of the Higgs boson to charge $-1/3$ quarks.}
\vspace*{.5in}
\label{cbbhfig}  
\end{figure}  

There are many analyses \cite{massloop}
which include a variety of two-loop effects, renormalization
group effects,
the effects of large mixing in the squark sector,
 etc., but the important point is that for
 given values  of $\tan\beta$ and the squark masses,
 there is an upper bound
on the lightest neutral Higgs boson mass.  
The maximum value of the lightest Higgs mass is shown
in Fig. \ref{mhmaxfig} including 2-loop and renormalization group
effects and we see that there is still a 
light Higgs boson even  when radiative corrections
are included.
From Figures \ref{mlight}
 and \ref{mhmaxfig}, we see that including 2-loop effects,
SUSY particle threshold effects, and renormalization group
group improvements lowers the upper bound on the neutral Higgs
boson mass.
  For large values of $\tan\beta$
the limit is relatively insensitive to the value of
$\tan\beta$ and with  a squark mass less than about $1~TeV$,
the upper limit on the Higgs mass is about $110~GeV$ if mixing in
the top squark sector is negligible ($A_T\sim 0$).
For large mixing, this limit is raised to
around $130~GeV$.
\begin{itemize}
\item  
The minimal SUSY model predicts a neutral Higgs
boson with a mass less than around $130~GeV$.
\end{itemize}
Such a mass scale will be accessible at LEPII or the
LHC and provides a definitive test of the MSSM.   

In a more complicated SUSY model with a richer Higgs
structure, this bound will, of course, be  changed.  However,
the requirement that the Higgs self coupling remain
perturbative up to the Planck scale gives an upper
bound on the lightest SUSY Higgs boson of around  
$150~GeV$ in all models with only singlets and doublets
of Higgs bosons.\cite{quiros}
This is a very strong statement.  It implies that either
there is a relatively light Higgs boson (which would
be accessible experimentally at LEPII or the LHC) or 
else there is some new physics between the weak scale
and the Planck scale which causes the Higgs
couplings to become non-perturbative.   

As an example of the effects of adding additional fields
to the MSSM, consider including a gauge singlet superfield, ${\hat N}$,
in the superpotential,
\beq
W= ...+\lambda_N{\hat H}_1 {\hat H}_2 {\hat N} +{\kappa
\over 3} {\hat N}^3
.
\eeq
There will now be 3 neutral Higgs boson described by a
$3\times 3$ mass matrix.  A limit on the lightest Higgs boson
mass can still be obtained, however, using the
fact that the smallest value of a real symmetric $n\times n$ matrix
is smaller than the smallest eigenvalue of the upper $2\times 2$
matrix.  Eq. \ref{mhlimcor} then becomes,\cite{singlet}
\beq
M_h^2< M_Z^2 \cos^2 2\beta +\epsilon_h+{2 M_Z^2 \lambda_N^2
\over g^2+g^{\prime 2}}\sin^2 2 \beta
.
\eeq
The bound on $M_h$ therefore grows with $\lambda_N$ and it is only
by making further assumptions that a numerical bound can
be obtained.  If we require that $\lambda_N$ be perturbative up
to the Planck scale, then we have roughly the same bound as in
the MSSM.  However, if we relax this assumption and require only that
$\lambda_N$ be perturbative up to say $10^5~GeV$, then the bound
increases to up to around $150~GeV$, depending on $\tan\beta$.

\subsubsection{Higgs Boson Couplings to Fermions}  
The Higgs boson couplings to fermions
are  dictated by
the gauge invariance of the superpotential
and  at lowest order
are  completely specified in terms of the
two parameters, $M_A$ and $\tan\beta$.   
 From Eq. \ref{superpot}, we see that the charge $2/3$ quarks
get their masses entirely from $v_2$, while the
charge $-1/3$ quarks receive their masses from $v_1$.
This is a consequence of the $U(1)_Y$ hypercharge assignments
for $H_1$ and $H_2$ given in Table 1.  In the Standard Model,
it is possible to give both the up and down quarks mass using
a single Higgs doublet.  This is because
in the Standard Model 
 the up quarks can
get their masses from the charge conjugate of the Higgs 
doublet.  Terms involving the charge conjugates of the superfields
are not allowed in SUSY models, however, and so a second Higgs
doublet with opposite $U(1)_Y$ hypercharge
from the first Higgs doublet is necessary
in order  to give the up quarks mass.  
Requiring that the fermions have their observed masses fixes
the couplings in the superpotential
 of Eq. \ref{superpot},\cite{habergun} 
\beqn
\lambda_D &=&{g M_d\over \sqrt{2}M_W \cos\beta}\nonumber \\
\lambda_U&=&{g M_u\over \sqrt{2}M_W \sin\beta}\nonumber \\
\lambda_L &=&{g M_l\over \sqrt{2}M_W \cos\beta}
\quad ,   
\label{yukdefs}
\eeqn 
where $g$ is the $SU(2)_L$ gauge coupling, $g^2=4 \sqrt{2} G_F M_W^2$.  
We see that the only free parameter in the superpotential 
now is the Higgs mass parameter, $\mu$,  (along with 
the angle $\beta$
in the $\lambda_i$ couplings).

In the MSSM, the $\mu$ parameter is a source of concern. 
It cannot be set to zero (see Eq. \ref{higgspot})
 because then
there would be no symmetry breaking.   The $Z$-boson
mass can be written in terms of the radiatively corrected neutral
Higgs boson masses and $\mu$:
\beq
M_Z^2=2\biggl[ {M_h^2-M_H^2\tan^2\beta\over 
\tan^2 \beta-1}\biggr] - 2 \mu^2
.
\eeq
In order to get the observed value of $M_Z$,
a delicate cancellation between the Higgs masses
and $\mu$ is required.  This is sometimes called the
$\mu$ problem.\cite{cvetic}

It is convenient to write  
  the couplings
for the  neutral Higgs  bosons to the fermions
in terms of the Standard  Model Higgs couplings,
\beq
{\cal L}=-{g m_i\over 2 M_W} \biggl[C_{ffh}{\overline f}_i f_i h
+C_{ffH} {\overline f}_i f_i H
+C_{ffA}{\overline f}_i \gamma_5 f_i A\biggr],  
\eeq
where $C_{ffh}$ is $1$ for a Standard Model Higgs
boson.
The $C_{ffi}$ are given in Table 3 and plotted in Figs.
\ref{cbbhfig}  and \ref{ctthfig}
as a function of $M_A$. 
We see that for small $M_A$ and large $\tan\beta$, the couplings of
the neutral Higgs boson to fermions can be significantly different
from the Standard Model couplings; the $b$-quark coupling becomes
enhanced, while the $t$-quark coupling is suppressed.  
It is obvious from Figs. \ref{cbbhfig}
 and \ref{ctthfig} that when $M_A$ becomes large
the Higgs-fermion couplings approach their standard model
values, $C_{ffh}\rightarrow 1$. In fact even for $M_A\sim 300~GeV$,
the Higgs-fermion couplings are very close to their 
Standard Model values.

\begin{table}[tb]
\begin{center}
{Table 3: Higgs Boson  Couplings to fermions}\vskip6pt
\renewcommand\arraystretch{1.2}
\begin{tabular}{|lccc|}
\hline
\multicolumn{1}{|c}{$f$}& $C_{ffh}$& $C_{ffH}$
 & $C_{ffA}  $
\\
\hline
$u$   &    ${\cos\alpha\over \sin\beta}$ &
    ${\sin\alpha\over\sin\beta}$
& $\cot\beta$ \\ 
$d$   &    $-{\sin\alpha\over\cos\beta}$ & 
     ${\cos\alpha\over\cos\beta}$ 
& $\tan\beta$  \\
\hline
\end{tabular}
\end{center}
\end{table}

\begin{figure}[tb]
\vspace*{.3in}   
\centerline{\epsfig{file=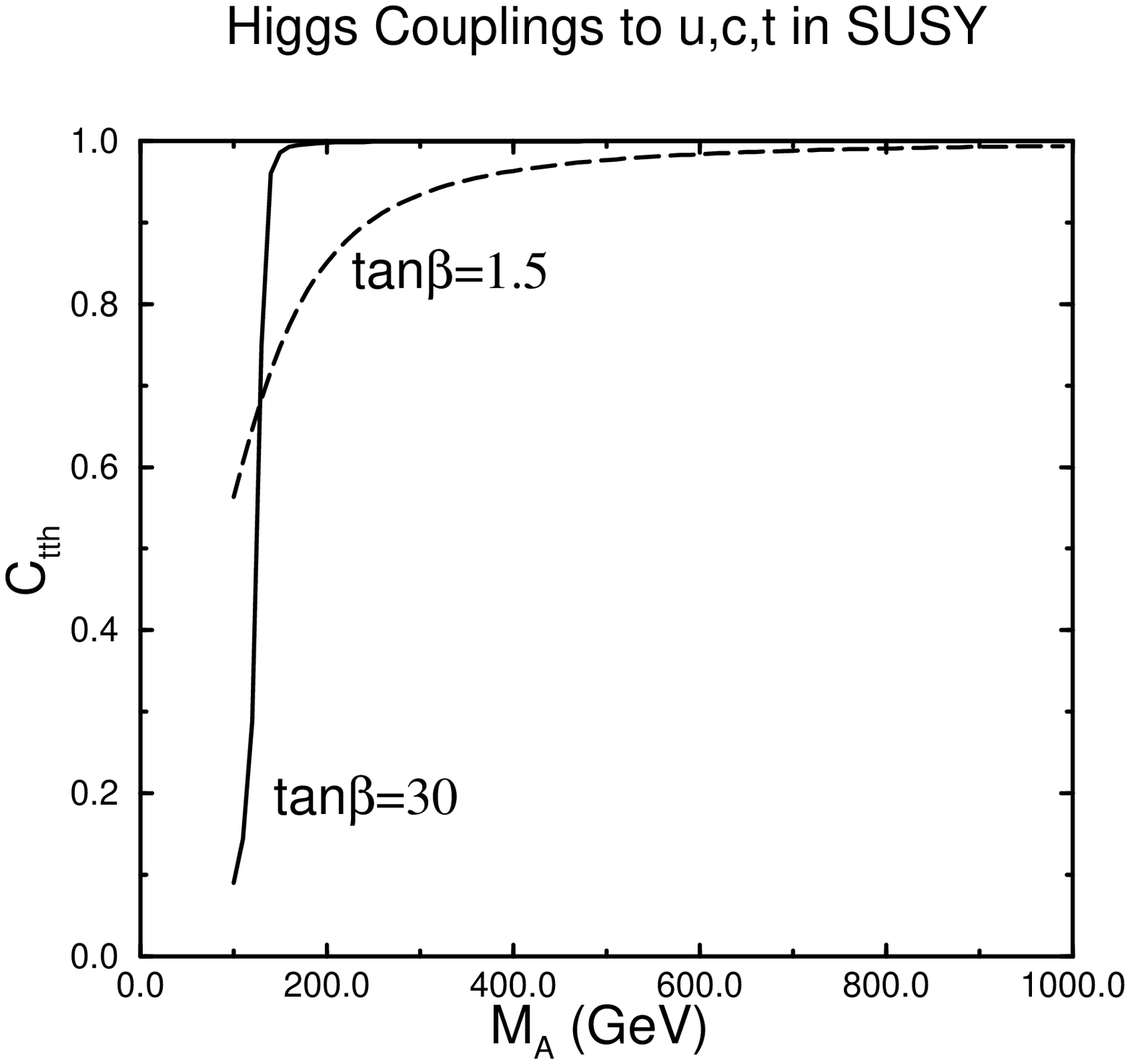,height=3.54in}}
\caption{Coupling of the lightest Higgs boson
to charge $2/3$ quarks including radiative
corrections [34]  in terms of the couplings
defined in Eq. 77.
The value  
$C_{tth}=1$ yields the Standard Model coupling of the Higgs
boson  
to charge $2/3$ quarks.}
\vspace*{.5in}
\label{ctthfig}
\end{figure}  

\begin{figure}[tb]
\vspace{.3in}
\centerline{\epsfig{file=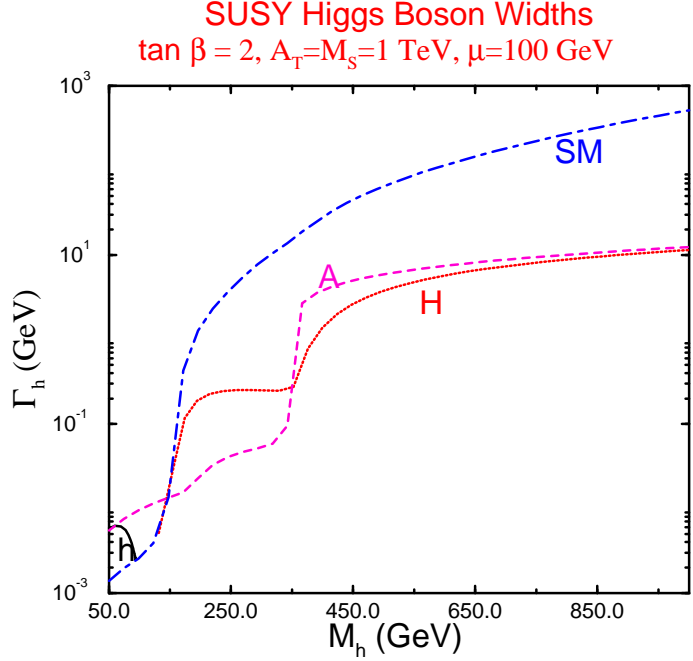,height=3.5in}}
\caption{\protect Total SUSY Higgs boson decay widths including
two-loop radiative corrections as a function of the Higgs
masses. 
The width for the Standard Model Higgs boson is shown for 
comparison.
The curve for the lightest Higgs boson is cut off
at the maximum $M_h$. The program HDECAY [34] was used to obtain
this plot.}
\vspace*{.5in}  
\label{hwidfig}
\end{figure}  

The Higgs boson couplings to gauge bosons are fixed by the
$SU(2)_L\times U(1)_Y$ gauge invariance.  
Some of the  phenomenologically important 
couplings  
 are:
\beqn
Z^\mu Z^\nu h:&&{igM_Z\over\cos\theta_W} \sin(\beta-\alpha)g^{\mu\nu}
\nonumber \\   
Z^\mu Z^\nu H:&&{igM_Z\over \cos\theta_W}\cos(\beta-\alpha) g^{\mu\nu}
\nonumber \\
W^\mu W^\nu h:&& igM_W \sin(\beta-\alpha)g^{\mu\nu}
\nonumber \\   
W^\mu W^\nu H:&& igM_W \cos(\beta-\alpha)g^{\mu\nu}
\nonumber \\
Z^\mu h(p)A(p^\prime):&&{g\cos(\beta-\alpha)\over 2 \cos\theta_W}
(p+p^\prime)^\mu\nonumber \\
Z^\mu H(p)A(p^\prime):&&-{g\sin(\beta-\alpha)\over 2 \cos\theta_W}
(p+p^\prime)^\mu
\quad . 
\label{vvhcoup}  
\eeqn  
We see that the couplings of the Higgs bosons to the gauge bosons
all depend on the same angular factor, $\beta-\alpha$. 
  The pseudoscalar, $A$, has no tree level coupling to pairs of
gauge bosons.   The angle $\beta$ is a
free parameter while the neutral Higgs mixing angle,
$\alpha$, which enters
into many of the couplings, can be found in terms of the
$M_A$ and $\beta$ 
masses:
\beq
\tan 2 \alpha={(M_A^2+M_Z^2)\sin 2\beta
\over (M_A^2-M_Z^2)\cos 2 \beta+\epsilon_h/\sin^2\beta}
\quad . 
\eeq
With our conventions, $-{\pi\over 2}\le \alpha\le 0$.  
 It is clear from Eq. \ref{vvhcoup} that 
the couplings of the SUSY Higgs bosons to gauge bosons are
always  suppressed
relative to those of the Standard Model.

  A complete set of couplings
for the Higgs bosons (including the charged and pseudoscalar Higgs)
 at tree level 
can be found in Ref.~\citenum{hhg}. 
These couplings completely determine the decay modes
of the SUSY Higgs bosons and their  experimental signatures.
 The important point is that 
(at lowest order)  all of the couplings are completely determined
in terms of $M_A$ and $\tan\beta$.  When radiative corrections are
included there is a   dependence on the squark masses and 
the mixing parameters.
This dependence is explored
in detail in Ref.~\citenum{lephiggs}.

It is an important feature  of the MSSM  that for large $M_A$,
the Higgs sector looks exactly  like that of the Standard Model.  As
$M_A\rightarrow \infty$, the masses of the charged Higgs
bosons, $H^\pm$,
and the heavier neutral Higgs, $H$, become large leaving
only the lighter Higgs boson, $h$, in the spectrum.  In this limit,
the couplings of the lighter Higgs
boson, $h$, to fermions and gauge bosons take
on their Standard Model values. We have, 
\beqn
\sin(\beta-\alpha)&& 
\rightarrow 1~ {\hbox {for}}~M_A\rightarrow \infty
\nonumber \\ 
\cos(\beta-\alpha)&&\rightarrow  0 
\quad . 
\eeqn
From Eq. \ref{vvhcoup}, we see that the heavier Higgs
boson, $H$, decouples from the gauge bosons in the heavy $M_A$ limit,
while
the lighter Higgs boson, $h$, has Standard Model couplings.
 Figs. \ref{cbbhfig} and \ref{ctthfig} demonstrate  that
the Standard Model limit is 
 rapidly approached in the fermion-Higgs couplings for $M_A > 300~GeV$.
  In the
limit of large $M_A$, it will thus 
 be exceedingly difficult to differentiate a SUSY
Higgs sector from the  Standard Model Higgs boson.
\begin{itemize}
\item
The SUSY Higgs sector with large $M_A$ looks like the
Standard Model Higgs sector.
\end{itemize}                                 
In this case, it will be difficult to discover SUSY through the
Higgs sector.  Instead, it will be necessary to find some
of the other SUSY partners of the observed particles.

The total width of the Higgs boson depends sensitively on $\tan\beta$
and is illustrated in Fig. \ref{hwidfig}
 for $\tan\beta=2$.\cite{squrad}
  We see that
the lightest Higgs boson has a width $\Gamma_h\sim 10-100~MeV$,
while the heavier Higgs boson has a width $\Gamma_H\sim .1-1~GeV$,
which is considerably narrower than the width of the Standard Model
Higgs boson with the same mass.  (The curve for the lighter Higgs
boson is cut off at the kinematic upper limit.)  
The pseudoscalar, $A$, is also narrower than a Standard
Model Higgs boson with the same mass.

\subsection{The Squark and Slepton Sector}

We turn now to a discussion of the scalar partners of the quarks
and leptons.  The left-handed $SU(2)_L$ quark doublet has
scalar partners,
\beq
{\tilde Q}= 
\left( \begin{array}{c}
 {\tilde u}_L\\
{\tilde d}_L \end{array}\right)
\quad .
\eeq
The right-handed quarks also have scalar partners, ${\tilde u}_R$
and ${\tilde d}_R$.  The L and R subscripts denote which  helicity quark
the scalars are partners of--
{\bf they are for identification 
purposes only.
These are ordinary complex scalars}.  Before SUSY is broken
the fermions and scalars have the same masses and this mass
degeneracy is split by the soft mass terms of Eq. \ref{lagsoft}.
The tri-linear $A$ terms allow the scalar partners of the
left- and right-handed fermions to mix to form the mass
eigenstates.  
  In the
top squark sector, the mixing between the scalar
partners of the left- and right handed top (the stops), ${\tilde t}_L$
and ${\tilde t}_R$,  is given by
\beq
M_{{\tilde t}}^2=\left(
\begin{array}{ll}
M_{\tilde Q}^2+M_T^2 & M_T(A_T-\mu \cot\beta)\\
~~~~+M_Z^2({1\over 2}-{2\over 3}
\sin^2\theta_W)\cos 2 \beta & ~ ~ ~\\
M_T(A_T-\mu\cot\beta)& M_{\tilde u}^2+M_T^2
\\
~~ &~~~~+{2\over 3}M_Z^2\sin^2\theta_W \cos 2 \beta
\end{array}\right)
\quad ,
\label{squarkm}
\eeq 
while in the $b-$ squark system the mass-squared matrix is
\beq
M_{{\tilde b}}^2=\left(
\begin{array}{ll}
M_{\tilde Q}^2+M_b^2 & M_b(A_b-\mu \tan\beta)\\
~~~~-M_Z^2({1\over 2}-{1\over 3}
\sin^2\theta_W)\cos 2 \beta & ~~\\
M_b(A_b-\mu\tan\beta)& M_{\tilde d}^2+M_b^2
\\
~~ & ~~~~
-{1\over 3}M_Z^2\sin^2\theta_W \cos 2 \beta
\end{array}\right)
\quad .
\label{squarkb}
\eeq 
For the scalars associated with the lighter
quarks, the mixing effects will be negligible if we
assume that all of the $A_i$ are of similar size (as often happens
in GUT models),
since the mixing is assumed to be
proportional to the quark mass.
If $\tan\beta >> 1$, then ${\tilde b_L}-
{\tilde b_R}$ mixing could be large and so be phenomenologically
relevant.
In principle, these mixing matrices could involve all
three  generations of squarks and so could be $6\times
6$ matrices in the $({\tilde q}_{iL}, {\tilde q}_{iL})$
basis, $(i=1,2,3)$.  
   
From Eqs. \ref{squarkm} and \ref{squarkb},
 we see that there are two important cases
to consider.  If the soft breaking occurs at a large scale
with all the soft masses roughly equal and
much greater than $M_Z$, $M_T$,  and $A_T$,
then  all the soft masses
will be   approximately
equal, and we will have $12$ degenerate squarks with
masses ${\tilde M}\sim M_{\tilde Q}
\sim M_{\tilde u}\sim
M_{\tilde d}$.  On the other hand, if the 
soft masses
and the tri-linear mixing term, $A_T$,  
are on the order of the electroweak scale, then mixing
effects become important.  
  
\begin{figure}[tb]
\vspace*{.3in}  
\centerline{\epsfig{file=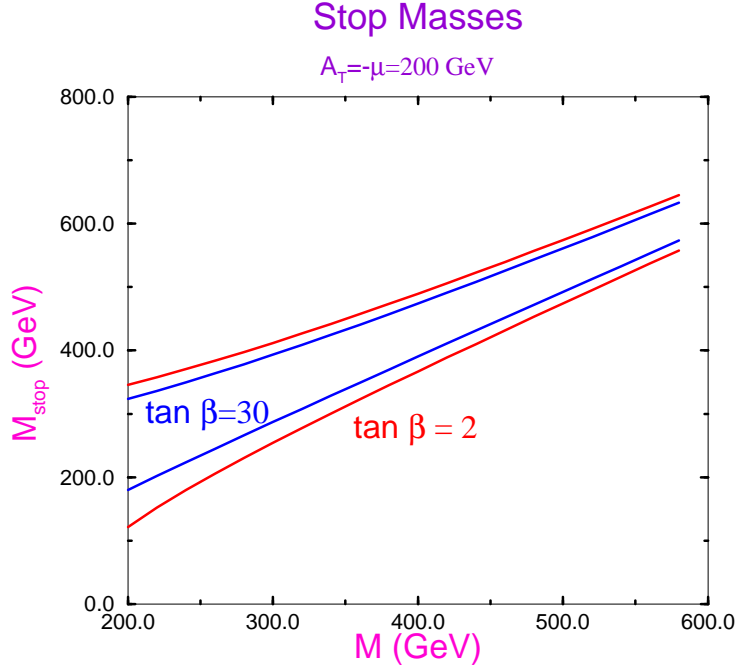,height=3.5in}}
\caption{Stop squark masses for large mixing parameters,
$A_T=\mu=200~GeV$, and for $\tan\beta=2$
and $\tan\beta=30$.  $M\equiv M_{\tilde Q}=M_{\tilde u}$ are  the
squark mass parameters of Eq. 52.} 
\vspace*{.5in}
\label{stopfig} 
\end{figure}  
If mixing effects are large, then  one of the stop squarks will become
the lightest squark, since the mixing effects are proportional
to the relevant quark masses and hence will be largest in this
sector.
The case where the lightest squark is the stop is particularly interesting
phenomenologically.
In Fig. \ref{stopfig}, we show the stop squark masses
for $M_{\tilde Q}=M_{\tilde u}
\equiv {\tilde M}$ and
for several values of $\tan\beta$.
  Of course the mixing effects 
cannot be too large, or the stop squark mass-squared will
be driven negative, leading to a breaking of the 
color  $SU(3)$ gauge symmetry.  
Typically, the requirement that the correct vacuum be chosen
leads to a restriction on the mixing parameter on the order
of $\mid A_T\mid <  {\tilde M}$.

The couplings of the squarks to gauge bosons are
completely  fixed by 
gauge invariance, with no free parameters.   A few
examples of the couplings are:
\beqn
\gamma^\mu ~{\tilde q}_{L,R}(p)~{\tilde q}_{L,R}^*(p^\prime):   
&&-i e Q_q (p+p^\prime)^\mu \nonumber \\
W^{\mu -} ~ {\tilde u}_L(p) ~ {\tilde d}_L^*(p^\prime): 
&&-{i g \over \sqrt{2}}(p +p^\prime)^\mu \nonumber \\
Z^\mu ~ {\tilde q}_{L,R}(p) ~ {\tilde q}_{L,R}^*(p^\prime) : 
&&
-{i g \over \cos\theta_W} \biggl[T_3 - Q_q \sin^2
\theta_W \biggr] (p + p^\prime)^\mu
\quad , 
\eeqn  
where $T_3$ and $Q_q$ are the quantum numbers of 
the corresponding quark.
The strength of the interactions are clearly given by the
relevant gauge coupling           constants.
  A complete set of Feynman rules can
be found in Ref.~\citenum{hkrep}.  

The mixing in the slepton sector is analogous to that in
the squark sector and we will not pursue it further.    
From Table 1, we see that the scalar partner of the $\nu_L$,
${\tilde \nu}_L$, has the same gauge quantum numbers
 as the $H_2^0$ 
Higgs boson.  It is possible to give ${\tilde \nu}_L$ a 
vacuum expectation value and use it to break the electroweak
symmetry.  Such a vacuum expectation value would break lepton
number (and $R$ parity) 
thereby giving the neutrinos a mass and so its magnitude
is severely restricted. \cite{rp}

\subsection{The Chargino Sector}
  
There are four  charge- $1$, spin- ${1\over 2}$ Majorana
fermions; ${\tilde \omega}^\pm$, the fermionic partners
of the $W^\pm$ bosons, (winos),  and ${\tilde h}^\pm$, the charged
fermion partners of the Higgs bosons, termed the Higgsinos.
Since they have the same quantum numbers,
these fermions can 
mix to form the $4-$ component Dirac spinors,
\beqn
\psi^+ &=& (-i {\tilde \omega}^+, {\tilde h}_2^+)
\nonumber\cr
\psi^- &=& (-i {\tilde \omega}^-, {\tilde h}_1^-)
\quad .
\eeqn 
Note that $\psi^+$ and $\psi^-$ are two independent Dirac
spinors and so it will take two different mixing matrices
to find the mass eigenstates. 
The physical mass states, $ \chi_{1,2}^\pm$, are
linear combinations formed by diagonalizing the mass
matrix and are usually called charginos.
We define these $2-$ component states in terms of the
mixing matrices, 
\beq
\begin{array}{ll} 
\chi_i^+ \equiv & V_{ij} \psi_j^+  \\
\chi_i^- \equiv & U_{ij} \psi_j^-, \qquad i=1,2 
\end{array}
\eeq 
(with $\psi^+_1\equiv -i {\tilde \omega}^+,~~\psi^+_2
\equiv {\tilde h}^+$, etc.). 
 The $4-$ component Majorana chargino mass eigenstates can be
formed as in Eq. \ref{majdef},
\beq
{\tilde \chi}_i^+=\left(\begin{array}{c}
\chi_i^+ \\
{\overline{\chi}}^+_i\end{array}\right)
\quad .  
\eeq

  The mass matrix for the charginos
 can be found from Eqs. \ref{lagw} and 
\ref{superpot},\footnote{ Note that some older
references define $\tan\beta=v_1/v_2$.  In comparing with the
literature, it is also important to check the definition of the
sign($\mu$).} 
\beq
{\cal L}=-{1\over 2} (\psi^+, \psi^-)
\left( \begin{array}{cc}
0 & M^T \\
M & 0 
\end{array}\right) \left(
\begin{array}{c}
\psi^+\\
\psi^-
\end{array}\right) +{\hbox{h.c.}}
\label{charmass}
\eeq
where
\beq
M=\left(
\begin{array}{cc}
M_2 & \sqrt{2}M_W\cos\beta\\
\sqrt{2}M_W\sin\beta & \mu\\
\end{array}\right) \quad .
\eeq   
It is clear that ${\tilde \omega}^\pm$ and ${\tilde h}^\pm$ are
not mass eigenstates, although for $M_2,\mu>>M_W$, the
mass of the lightest chargino is approximately $min(M_2,\mu)$.
For $M_2>>\mid\mu\mid$, the chargino is termed ``Higgsino-like'',
while for $M_2 << \mid \mu \mid$ it is called ``gaugino-like''.
The properties of the chargino in these two regimes are significantly
different.  
 
  The diagonal chargino mass matrix, $M_{\chi^+}$,  can 
be found by diagonalizing the mass matrix using the $2$
unitary matrices, $U$ and $V$, 
\beq
M_{\chi^+}=U^*MV^{-1}
\eeq
with,           
\beq
V_{ij}= \left(  
\begin{array}{rr}
\cos\phi_+ & \sin\phi_+ \\
\sin\phi_+& -\cos\phi_+
\end{array}\right) 
, \qquad \qquad 
U_{ij}=\left(
\begin{array}{rr}
\cos\phi_- & \sin\phi_- \\
-\sin\phi_- & \cos\phi_-
\end{array}
\right) 
\quad .  
\eeq 
It is straightforward to find analytic expressions for the
mixing angles :
\beqn
\tan 2\phi_+&= {2\sqrt{2}M_W(\mu\sin\beta+M_2\cos\beta)\over
M_2^2-\mu^2-2M_W^2\cos 2\beta}\nonumber \\
\tan 2\phi_-&= {2\sqrt{2}M_W(\mu\cos\beta+M_2\sin\beta)\over
M_2^2-\mu^2+2M_W^2 \cos 2\beta}
.
\eeqn
Useful expressions for the mixing angles in terms
of the chargino mass eigenstates are
 given in Refs.~\citenum{eech1} and \citenum{eech2}.  
One of the mass eigenstates (say $M_{{\tilde\chi}_2^+}$) 
will be negative if
\beq
Det(M)=M_2\mu-M_W^2\sin(2\beta)>0  
\quad .
\eeq
The easiest way to deal with a negative mass eigenstate is
to define a sign factor, $\eta_i$,  which is $1$ for positive
mass and $-1$ for negative mass.  If we consistently replace
$V_{2i}$ by $\eta_i V_{2i}$ in all couplings, then the
correct Feynman rules are obtained.
$U$ and $V$ are always assumed to be
chosen such that the mass eigenstates are
real and positive.

The mass eigenstates are  given by,
\beqn
M^2_{{\tilde \chi}^\pm_{1,2}}&=&{1\over 2}\biggl\{
M_2^2+2 M_W^2+\mu^2\mp
\biggl[ (M_2^2-\mu^2)^2+4M_W^4\cos^2 2 \beta
\nonumber \\
&&
+ 4M_W^2(M_2^2+\mu^2+2M_2\mu\sin^2\beta)
\biggr]^{1/2}\biggr\}  .
\label{chargmass} 
\eeqn   
By convention $M_{{\tilde \chi}_1^\pm}$ is the
lighter chargino.
 In Fig \ref{chmassfig},
 we show the mass of the lightest chargino
as a function of $\mu$ for several values of $M_2$ 
and $\tan\beta$.  
 From Eq. \ref{chargmass}, it is clear that for $\mu\rightarrow 0$, there
will be an almost massless chargino, which is clearly
seen in Fig \ref{chmassfig}.  This possibility is
excluded by experiment.

Combining our results, we find
the interaction Lagrangian of the $4-$ component 
charginos with the gauge bosons,
\beqn
{\cal L}&=&  e {\overline {\tilde \chi}}_i^+\gamma^\mu 
{\tilde \chi}_j^+ A_\mu \delta_{ij}  
           + \biggl({g\over
\cos\theta_W}\biggr) 
  {\overline {\tilde \chi}}_i^+\gamma^\mu\biggl[
C_{ij}^+P_+ + C_{ij}^- P_-\biggr] 
{\tilde \chi}_j^+ Z_\mu  \nonumber \\
& &-g\biggl[ {\overline e}_L
P_+ {\tilde \chi}^{+c}_i
{\tilde \nu}_L V_{i1} +
{\overline \nu}_LP_+ {\tilde \chi}_i^+
{\tilde e}_L U_{i1} + h.c.\biggr]\quad ,
\label{charglag}
\eeqn
where $e<0$, $P_\pm={1\over 2} (1\pm \gamma_5)$ and
\beqn
C_{ij}^+& =& -V_{i1}V_{j1}^*-{1\over 2}V_{i2}V_{j2}^*+\delta_{ij}
\sin^2\theta_W
\nonumber \\
C_{ij}^-&=& -U_{i1}^*U_{j1}-{1\over 2}U_{i2}^*U_{j2}^*+\delta_{ij}
\sin^2\theta_W
\quad . 
\nonumber 
\label{cchargdef}
\eeqn 
The Feynman rules derived from Eq. \ref{charglag} can be used to
compute processes of physical interest and can
be found in Ref.~\citenum{hkrep}.  As an example, in the
Appendix we compute $e^+e^-\rightarrow {\tilde \chi}_1^+ {\tilde
\chi}_1^-$, where there is an interesting
interplay between sneutrino exchange
and the $s-$ channel $\gamma$- and $Z$ -exchange.

\begin{figure}[tb]
\vspace*{.3in} 
\centerline{\epsfig{file=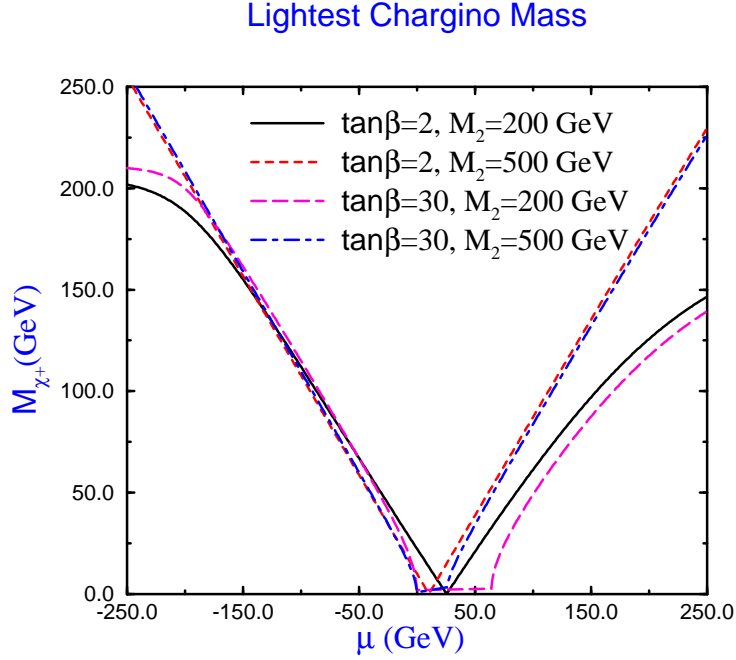,height=3.5in}}
\caption{Lightest chargino mass for fixed $\tan\beta$
and $M_2$.}
\label{chmassfig}
\end{figure}

\subsection{The Neutralino Sector} 
  
In the neutral fermion sector, the
neutral fermion partners of
the $B$ and $W^3$ gauge bosons, ${\tilde b}$ and
${\tilde \omega}^3$, can mix with the neutral
fermion partners of the Higgs bosons, ${\tilde h}_1^0,
{\tilde h}_2^0$ to form the mass eigenstates.  Hence the physical states,
${\tilde \chi}_i^0$, termed neutralinos, are
found by diagonalizing the $4\times 4$ mass
matrix, 
\beq
M_{\tilde \chi_i^0} =
\left(\begin{array}{cccc}
M_1 & 0& -M_Z \cos\beta \sin\theta_W & M_Z \sin\beta\sin\theta_W\\
0 & M_2 & M_Z \cos\beta\cos\theta_W & -M_Z \sin\beta\cos\theta_W\\
-M_Z \cos\beta\sin\theta_W & M_Z \cos\beta\sin\theta_W & 0 & -\mu \\
M_Z \sin\beta\sin\theta_W & -M_Z \sin\beta\cos\theta_W & -\mu & 0 
\end{array}
\right)
\eeq
where $\theta_W$ is the electroweak mixing angle
and we work in the ${\tilde b}, {\tilde \omega}^3,
{\tilde h}_1^0, {\tilde h}_2^0$ basis.  The physical
masses can be defined to be positive and by convention,
$M_{{\tilde \chi}_1^0}
< M_{{\tilde \chi}_2^0}
< M_{{\tilde \chi}_3^0}
< M_{{\tilde \chi}_4^0}$.
In general, the mass eigenstates do not correspond to
a photino, (a fermion partner of the photon), or a 
zino, (a fermion partner of the $Z$), but are complicated
mixtures of the states. The photino is only a mass eigenstate
if $M_1=M_2$.  Physics involving the neutralinos therefore
depends on $M_1$, $M_2$, $\mu$, and
$\tan\beta$.  
The lightest neutralino, ${\tilde \chi}_1^0$, is usually assumed
to be the LSP.  
 
\section{WHY DO WE NEED SUSY?}
 Having introduced the MSSM as an effective
theory at the electroweak scale  and briefly
discussed the various new particles and interactions  of the model, we
turn now to a consideration of the reasons for constructing 
a SUSY theory in the first place.
  We have already considered the cancellation
of the quadratic divergences, which is automatic in
a supersymmetric model.   
There are, however, many other reasons why theorists are excited about
supersymmetry.

\subsection{Coupling constants run!}

In a gauge theory, coupling constants scale with energy
according to the relevant $\beta$-function.  Hence, having
measured a coupling constant at one energy scale, its value at any
other energy can be predicted.  At  one loop, 
\beq
{1\over \alpha_i(Q)}={1\over \alpha_i(\mu)}+
{b_i\over 2 \pi} \log\biggl({\mu\over Q}\biggr)
\quad .
\label{smbeta}
\eeq
In the Standard (non-supersymmetric) Model, the coefficients
$b_i$ are given by,
\beqn
b_1&=& {4\over 3}N_g+{N_H\over 10}\nonumber \\
b_2&=& -{22\over 3}+{4\over 3}N_g+{N_H\over 6}
\nonumber \\
b_3&=&-11+{4\over 3}N_g \quad ,
\eeqn
where $N_g=3$ is the number of generations and $N_H=1$ is
the number of Higgs doublets. 
The evolution of the coupling constants is seen to be sensitive to
the particle content of the theory.  
  We can take $\mu=M_Z$ in Eq. \ref{smbeta}, input the
measured values of the coupling constants at the $Z$-pole and
evolve the couplings to high energy.  The result is shown in
 Fig. \ref{smccfig}.  
There is obviously no meeting of the coupling constants at high
energy.  

Suppose we assume that the unifying gauge group is $SU(5)$.  This requires
that the $SU(3)$, $SU(2)_L$, and $U(1)_Y$ generators, $T_i$, be
normalized in the same manner.  Each generation of fermions
is contained in a ${\overline 5}$ and $10$ of $SU(5)$
with 
the ${\overline 5}$ given by,
\beq
{\overline 5}=({\overline d}, {\overline d},
{\overline d},  e^-, \nu)
.
\eeq 
The $SU(3)$ and $SU(2)_L$ generators both satisfy $Tr(T_i)^2={1\over 2}$
for the ${\overline 5}$.  The $U(1)_Y$ generator for the
${\overline 5}$ must  have the same normalization,
\beq
Y_{GUT}=\sqrt{{3\over 5}} diag({1\over 3},{1\over 3},{1\over 3},
-{1\over 2}, -{1\over 2})
.
\eeq
By comparing with Table 1, we see that $Y_{GUT}=\sqrt{{3\over 5}}Y$
and so we must have
\beq
g^\prime=g^*\sqrt{{3\over 5}}
,
\eeq
where $g^*$ is the GUT coupling constant, in order to
obtain the correct couplings of the fermions
to the gauge bosons.

\begin{figure}[tb]
\vspace*{.3in}  
\centerline{\epsfig{file=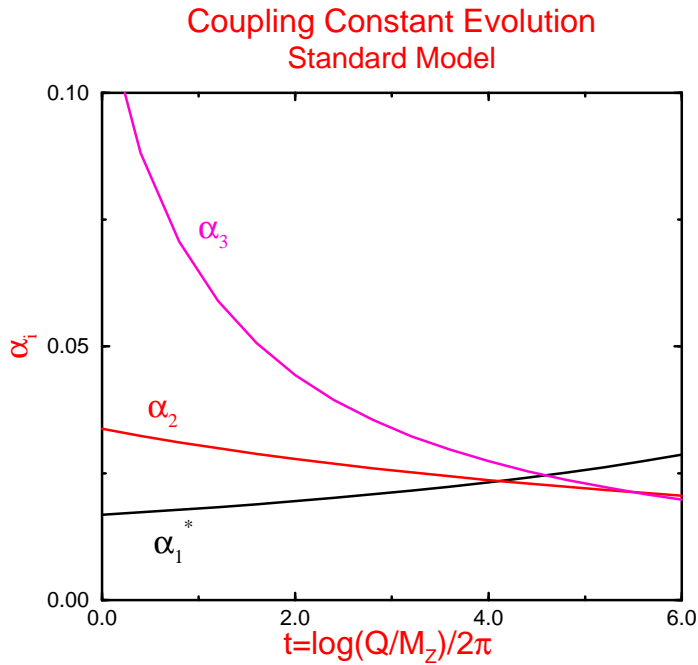,height=3.5in}}
\caption{Evolution of the gauge coupling
constants in the Standard Model from the experimentally
measured values at the $Z$-pole.  
$\alpha_1^*\equiv 5/3 \alpha_1$, since this is the relevant
coupling in SU(5)-like Grand Unified Theories. }
\label{smccfig}
\vspace*{.5in}  
\end{figure}

\begin{figure}[tb]
\vspace*{.3in}    
\centerline{\epsfig{file=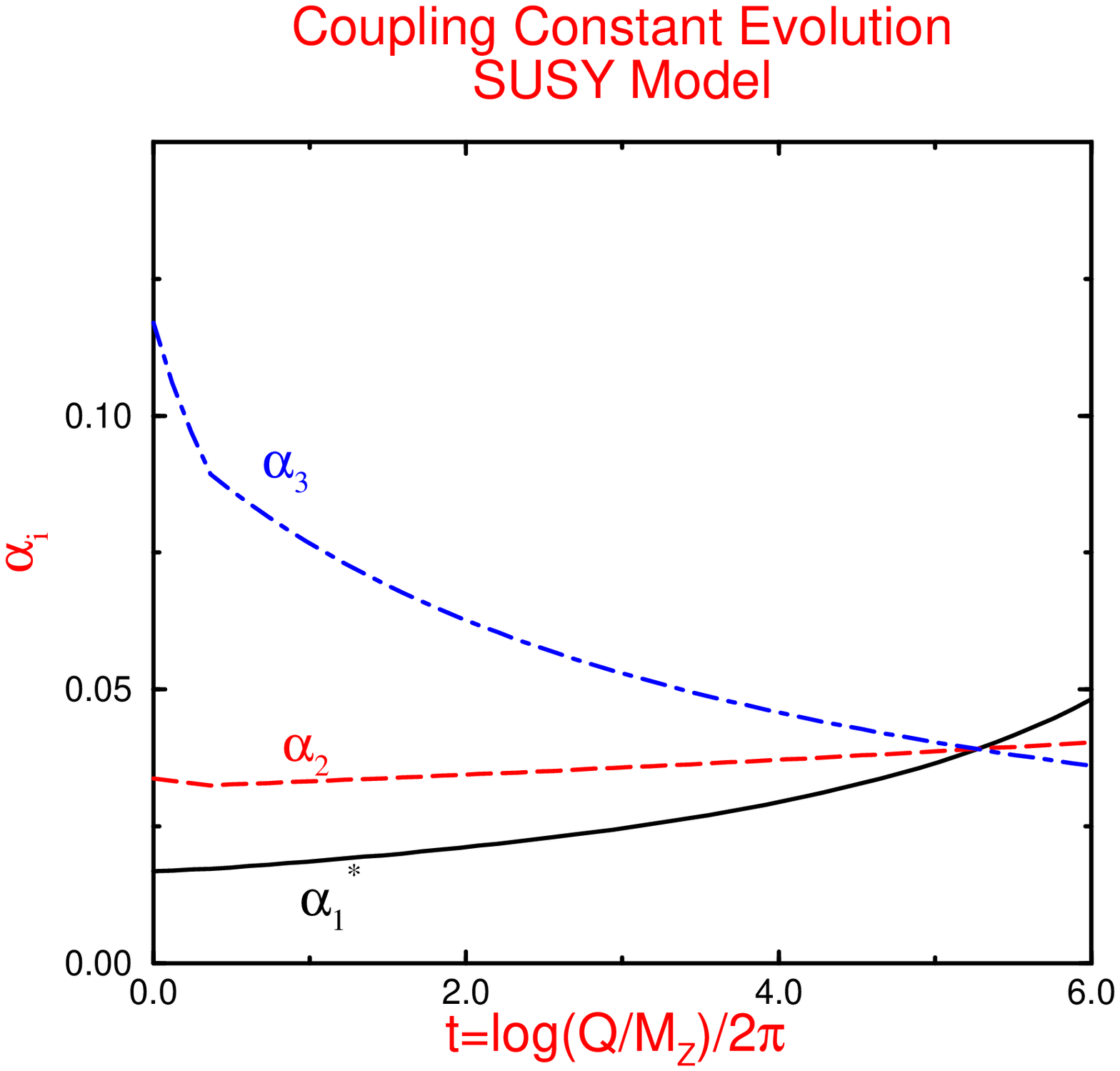,height=3.5in}}
\caption{Evolution of the coupling constants
in a low energy SUSY model
from the experimentally measured values at the $Z$-pole.
The SUSY thresholds are taken to be at
$1~TeV$. 
$\alpha_1^*\equiv 5/3 \alpha_1$, since this is the
relevant coupling in Grand Unified Theories.  }  
\vspace*{.5in}
\label{susyccfig}  
\end{figure}

If the theory is supersymmetric, then the spectrum is different
and the new particles contribute to the evolution of the coupling constants.
In this case we have at one loop,\cite{betafuns} 
\beqn
b_1&=& 2 N_g+{3\over 10}N_H\nonumber \\
b_2&=& -6+2 N_g+{N_H\over 2}\nonumber \\
b_3&=& -9+ 2 N_g
\quad .
\eeqn
Because a SUSY model of necessity contains two
Higgs doublets, we have $N_H=2$.  If we assume that the mass of
all the SUSY particles is around $1~TeV$, then the coupling
constants scale as shown in Fig.\ref{susyccfig}.  We see that the coupling
constants meet at a scale around $10^{16}$ GeV.\cite{early,unif,ccunif} 
This meeting of the coupling constants is a necessary feature of 
a Grand Unified Theory (GUT).  

\begin{itemize}
\item
SUSY theories can be naturally incorporated
into Grand Unified Theories.
\end{itemize}

It is interesting to use the requirement of unification to
predict $\alpha_s(M_Z)$.  This requirement gives the prediction:
\beq
{b_1-b_2\over \alpha_3(\mu)} +{b_2-b_3\over \alpha_1^*(\mu)}+
{b_3-b_1\over \alpha_2(\mu)}=0,
\eeq
valid at any scale $\mu$
between $M_X$ and $M_Z$. If we input the
SUSY $\beta$-functions, $\alpha_1(M_Z)=1/128$, and
$\sin^2 \theta_W(M_Z)=.2315$, we obtain a prediction from the
MSSM at $1-$loop:
\beq
\alpha_s(M_Z)=.116,
\eeq  
in reasonable agreement with the measured value at LEP, $\alpha_s(M_Z)
=.118\pm .003$.

Unfortunately, the picture is not so rosy when we attempt to take 
the MSSM seriously and include  two
loop beta functions, effects from passing
through SUSY particle thresholds, etc.\cite{ccunif} Typically,
the prediction for $\alpha_s(M_Z)$ becomes significantly larger
that the experimental value.  The goal is to learn something
about the underlying GUT theory by computing the threshold
corrections at the GUT scale and seeing which models are
consistent with the data.\cite{pierce}   
  
\subsection{SUSY GUTS}

The observation that the measured coupling constants tend to meet at
a point when evolved to high energy 
assuming the $\beta$-function of a low energy SUSY model
has led to widespread acceptance of
a standard SUSY GUT model.  We assume that the 
$SU(3)\times SU(2)_L\times U(1)_Y$ gauge coupling
constants are unified at a high scale $M_X\sim 10^{16}~GeV$:
\beq
\sqrt{{5\over 3}}g_1(M_X)=g_2(M_X)=g_s(M_X)\equiv g^*
\quad .
\eeq
\subsection{CMSSM}

We will describe
 two types of possible GUT models which differ in the source of
the soft SUSY breaking terms.  The first model is sometimes called the
constrained MSSM (CMSSM) or ``super-gravity inspired'' (SUGRA) MSSM.

Along with the coupling constants,
the gaugino masses, $ M_i$, are  assumed to unify,
\beq  M_i(M_X)\equiv m_{1/2}
\quad .
\eeq
At lowest order, the gaugino masses  then scale in the same
way as the corresponding coupling constants,
\beq
M_i(M_W)=m_{1/2}{g_i^2(M_W)\over g^{*2}}
\label{gaugmass}
\eeq
yielding  
\beqn
 M_2(M_W)&=& \biggl({\alpha(M_W)\over \sin^2\theta_W(M_W)}
\biggr)\biggl({1\over
\alpha_s(M_W)}\biggr)  M_3(M_W)\nonumber \\
 M_1(M_W)&=& {5\over 3} \tan^2\theta_W(M_W)  M_2(M_W)
\quad .
\eeqn
The gluino mass is therefore always the heaviest of the
gaugino masses.  
This relationship between the gaugino masses is a fairly robust
prediction of SUSY GUTS and as we will
see in the next section persists in  models where
the supersymmetry is broken through the gauge interactions.
 
Typical SUSY GUTS of this type also assume that there is a common
scalar mass at $M_X$,
\beqn
&m_1^2(M_X)=&m_2^2(M_X)\equiv m_0^2\nonumber \\
M_{\tilde Q}^2(M_X)= M_{\tilde d}^2(M_X)=& 
M_{\tilde u}^2(M_X)=&M_{\tilde L}^2(M_X)=
M_{\tilde e}^2(M_X)\equiv m_0^2
 .
\label{scalgut}
\eeqn  
The neutral Higgs boson
masses at $M_X$ are then $M_{h,H}^2=m_0^2+\mu^2$.

It is instructive to study the scalar masses within this
scenario.  
  The evolution of the sleptons between $M_X$ and $M_W$
 is small and
we have the approximate result
for the slepton masses,\cite{xerxes,coups2}  
\beq
M_{\tilde L}^2(M_W)\sim M_{\tilde e}^2(M_W)\sim m_0^2,
\eeq
while the squark masses are roughly
\beq
M_{\tilde q}^2 (M_W)\sim m_0^2+4 m_{1/2}^2
\quad .
\eeq  
Since the squarks 
have strong interactions, (which drives the masses upwards),
their masses at the weak scale tend to be larger than the sleptons.  
Once all the particle masses have been computed in this
scheme, then their production cross sections and
decay rates at any given accelerator can be computed
unambiguously.

As a final simplifying assumption, a common $A$ parameter is
assumed,
\beq
A_T(M_X)=A_b(M_X)=....\equiv A_0
\quad .
\eeq
With these assumptions, the SUSY sector is completely described
by 5 input parameters at the GUT scale,\cite{fp}
\begin{enumerate}
\item
A common scalar mass, $m_0$.
\item
A common gaugino mass, $m_{1/2}$.
\item
A common trilinear coupling, $A_0$.
\item
A Higgs mass parameter, $\mu$.
\item
A Higgs mixing parameter, $B$.
\end{enumerate} 

The picture is that there is a hidden sector of the
theory containing fields which do not transform under the
$SU(3)\times SU(2)_L\times U(1)_Y$ gauge group.
We assume that the supersymmetry is broken in this
hidden sector and communicated to the fields of the MSSM by
gravitational interactions.\cite{wein}
When the supersymmetry is  broken at the scale $M_{SUSY}$, the
gravitino will obtain a mass,
\beq
M_{3/2}\sim {M_{SUSY}^2\over M_{pl}} 
\quad .
\eeq
The soft mass terms of Eq. \ref{lagsoft} will then be,
\beq
m_{soft}\sim M_{3/2}\sim {M_{SUSY}^2\over M_{pl}} 
\quad .
\eeq
To obtain $M_{soft}\sim 1~TeV$ ( which we argued was
necessary in order that supersymmetry solve the hierarchy
problem), we need to break supersymmetry at a scale,
\beq
M_{SUSY}\sim 10^{11}~GeV
\quad .
\eeq
  If the fields of the hidden
sector have canonical kinetic energy terms, then the scalar 
masses will satisfy the relationships of Eq. \ref{scalgut}.
Although this framework is somewhat
${\it ad~hoc}$, it does provide guidance to reduce the immense
parameter space of a SUSY model.

  The strategy is now to 
input the $5$ parameters given above   at $M_X$  and to  use the
renormalization group equations to evolve the parameters to $M_W$. 
In fact, the requirement that the $Z$ boson obtain its measured
value  when the parameters are evaluated   at 
low energy can be used to restrict $\mid \mu B\mid$, leaving
the $sign(\mu)$ as a free parameter.  We can also trade
the parameter $B$ for $\tan\beta$.  In this way the parameters
of the model 
become
\beq
m_0, m_{1/2}, A_0, \tan\beta, {\rm sign}(\mu)
\quad .
\eeq
  This
form of a SUSY theory is extremely predictive, as
the entire low energy spectrum is predicted in terms of a 
few input parameters.   Also, all phenomenological limits can be expressed
as limits on these parameters.  
Within this scenario, contours for the
various SUSY particle masses can be found as a function
of $m_0$ and $m_{1/2}$ for given values of $\tan\beta$, 
$A_0$ and ${\rm sign}(\mu)$.\cite{coups2,fp}

  Changing the input parameters at 
$M_X$ (for example, assuming non-universal scalar masses)
of course changes the phenomenology at the weak scale.  A
preliminary investigation of the sensitivity of the low energy
predictions to these assumptions has been made in Ref.~\citenum{snow}.
For now, we will consider the Grand Unified Model described
above as a starting point for phenomenological investigations
into SUSY.

\subsection{GMSB}
An alternative picture of the SUSY breaking is gauge mediated
SUSY breaking.\cite{mess,kolda}  In this type of model,
the SUSY breaking again occurs in a hidden sector.  The hidden sector
is assumed to contain
 new chiral supermultiplets, called messenger fields, which
transform under the $SU(3)\times SU(2)_L\times U(1)_Y$ gauge group.
When supersymmetry is broken, the messengers obtain
a mass, $\Lambda$.

The SUSY breaking is communicated to the MSSM particles through
the gauge interactions.  The gauginos obtain
mass at $1-$ loop,
\beq
M_i\sim {\alpha_i\over 4 \pi}\Lambda
\quad .
\eeq
  We see that the gauginos satisfy
Eq. \ref{gaugmass}.  Similarly, the scalars of the MSSM obtain masses at two
loops from diagrams involving the gauge fields and the messenger fields.
The scalar masses are,
\beq
{\tilde M}^2\sim \biggl({\Lambda\over 4 \pi}\biggr)^2\biggl\{
\alpha_s^2 C_3 +\alpha_2^2 C_2 +\alpha_1^2 C_1
\biggr\},
\eeq
where $C_i$ are the quadratic Casimir operators for the
$SU(3)\times SU(2)_L\times U(1)_Y$ gauge groups.  
The squarks and sleptons with the same gauge quantum
numbers will  automatically have the same masses.  For example,
\beq
M_{\tilde e}=M_{\tilde \mu}=M_{\tilde \tau},
\eeq
etc.
In order to obtain soft masses for the gauginos and scalars
around $1~TeV$, we need
\beq
\Lambda\sim 10^4-10^5~GeV.
\eeq
  Note
that this scale is much smaller than the intermediate
scale found in the CMSSM.  

The most important difference between the CMSSM and the GMSB
models is that here the LSP is the gravitino, ${\tilde G}_{3/2}$.
In this case the gravitino mass is,
\beq
m_{3/2}\sim {\Lambda^2\over M_{pl}}\sim 10^{-10} ~GeV
\quad.  
\eeq
This leads to strikingly different phenomenology
from the CMSSM since this
model allows
\beq
{\tilde \chi}_1^0\rightarrow \gamma {\tilde G}_{3/2},
\eeq
giving a signal for SUSY of missing $E_T$ plus  photons.
A review of GMSB models can be found in Ref.~\citenum{kolda}.   

\subsection{Electroweak Symmetry Breaking}

The CMSSM model has the appealing feature
that it explains the mechanism of electroweak symmetry breaking.
In the Standard Model (non-supersymmetric) with 
a single Higgs doublet, $\phi$,
the scalar potential is given by:
\beq
V(\phi)=\mu^2 \mid \phi\mid ^2+\lambda (\mid \phi\mid^2)^2 
\quad .
\eeq
By convention, $\lambda>0$.  If $\mu^2>0$, then $V(\phi)>0$
for all $\phi$ not equal to $0$
 and there is no electroweak symmetry breaking.
If, however, $\mu^2<0$, then the minimum of the potential is not
at $\phi=0$ and the potential has the familiar Mexican hat shape.
When the Lagrangian is expressed in terms of the physical field,
$\phi^\prime\equiv (\phi-v)/\sqrt{2}$, which
has zero vacuum expectation value, then
the electroweak symmetry is broken and the $W$ and $Z$ gauge bosons
acquire non-zero masses. 
We saw  in the previous sections
that this same mechanism gives the $W$ and $Z$ gauge
bosons their masses in the MSSM.
 This simple picture leaves one looming question: 
\beq
 \bf{\rm Why~ is~ 
\mu^2< 0? } 
\nonumber
\eeq
  It is this question which the  SUSY GUT  models can
answer.

In the minimal CMSSM
 model,  the neutral Higgs
bosons both 
have masses, $M_{h,H}^2=m_0^2+\mu^2$,
at $M_X$,  while the squarks and sleptons have
mass $m_0$ at $M_X$.   Clearly, at $M_X$, the
electroweak symmetry is not broken since the Higgs
 bosons have positive mass-squared.
 The masses scale with energy according to the 
renormalization group equations.\cite{yukren}
  If we neglect gauge couplings and 
consider only the scaling of the third generation scalars
we have,\cite{ewsbtop}
\beqn
{d\over d\log(Q)}\left(\begin{array}{c}
M_h^2\\
M_{\tilde t_R}^2\\
M_{\tilde Q_L^3}^2\end{array}\right)&=&-{8 \alpha_s\over 
3 \pi}  M_3^2
\left(\begin{array}{c}
0\\
1\\
1\end{array}\right)
\nonumber \\
&
+&{\lambda_T^2\over 8 \pi^2}
\biggl(M_{\tilde Q_L^3}^2+M_{\tilde t_R}^2
+M_h^2+A_T^2\biggr)
\left(\begin{array}{c}
3\\
2\\
1\end{array}\right) ,
\label{scaling}
\eeqn 
where ${\tilde Q}_{L}^3$ is the $SU(2)_L$ doublet containing
${\tilde t}_L$ and ${\tilde b}_L$, $h$ is the lightest neutral Higgs
boson, $\lambda_T$ is the top quark Yukawa coupling
constant given in Eq. \ref{yukdefs},  
 and $Q$ is the effective scale at which
the masses are measured.  The signs are such that the Yukawa
interactions (proportional to $M_T$) decrease the masses, while the
gaugino interactions increase the masses.  Because of the $3-2-1$ structure of
the last term in Eq. \ref{scaling},
 the Higgs mass decreases faster than the
squark masses and it is possible to drive $M_h^2<0$ at low energy,
while keeping $M_{\tilde Q_L^3}^2$ and
$M_{\tilde t_R}^2$ positive. 
A generic set of scalar masses in a typical SUSY GUT model is shown
in Fig. \ref{specfig}.  We can clearly see that the lightest Higgs boson
mass becomes negative around the electroweak scale.\cite{masssamp}    

 For large $\lambda_T$, we have
the approximate solution,
\beq
M_h^2(Q)=M_h^2(M_X)-{3\over 8\pi^2}\lambda_T^2
(M_{\tilde Q_L^3}^2+M_{\tilde t_R}^2
+M_h^2+A_T^2)\log\biggl({M_X\over Q}\biggr)
\quad .
\eeq
Hence the larger $M_T$ is, the faster $M_h^2$ goes negative.
This of course generates electroweak symmetry breaking.
If $M_T$ were light, $M_h^2$ would remain positive.\cite{ewsbtop} 
This observation was made fifteen
 years ago when we thought
the top quark was light, ($\sim 40~GeV$).  At that
time it was ignored as not being phenomenologically relevant.
In fact, this mechanism only works for $M_T\sim 175~GeV$!  
\begin{itemize}
\item
SUSY GUTS can explain electroweak symmetry breaking.
The lightest Higgs boson mass is negative,
$M_h^2<0$, because $M_T$ is large.
\end{itemize}
\begin{figure}[tb]
\vspace{.3in}
\centerline{\epsfig{file=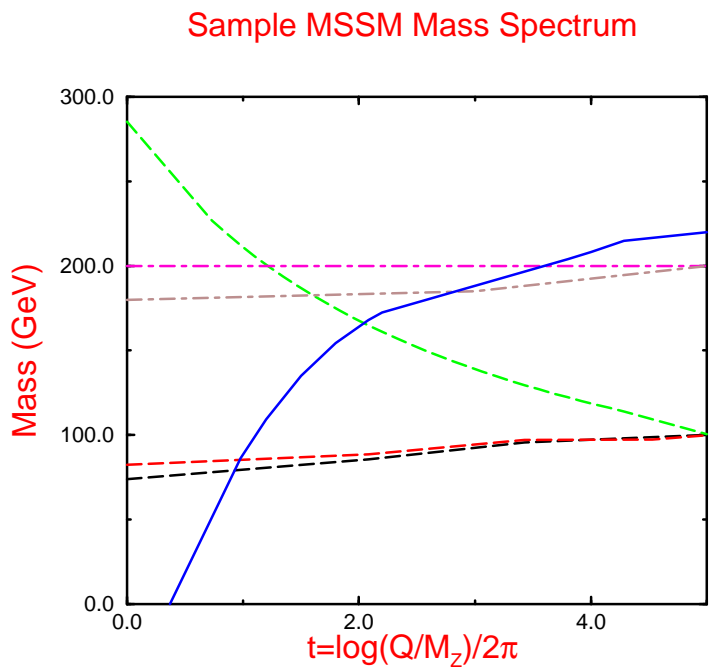,height=3.5in}}
\caption{Sample masses of SUSY particles in a SUSY GUT.
At the GUT scale $M_X$, we have taken $m_0=200~GeV,
m_{1/2}=100~GeV, \mu=100~GeV$ and $A_i=0$.  The solid
line is the lightest neutral Higgs boson mass.  The dashed
lines are the gaugino masses (the largest is the gluino) and the
dot-dashed lines are typical squark masses. }  
\vspace*{.5in}
\label{specfig}   
\end{figure}  
The requirement that the electroweak symmetry breaking occur
through the renormalization group scaling of the Higgs boson
mass also restricts the
allowed values of $\tan\beta$ to $\tan \beta > 1$. 
(Remember that $\lambda_T$ depends on $\beta$ through Eq. \ref{yukdefs}.)

\subsection{Fixed Point Interactions}

In the previous subsection we saw that a large top
quark mass could generate electroweak symmetry breaking
in a SUSY GUT  model.  
The top quark mass is determined in terms of its Yukawa
coupling and scales with energy, $Q$,\cite{btau} 
\beq
\lambda_T(Q)={M_T(Q)\over M_W}
{g\over \sqrt{2}\sin\beta}
\quad .
\eeq
Including both the gauge couplings and the Yukawa couplings
to the $t$- and $b$- quarks, the scaling is:
\beq
{d \lambda_T\over d \log(Q)}=
{\lambda_T\over 16 \pi^2}
\biggl\{
-{13\over 9} g^{\prime 2}-3 g^2-{16\over 3} g_s^2
+6 \lambda_T^2+\lambda_B^2\biggr\}
\quad .
\eeq
To a good approximation, we can consider only the contributions
from the strong coupling constant, $g_s$, and the top
quark Yukawa coupling,
$\lambda_T$.  If we begin our scaling at 
$M_X$ and evolve $\lambda_T$ to lower energy, we will come to a
point where the evolution of the Yukawa coupling stops,
\beq
{d \lambda_T\over d \log(Q)}=0
\quad .
\eeq
At this point we have roughly, 
\beq
-{16\over 3} g_s^2+6 \lambda_T^2=0
\eeq
which gives,
\beq
\lambda_T\sim{4\over 3}\sqrt{2 \pi\alpha_s}\sim 1
,
\eeq
or
\beq
M_T\sim (200~GeV) \sin\beta \quad . 
\eeq
This point where the top quark mass stops evolving is called
a ${\it fixed~point}$.
What this means is that no matter what the initial condition
for $\lambda_T$ is at $M_X$, it will always evolve to
give the same value at low energy.  
  For $\tan\beta\sim2$, the fixed point
value for the top quark mass is close to the experimental value.  
More sophisticated analyses do not change this picture substantially.  
\begin{itemize}
\item
SUSY GUTS can naturally accommodate a large top quark mass
for $\tan\beta\sim 1-3.$
 
\end{itemize}

\subsection{$b-\tau$ Unification}

The unification of the $b$- and $\tau$- Yukawa coupling
constants,
$\lambda_B$ and $\lambda_\tau$, 
at the GUT scale is a concept much beloved by
theorists since
\beq
\lambda_B(M_X)=\lambda_\tau(M_X)
\label{btau} 
\eeq
occurs naturally in many GUT models (such  as the $SU(5)$ GUT).
Imposing the boundary condition of Eq. \ref{btau} and  requiring that
the $b$ quark have its experimental value  at low energy leads to
a prediction for the top quark mass in terms of $\tan\beta$.
There are two solutions which yield $M_T=175~GeV$,\cite{btau}
\beqn
\tan\beta&\sim& 1\nonumber \\
{\rm or}\quad \tan\beta&\sim& {M_T\over M_b}
\quad .  
\eeqn 
The first solution roughly corresponds to the fixed point solution
of the previous subsection. 
The second solution with $\tan\beta\sim 35$  has interesting
phenomenological consequences, since for large $\tan\beta$
the coupling of the lightest Higgs boson to $b$ quarks is enhanced
relative to the Standard Model. (See Fig. \ref{cbbhfig}). 
 The  values in the $\tan\beta -
M_T$ plane
allowed by $b-\tau$ unification
 depend sensitively on the exact value of the strong coupling
constant, $\alpha_s$, used in the evolution and so there is a
significant uncertainly in the prediction.\cite{largeb}    
\begin{itemize}
\item
SUSY GUTs allow for 
the unification of the $b-\tau$
Yukawa coupling constants at the GUT
scale  along with the experimentally
observed value for the top quark mass.
\end{itemize}
Similar relationships to Eq. \ref{btau} involving the first two generations
do not work.  
\subsection{Comments}
We see that SUSY plus grand unification has many desirable
features which are not sensitive to the exact mechanism of 
SUSY breaking or to the  details of the underlying GUT:
\begin{enumerate}
\item 
There are no troubling  quadratic divergences requiring
disagreeable cancellations.
\item
$M_T$ is large because $\lambda_T$ evolves from the GUT
scale to its fixed point.
\item
Electroweak symmetry is broken, $M_h^2<0$, because $M_T$ is
large. 
\item
$b-\tau$ unification can be incorporated, leading to the 
experimentally observed value for the top quark mass.
\end{enumerate}

\section{SEARCHING FOR  SUSY}

\subsection{Indirect Hints for SUSY}

One might hope that the precision measurements at the
$Z$-pole could be used to garner information on the SUSY
particle spectrum.  Since the precision electroweak measurements
are overwhelmingly in excellent agreement with the predictions of
the Standard Model, it would appear that stringent
limits could be placed on the existence of SUSY particles at
the weak scale.  There are two reasons why this is not the 
case.

  The first is that SUSY is a ${\it decoupling~theory}$.
With the exception of the Higgs particles,
 the effects of SUSY particles at the weak scale are
suppressed by powers of $M_W^2/M_{SUSY}^2$, where $M_{SUSY}$ is
the relevant SUSY mass scale,  and so
for $M_{SUSY}$ larger than a few hundred $GeV$,
the SUSY particles  give negligible contributions to 
electroweak processes.
The second reason why there are not
 stringent limits
from precision results at LEP has to do with the Higgs 
sector.  The Higgs bosons are the only particles
in the spectrum  which  do not decouple
from the low energy physics when they are very massive.
The
fits to electroweak data tend to prefer a Higgs boson in
the $100~GeV$ mass  range.\cite{pdg}  Since the MSSM requires a light
Higgs boson with a mass in this region
anyways,   the electroweak data is completely consistent
with a SUSY model with a light Higgs boson and all other SUSY
particles significantly heavier.

Attempts have been made to perform global fits to the electroweak
data and to fix the SUSY spectrum this way.\cite{deb,ewsusy}
It is possible to obtain a fit where the $\chi^2$/degree of
freedom is roughly the same as in the Standard Model fit. 
Although the fits do not yield stringent limits
on the SUSY particle masses, they do    
exhibit several interesting features.
  They tend to prefer either small $\tan\beta$, $\tan\beta \sim 2$, or
 relatively large values, $\tan\beta\sim 30$.  
In addition, the fitted values for the strong coupling constant
at $M_Z$,
$\alpha_s(M_Z)$,  are slightly smaller in  SUSY  models than
in the Standard Model.  (For $\tan\beta=1.6$, Ref.~\citenum{deb}
finds $\alpha_s(M_Z)=.116\pm.005$ and for $\tan\beta=34$, they
find $\alpha_s(M_Z)=.119\pm.005$.)   
It is clear that all precision electroweak measurements can
be accommodated within a SUSY model, but the data show
no clear preference for  these models.

\subsection{Limits from the $\rho$ Parameter}
One of the most precisely measured electroweak quantities
is the $\rho$ parameter,
\beq
\rho={M_W^2\over M_Z^2\cos^2\theta_W}
\quad .
\eeq
A large mass splitting between the stop and sbottom
squarks can give a significant contribution
to the $\rho$ parameter, just as does the $t-b$ mass splitting.
If we define $\theta_t$ to be the mixing angle associated
with the stop mass matrix, Eq. \ref{squarkm},
 and neglect the mixing
in the ${\tilde b}_L-{\tilde b}_R$ sector, there is a contribution
to the $\rho$ parameter from the squark sector of \cite{rhosusy} 
\beqn
\delta \rho^{SUSY}&=&{3 G_F\over 8\sqrt{2}\pi^2}
\biggl\{ -{1\over 4} \sin^2 2\theta_t f(M_{\tilde {t_1}}^2,
M_{\tilde {t_2}}^2)
\nonumber \\
&&+\cos^2\theta_t f(M_{\tilde {t_1}}^2,
M_{\tilde {b_L}}^2)+\sin^2\theta_tf(M_{\tilde {t_2}}^2,
M_{\tilde {b_L}}^2)\biggr\}
,
\eeqn
where $t_1$ and $t_1$ are the stop mass eigenstates and
\beq
f(m_1^2,m_2^2)=m_1^2+m_2^2-{2 m_1^2 m_2^2\over m_1^2-m_2^2}
\log\biggl({m_1^2\over m_2^2}\biggr) \quad .
\eeq 
The function,$f(m_1^2,m_2^2)$, has the  
property that it vanishes for degenerate squark masses, 
\beq
f(m^2,m^2)=0\quad .
\eeq
  If one of the masses is much heavier than the other, then
we have 
\beq
f(m^2,0)\rightarrow m^2\quad .
\eeq
  Hence the contribution
of a squark doublet can be very large if the mass splitting
is large.  In the limit in which there is no mixing in
the stop
sector and $M_{\tilde {t}}>>M_{\tilde b}$,
the $\rho$  parameter becomes,
\beq
\delta \rho^{SUSY}\rightarrow {3 G_F\over 8\sqrt{2}\pi} 
M_{\tilde t}^2
.
\eeq
For squark masses in the $200~GeV$ region, the contribution
to the $\rho$ parameter is typically in the range of 
$10^{-3}$, depending on assumptions about $\tan\beta$
and the mixing parameters.  This can be used to limit the allowed values of
the squark masses and the squark mass splittings.\cite{rhosusy}

\subsection{Flavor Changing Neutral Currents}  
From the squark mass matrices of Eq. \ref{squarkm}, it is
apparent that the unitary matrices, ${\tilde U}$, which
diagonalize the squark mass matrices are not, in general,
the
same as the CKM matrix, $V$, which diagonalizes the quark mass
matrix.  The physical mass eigenstates are given by,
\beqn
q_i^p&=&\sum_j V_{ij}q_j
\nonumber \\
{\tilde q}^p_i&=&\sum _j {\tilde U}_{ij} {\tilde q}_j
\quad .
\label{squmix}
\eeqn
These matrices work their way into the various squark-couplings
and introduce flavor off-diagonal interactions, which
are severely restricted by limits on rare decays.

One
of the most restrictive  limits on flavor
off-diagonal interactions is from
the CLEO measurement of the
inclusive decay  $B\rightarrow X_s\gamma$,\cite{cleo} 
\beq
BR(B\rightarrow X_s\gamma)=(2.32\pm .67)\times 10^{-4}
, \qquad {\rm CLEO}  
\eeq  which
is sensitive to
loops containing the new particles of a SUSY model.
The contribution from $tH^\pm$ loops always  adds constructively
to the Standard Model result. There 
are additional contributions from squark-chargino loops, squark-
neutralino loops, and squark-gluino loops.  The contributions
from the squark-neutralino and squark-gluino loops are small and
are typically neglected.  The 
dominant contribution from the squark-chargino
loops  is proportional to $A_T\mu$
 and thus 
can have either sign relative to  the Standard Model and
charged Higgs loop contributions.
  There will therefore   be regions of SUSY parameter
space which are excluded depending upon whether there is
constructive or destructive interference between the Standard
Model/ charged Higgs contributions and the squark-chargino
contribution.\cite{bsg}
At small $\tan\beta$, the $b\rightarrow s \gamma$ branching
ratio is close to the Standard Model value for most
of the parameter space.     
For large
$\tan\beta$,  the squark-chargino contribution
is  completely dominant and  
 the limit which  can be obtained is very sensitive to the
sign$(A_T\mu)$.
Neglecting QCD corrections (which
are significant) we have,\cite{bsg2}
\beq  
{BR(b\rightarrow s\gamma)\over BR(b\rightarrow c e {\overline \nu})}
\sim{\mid V_{ts}^*V_{tb}\mid^2\over \mid V_{cb}\mid^2}
{6 \alpha\over \pi}
\biggl\{
C+{M_T^2 A_T\mu\over M_{\tilde t}^4}\tan\beta\biggr\}^2
\quad ,
\eeq
where $C$ (positive) is the contribution from the Standard
Model and charged Higgs loops and $M_{\tilde t}$ is the stop mass.
For $A_T\mu$ positive, this leads to a larger branching
ratio, $BR(b\rightarrow s\gamma)$, than in the Standard Model.
Since the Standard Model prediction is already somewhat
above the measured value,
we require $A_T\mu<0$ to avoid conflict
with the experimental measurement 
if $M_{\tilde t}$ is at the electroweak scale 
and $\tan\beta$ is large.
 Detailed plots of the allowed regions for various 
assumptions about $\tan\beta$, $\mu$, and $A_T$ are given in Refs.~
\citenum{deb} and \citenum{bb}.  
Depending on $\tan\beta$ 
and the sign of $A_T\mu$,
this process probes stop masses in the $100-300~GeV$
region. 

\begin{figure}[tb]
\vspace*{-2.2in}  
\centerline{\epsfig{file=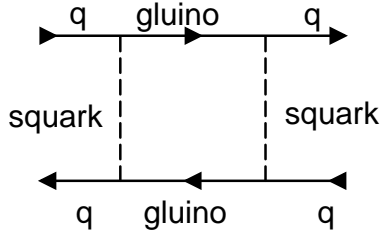,height=3.5in}}
\caption{Flavor changing neutral current effects
in $K^0- {\overline K}^0$ mixing from off-
diagonal $q {\tilde q} {\tilde g}$ couplings.
}
\vspace{.5in} 
\label{kkfeyn}
\end{figure}

Another class of important indirect limits on SUSY models comes
from flavor changing neutral current (FCNC)  processes
such as $K^0-  {\overline K}^0$ mixing.
Consider the squark-squark-gluino coupling resulting
from Eq. \ref{squmix},
\beqn
{\cal L}&=&g_s {\tilde g}\sum_{i=1}^3 q_i {\tilde q}_i^*+{\hbox{h.c.}}
\nonumber \\
&\sim & g_s {\tilde g}\sum_{l,j}
\biggl({\tilde U}V^\dagger\biggr)_{lj}
 q_j^p {\tilde q}_l^{p*}+{\hbox{h.c.}} 
\quad ,
\eeqn
where $i$ is a flavor label.

Diagrams such as Fig. \ref{kkfeyn},
 can  then introduce large contributions.
The amplitude from Fig. \ref{kkfeyn} will be schematically,
\beqn
{\cal A}\sim \alpha_s^2 \sum_{\alpha,\beta}
 \biggl(V {\tilde U}^\dagger\biggr)_{i,\alpha}
\biggl(V {\tilde U}^\dagger\biggr)^*_{j,\alpha}
\biggl(V {\tilde U}^\dagger\biggr)_{i,\beta}
\biggl(V {\tilde U}^\dagger\biggr)^*_{j,\beta}F(M_{\tilde \alpha}^2,
M_{\tilde \beta}^2)                           
,
\label{kkmix}
\eeqn
where $F(M_{\tilde \alpha}^2, M_{\tilde \beta}^2)$ 
is a complicated function of the squark masses in general.
If the squarks are degenerate, however, then $F$ is independent of 
$\alpha$ and $\beta$ and the sum of Eq. \ref{kkmix} can
be performed since,
\beq
\sum_\alpha
 \biggl(V {\tilde U}^\dagger\biggr)_{i,\alpha}
\biggl(V {\tilde U}^\dagger\biggr)^*_{j,\alpha}=0
\qquad {\hbox{for~}} i\ne j
\eeq
The contributions
from the off-diagonal squark-quark-gluino couplings
 vanish if the squarks have 
degenerate masses and so the limits are of the form:
\beq
{\Delta{\tilde M}^2\over {\tilde M}^2}< {\cal O}(10^{-3})
\qquad ,  
\eeq
where $\Delta {\tilde M}^2$ is the mass-squared splitting
between the different squarks and ${\tilde M}$ is the
average squark mass.  
A detailed discussion of FCNCs in SUSY models and references
to the literature is given in Refs.~ \citenum{fcnc}
and \citenum{halltasi}.
   As a practical
matter, the assumption is often made that there are $10$ degenerate
squarks, corresponding to the scalar partners of the $u,d,c,s,$ and
$b$ quarks, while the stop squarks  are allowed to have different 
masses from the others.
  This avoids phenomenological problems with FCNCs.
In both GUT models which we have considered, the CMSSM and the GMSB,
the squarks are degenerate at the GUT scale and so flavor
changing effects are introduced only through renormalization
group effects and are therefore small.

\section{Experimental Limits and Search Strategies}

We turn now  to  a discussion of some of the
existing experimental limits  on
the various SUSY particles and also to the search 
strategies applicable at present and future accelerators.
We focus primarily on the Higgs sector.
Detailed discussions are given in the contributions
of Refs.~\citenum{lamhoo} and \citenum{paigetasi}.

\subsection{Observing SUSY Higgs Bosons}

The goal in the Higgs sector is to observe the
$5$ physical Higgs particles, $h,H,A,H^\pm$, and to measure as
many couplings as possible to verify that the couplings are those
of a SUSY model.
  The lightest neutral Higgs boson in 
the minimal SUSY model is unique in the SUSY spectrum because there
is  an upper bound to its mass,
\beq
M_h<130~GeV.
\eeq
All other SUSY particles in the model
can be made arbitrarily heavy by adjusting the soft
SUSY breaking parameters in the model and so can  be
just out of reach of today's or
tomorrow's accelerators (although if they
are heavier than around $1~TeV$, much of the  motivation
for low energy SUSY disappears). The lightest SUSY Higgs
boson, however, cannot be much outside the range of LEPII
and can almost certainly be observed at the LHC .  Hence an
extraordinary theoretical effort has gone into the 
study of the reach of various accelerators in the SUSY
Higgs parameter
space since in this sector it will be possible to
experimentally exclude the MSSM if a light Higgs boson is
not observed.

 If we find
a light neutral Higgs boson, then we want to map out the
parameter space to see if we can distinguish it
from a Standard Model Higgs boson.
The only way to do this is to measure a variety of production
and decay modes and attempt to extract the various
couplings of the Higgs bosons to fermions and gauge
bosons.  
Since as $M_A\rightarrow \infty$, the $h$ couplings approach
those of the Standard Model, there will clearly be a region
where the SUSY Higgs boson  and the Standard Model Higgs boson are
indistinguishable.   
This is obvious from Figs. \ref{cbbhfig} and \ref{ctthfig}.

The search strategies for the SUSY Higgs boson depend
sensitively on the Higgs  boson branching ratios, which
in turn depend on $\tan\beta$.  In Figs. \ref{hbr1fig} and
\ref{hbr2fig},
we show the branching ratios for the lightest SUSY 
Higgs boson, $h$, into some interesting
decay modes assuming that there are no SUSY
particles light enough for the $h$ to decay into.
(These figures include radiative corrections to
the branching ratios, which can be important.\cite{squrad})  
For a Higgs boson below the $WW$ threshold, the
decay into $b {\overline b}$ is  completely dominant.  Unfortunately,
there are large QCD backgrounds to this decay mode and so it
is often necessary to look at rare decay modes. 
The branching ratios to $b {\overline b}$, $\tau^+\tau^-$,
and $\mu^+\mu^-$ are relatively insensitive to $\tan\beta$,
but the $W W^*$, $ZZ^*$, and $\gamma\gamma$ rates have
 strong dependences on $\tan\beta$ as we can see from
 Figs. \ref{hbr1fig} and \ref{hbr2fig}.

\begin{figure}[tb]
\vspace*{.3in}  
\centerline{\epsfig{file=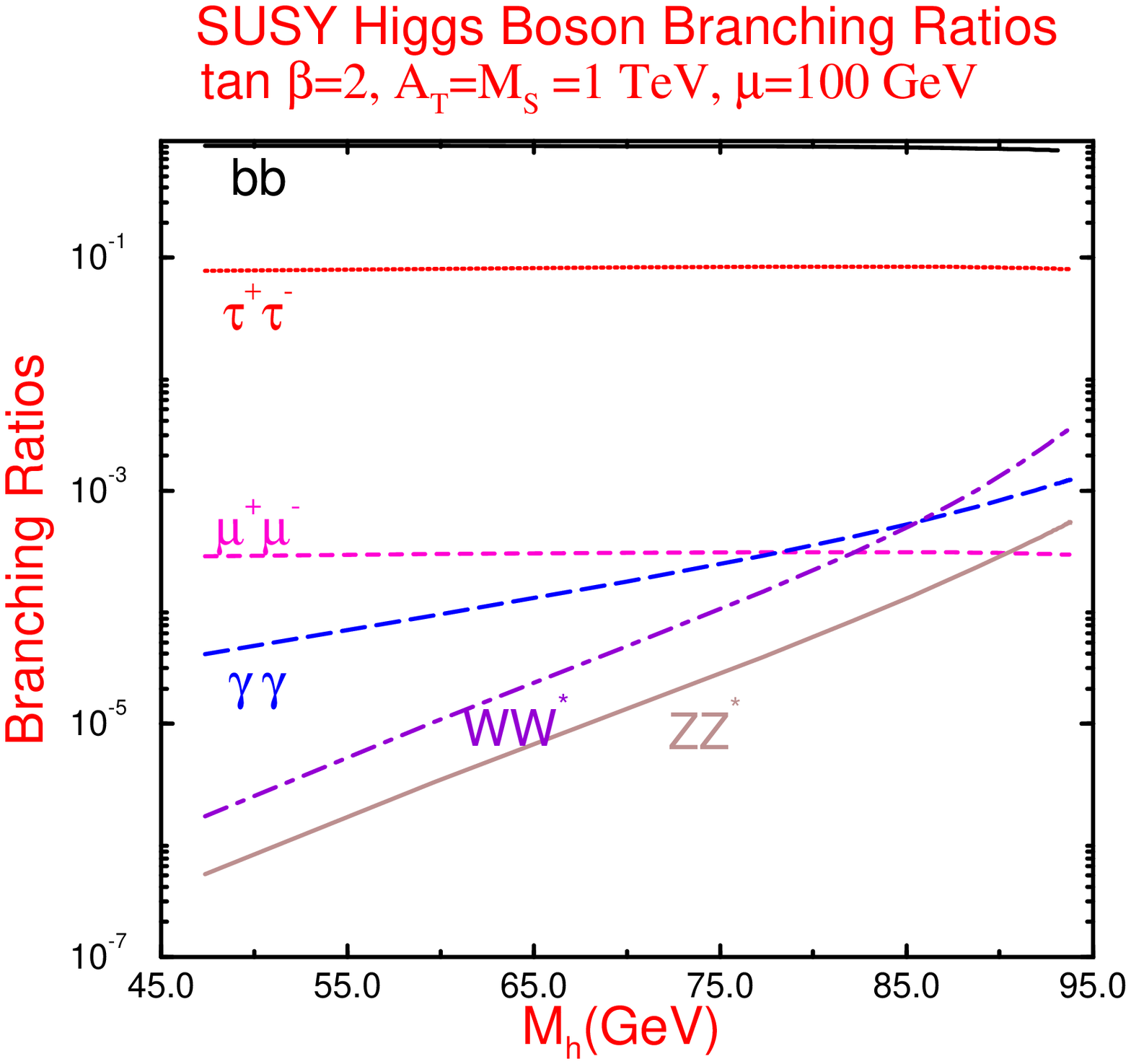,height=3.5in}}
\caption{Branching ratios of the lightest Higgs boson
assuming  decays into other SUSY particles are
kinematically forbidden.  $WW^*$ and $ZZ^*$ denote
decays with one off-shell gauge boson and $M_S$ is a typical squark
mass.}
\vspace{.5in} 
\label{hbr1fig}
\end{figure}  

\begin{figure}[tb]
\vspace*{.3in}  
\centerline{\epsfig{file=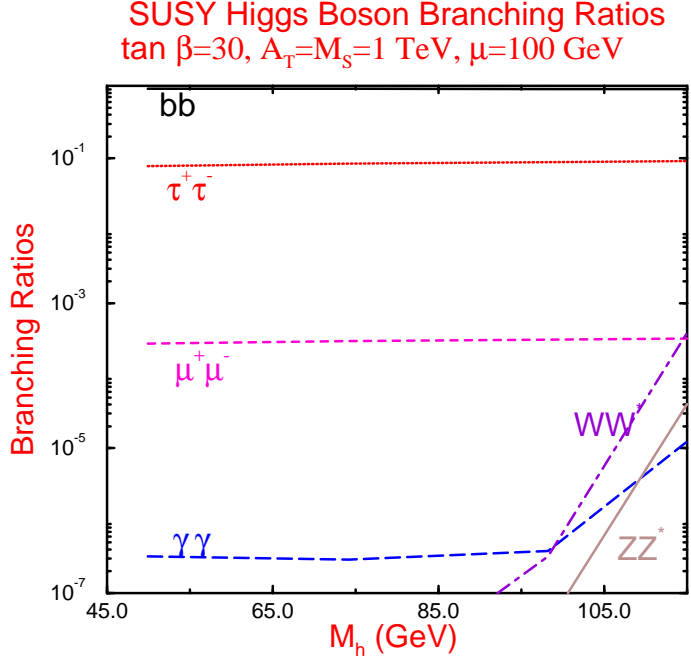,height=3.5in}}
\caption{Branching ratios of the lightest Higgs boson assuming
decays into other SUSY particles are  kinematically forbidden.}
\vspace*{.5in} 
\label{hbr2fig} 
\end{figure} 

\subsection{Higgs Bosons at LEP and LEPII}
Direct limits on SUSY Higgs production have
been obtained at LEP and LEPII by searching for the complementary
processes,\cite{lephiggs}
\beqn
e^+e^- &\rightarrow & Z h\nonumber \\
e^+e^-&\rightarrow & Ah
\quad .  
\eeqn 
From the couplings of Eq. \ref{vvhcoup},
 we see that the process $e^+e^-\rightarrow
Zh$ is suppressed by  $\sin^2(\beta-\alpha)$ relative
to the Standard Model Higgs boson production process, while
$e^+e^-\rightarrow Ah$  is proportional to  $\cos^2(\beta -\alpha)$.
  The moral is
that it is impossible to suppress both processes simultaneously
if both the $h$ and the $A$ are kinematically accessible!  
Because the Higgs sector (at lowest order)
can be described by the two parameters,
$M_h$ and $\tan\beta$, searches exclude a region in this plane.
At LEPII, the cross section for either $Zh$ (small
$\tan\beta$) or $Ah$ (large $\tan\beta$) is roughly
$.5~pb$.  With a luminosity of $150/pb/yr$, this leads to
$75$ events/yr before the inclusion of branching ratios.
Using preliminary data at $\sqrt{s}=183~GeV$,
ALEPH  and DELPHI exclude the region at $95\%$ confidence
level,\cite{sopc}
\beq
M_h> 73~GeV, ~{\rm for~any~}\tan\beta
\quad .
\eeq
For a given value of $\tan\beta$, there may be
 a stronger bound.
It is important to note that the LEP and LEPII
 searches do not leave any 
window for a very  light (on the order of a few
$GeV$) Higgs boson.
  The limit on a SUSY Higgs boson is weaker than the 
corresponding limit on the  Standard Model Higgs boson,\cite{sopc}
$M_h^{SM}>77.5~GeV$,
  due to the suppression in the couplings of the Higgs boson
to vector bosons.

\begin{figure}[tb]
\vspace*{.3in}  
\centerline{\epsfig{file=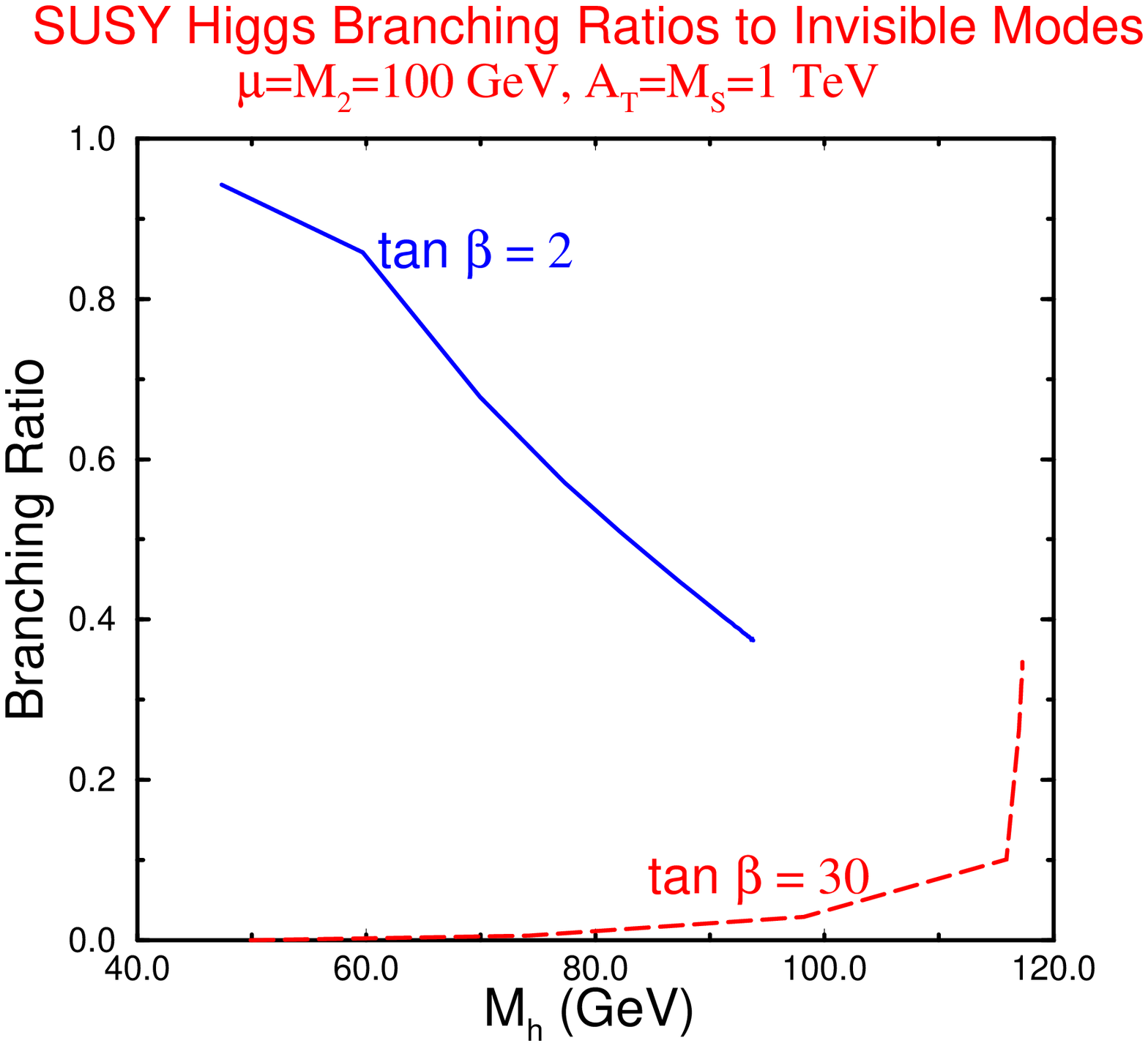,height=3.5in}}
\caption{Branching ratio of the lightest Higgs boson
to ${\tilde \chi}_1^0{\tilde \chi}_1^0$.
The curve with $\tan\beta=30$ has $M_{\chi_1^0}=33~GeV$,
while that with $\tan\beta=2$ has $M_{\chi_1^0}=7~GeV$.[34]}
\vspace*{.5in}
\label{hbrinvfig}  
\end{figure} 
The limits on the Higgs boson mass
 could be substantially altered if there is a significant
branching rate into invisible decay modes, such as the
neutralinos, 
\beq h,A\rightarrow
{\tilde \chi}_1^0{\tilde \chi}_1^0
\quad .
\eeq
These branching ratios  could be as high as $80\%$, but
are  extremely
model dependent since they depend sensitively on 
the parameters of the neutralino mixing matrix.
In Fig. \ref{hbrinvfig}, we show the branching ratio of the lightest
Higgs boson to ${\tilde \chi}_1^0{\tilde \chi}_1^0$ for
several choices of parameters.  For
$\tan\beta=2$, with the set of parameters which we have chosen, the
branching ratio is always greater than $40\%$.
  If the invisible decay modes are significant,
  a different search strategy  for the Higgs boson must
be utilized and LEPII can put a limit on the 
product of the Higgs boson mixing angles, $\beta-\alpha$, and
the branching ratio to invisible modes:
\beqn
R_1^2&\equiv&\sin^2(\beta-\alpha)BR(h\rightarrow {\hbox{visible}})
\nonumber \\
R_2^2&\equiv& \sin^2(\beta-\alpha
)BR(h\rightarrow {\hbox{invisible}})
\quad .
\eeqn
Studies of the expected limits on $R_1$ and $R_2$ at various
LEPII energies can be found in Ref.~\citenum{lephiggs}.

\subsection{Higgs Bosons at $\mu^+\mu^-$ Colliders} 

A $\mu^+\mu^-$ collider could in principle obtain stringent bounds
on a SUSY Higgs boson 
through its $s$-channel couplings to the Higgs.\cite{bargermm}
Since these couplings are proportional to the lepton mass,
the $s$-channel Higgs couplings will be much larger
at a $\mu^+\mu^-$ collider than at an $e^+e^-$ collider.   
For large $\tan\beta$, the lighter SUSY Higgs boson could be found in
the process $e^+e^-\rightarrow Zh$
 at LEPII or at an NLC. \cite{lephiggs,nlchiggs}
 However, for   
large $\tan\beta$, the coupling of the heavier Higgs boson to gauge
boson pairs is highly suppressed, (see Eq. \ref{vvhcoup}),
 so the $H$ can not be
discovered
 through $e^+e^-\rightarrow ZH$.  Instead the $H$ can be found
through $\mu^+\mu^-\rightarrow H\rightarrow b {\overline b}$, which
is enhanced by the factor $\tan^2\beta$ relative to
$\mu^+\mu^-\rightarrow h_{SM}\rightarrow b {\overline b}$.

A muon collider could also  be very useful for
obtaining precision measurements of the lighter Higgs boson mass
and couplings.
The idea is that the $h$ has been discovered through
either the process $e^+e^-\rightarrow Z h$ or $\mu^+
\mu^-\rightarrow Z h$ and so we have a rough idea of the
Higgs boson mass.
A muon collider could  be tuned to sit right on
the resonance, $\mu^+\mu^-\rightarrow h$.
  By doing an energy scan around the region
of the resonance, a precise value of the mass could be 
obtained due in large part to the narrowness of the muon beam
as compared to the beam in an electron collider.  (The
narrowness of the beam is
due to the suppression of synchrotron radiation in a muon collider.)

The narrowness of the $\mu^+\mu^-$ beam is parameterized in
terms of the beam energy resolution, $R$,  as
\beq
\delta_E=(7~MeV) \biggl({R\over .01\%}\biggr)
\biggl({\sqrt{s}\over 100~GeV}\biggr)
.
\eeq
If we compare the beam energy spread with the width of a SUSY
Higgs boson, we see that for $M_h\sim 100~GeV$ and $\delta_E 
<.01$  the energy spread is less than the
Higgs width, $\delta_E < \Gamma_h$.  In this limit the effective
cross section is given by,
\beq
\sigma_{eff}=\pi {\sqrt{2\pi}    \Gamma(h\rightarrow \mu^+\mu^-)
BR(h\rightarrow X)\over M_h^2\delta_E}
,
\eeq
making it clear that the smallest possible $R$ gives the best
measurement of the Higgs mass.  For $M_h=100~GeV$
and ${\cal L}=100 fb^{-1}$, Ref.~\citenum{bargermm} finds that a
$1~\sigma$ measurement of $60~MeV$ will be possible for the
Higgs mass.  In Fig. \ref{muhiggfig}, we show the cross section
for $\mu^+\mu^-\rightarrow h$ for several values of $R$.
This process requires that the beam energy be adjusted
to within $\delta_E$ of $M_h$.  
Both the lighter Higgs boson, $h$, and the heavier
neutral Higgs bosons, $H$, and $A$, can be studied at 
a muon collider through their $s$-channel production essentially
up to the kinematic limit over much of the parameter space. 
\begin{figure}[tb]
\vspace*{.3in}  
\centerline{\epsfig{file=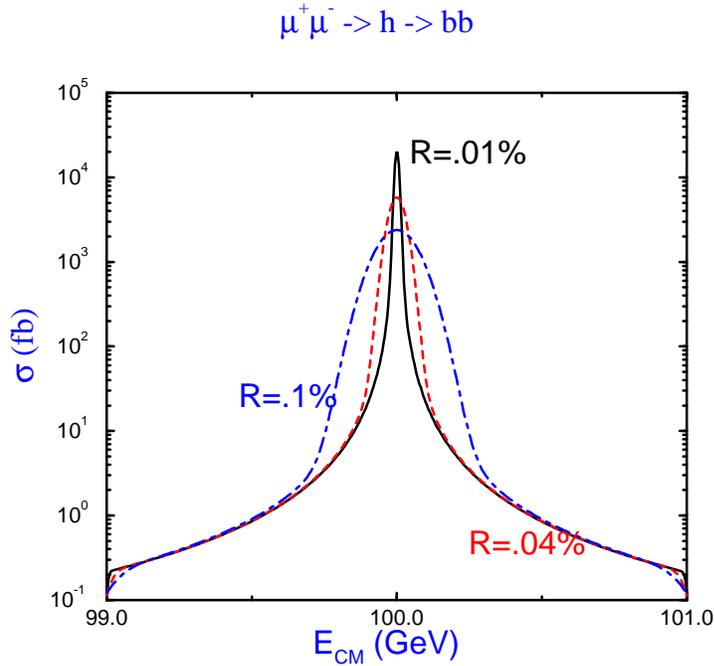,height=3.5in}}
\caption{Cross section for $\mu^+\mu^-\rightarrow
h$ for several values of $R$.  The cross
section must be multiplied by the model dependent
couplings $(C_{\mu\mu h}C_{bbh})^2$.}
\vspace*{.25in}  
\label{muhiggfig}
\end{figure}

\subsection{Higgs Bosons at the LHC}
 At the LHC,  for most Higgs masses the dominant
production mechanism is gluon fusion, $gg\rightarrow h,H$ or $A$.
These processes proceed through triangle diagrams with
internal $b$ and $t$
quarks and also through squark loops.  In the limit in which the
top quark is much heavier than the Higgs boson, the top
quark contribution is a constant, while the $b$ quark contribution
is suppressed by $(M_b/v)^2\log(M_h/M_b)$ and
so only the top quark contribution is numerically
important.  For large $\tan\beta$,
however, the dominance of the top quark loop is overtaken by the 
large ${\overline b}
bh$ coupling and the bottom quark contribution becomes
important, (as seen in Fig. \ref{cbbhfig}).
The production rate is therefore extremely sensitive to $\tan
\beta$.  
  Both QCD corrections and squark loops can also be
numerically important.\cite{spira} In fact, the QCD
corrections increase the rate by a factor between $1.5$ and $2$. 
  The rate for $pp\rightarrow h$
at the LHC is
shown in Fig. \ref{sushiggfig}
 as a function of $\tan\beta$ for $M_h=80~GeV$.
  We see that there are a relatively large number
of events.  For example, for $M_h\sim 80~GeV$, the LHC cross section is
roughly $100~pb$.  With a luminosity of $10^{33}/cm^2/sec$, this
yields $10^6$ events/year.

\begin{figure}[tb]
\centerline{\epsfig{file=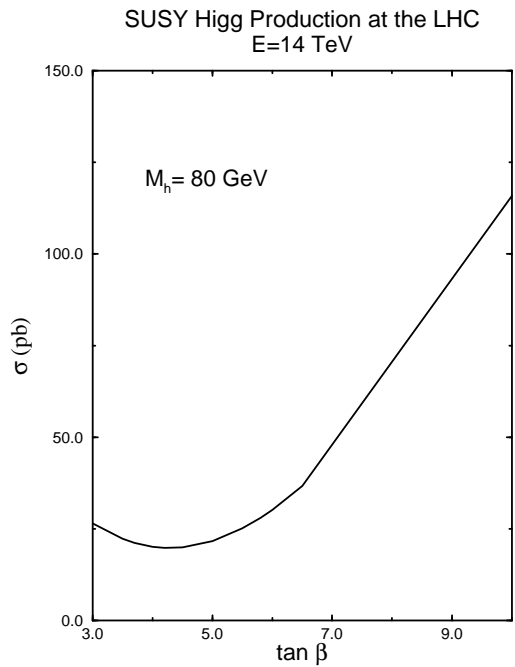,height=3.5in}}
\caption{Cross section for production of the lightest
SUSY Higgs boson at the LHC as a function of $\tan\beta$.}    
\vspace*{.25in}  
\label{sushiggfig}
\end{figure}  

Unfortunately, there are large backgrounds to the dominant decay
modes, 
( $b \overline{b}, \mu^+\mu^-$, and $\tau^+\tau^-$),
for a Higgs boson in the $100~GeV$ region.\cite{lhchiggs}
The decay $h \rightarrow 
Z Z^*$ will be useful, but its
branching ratio decreases rapidly with decreasing
Higgs mass.  In order to cover the region around $M_h\sim 80-100~GeV$,
 it will be 
necessary to look for the Higgs decay to $\gamma\gamma$,
\beq
gg\rightarrow h,H\rightarrow \gamma\gamma
\quad .
\eeq
(From Figs. \ref{hbr1fig} and \ref{hbr2fig},
 we see that the $BR(h\rightarrow
\gamma\gamma)$ is typically $< 10^{-3} - 10^{-5}$.)
This process will be extremely difficult to observe at
the LHC due to the small rate and the desire to observe
the $h\rightarrow \gamma\gamma$  decay
has been one of the driving forces behind the design
of both LHC detectors.\cite{lhcprop}
For large
$M_A$, the rate is roughly independent of $\tan\beta$
for $\tan\beta>3$ 
and can be used to exclude $M_A>150~GeV$ with
the full design luminosity of $3\times 10^{5}/pb$.
(With a smaller luminosity of $3\times 10^4/{\rm pb}$,
the $h\rightarrow \gamma\gamma$ process is sensitive to
roughly $M_A>270~GeV$.  See Fig. \ref{atlasfig}
 for the exact region.)

In order to exclude the region with smaller $\tan\beta$, the 
process $pp\rightarrow Wh\rightarrow l \nu {\overline b} b$
can be used.\cite{wm}
  This process can exclude a region with $M_A > 
100~GeV$ and $\tan\beta < 4$ (see Fig. 15)
 and demonstrates the crucial need for $b$-tagging
at the LHC in order to cover all  regions
of SUSY parameter space.
In Fig. \ref{atlasfig}, we see the excluded region formed by combining
potential LHC and LEP limits.\cite{dfr} 
A variety of Higgs production and decay channels can be
utilized in order to probe the entire $\tan\beta-M_A$ plane.
 The most striking feature of Fig. \ref{atlasfig} is the
region
 around $M_A\sim100~GeV$ for $\tan\beta > 5$ where 
the lightest Higgs boson cannot be observed.
In the region with
$M_A\sim 100-200~GeV$, both the $h {\overline t} t$ coupling
and the $h\rightarrow \gamma\gamma$ branching ratios are suppressed
relative to the Standard Model rates.  Furthermore, the dominant
decays, $h\rightarrow b {\overline b}$ and $h\rightarrow 
\tau^+\tau^-$, have large backgrounds from $Z$ decays.
It will be necessary to look for the decays of the heavier
neutral Higgs boson, $H$, or the pseudoscalar, $A$,
 to $\tau^+\tau^-$
pairs in order to probe this region,
\beq
H,A\rightarrow \tau^+\tau^-\rightarrow l \nu {\overline q} q
\quad .
\eeq
Detector studies by the ATLAS and CMS collaborations suggest
that these decay modes may be accessible.

\begin{figure}[tb]
\vspace*{.5in}  
\centerline{\epsfig{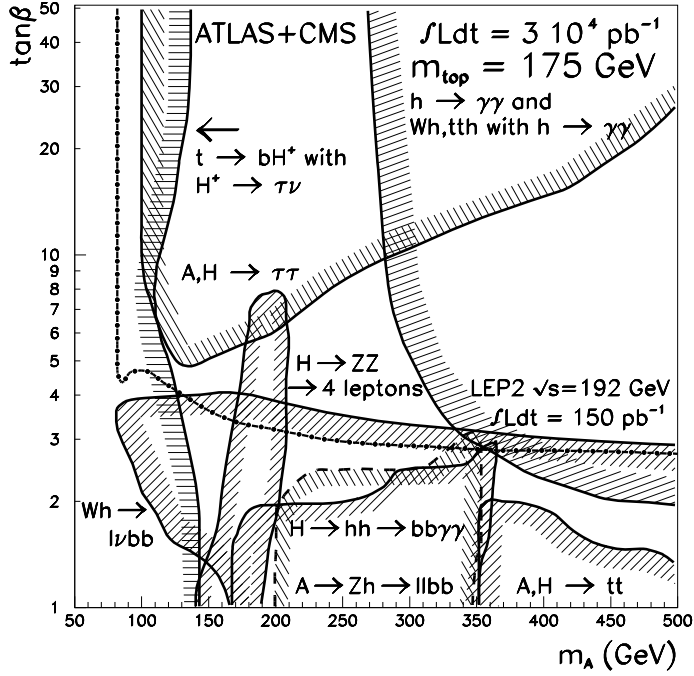}} 
\caption{LHC (with low luminosity)
  and LEPII discovery limits for SUSY
Higgs bosons. Figure from Ref. [69].}
\vspace*{.25in}  
\label{atlasfig}
\end{figure}

\section{Finding the Zoo of SUSY Particles}

In addition to the multiple Higgs particles associated
 with SUSY models, there is a whole zoo of other new
particles.  There are the squarks and gluinos which
are produced through the strong interactions and the sleptons,
charginos, and neutralinos which are produced weakly.

We begin by discussing some generic signals for supersymmetry.
All SUSY particles 
in a theory with $R$ parity conservation
eventually decay to the LSP, which is
typically taken to be the lightest neutralino, ${\tilde \chi}_1^0$, 
although in GMSB models it is the gravitino.\cite{wein}
  The LSP's interactions with matter are extremely
 weak and so it escapes detection leading to missing energy.  
 
\begin{itemize}
\item
A basic SUSY signature is missing energy, $E_T^{miss}$,
 from the undetected LSP.
\end{itemize}
A SUSY interaction
typically produces a cascade of decays, until the final state
consists of only the LSP plus jets and leptons.  Hence typical final 
states are:
\begin{itemize}
\item
$l^\pm+{\rm jets}+E_T^{miss}$
\item
$l^\pm l^\pm + {\rm jets}+E_T^{miss}
$
\item $l^\pm l^\mp+{\rm jets}+E_T^{miss}$ 
\quad .
\end{itemize} 
Because of the presence of the LSP in the final state
of an $R$ parity convserving theory, it is not possible to
completely reconstruct the masses of the SUSY particles, although
a significant amount of information about the masses
can be obtained from the event structure.
\begin{itemize}
\item
A combination of characteristic signatures
 may determine the SUSY model.
\end{itemize}

Because the gluinos are Majorana particles, they have some special
characteristics which may be useful for their experimental detection.
They have the property:
\beq
\Gamma({\tilde g}\rightarrow l^+ X)=
\Gamma({\tilde g}\rightarrow l^- X)\quad .
\eeq
Hence gluino pair production can lead to final states with same sign 
$l^\pm l^\pm$ pairs.\cite{fp}
  The standard model background for this
type of signature is rather small.
\begin{itemize}
\item 
Same sign di-lepton pairs are a useful signature for gluino pair
production.
\end{itemize} 

Another generic signature for SUSY particles is tri-lepton 
production.\cite{paigetasi}  If 
we consider the process of chargino-neutralino production,then it is 
possible to have the process:
\beq
{\tilde \chi}_1^\pm {\tilde \chi}_2^0\rightarrow
l \nu {\tilde\chi}_1^0+
{\overline l}^\prime l^\prime {\tilde \chi}_1^0
\quad .
\eeq
Again this is a signature with a small standard model background.

How these signatures can be observed at the LHC  is the subject 
of F. Paige's lectures at this school.\cite{paigetasi}

\subsection{Chargino and Neutralino Production}

As an example of SUSY particle searches, we consider the
search for chargino pair production at an electron-positron
collider,
\beq
e^+ e^- \rightarrow {\tilde \chi}_1^+ {\tilde \chi}_1^-
\quad , 
\eeq 
(where ${\tilde \chi}_1^\pm$ are the lightest charginos.)
The chargino mass matrix has a contribution 
from both  the fermionic partner
of the $W^\pm$,
 ${\tilde\omega}^\pm$, and from the fermionic partner of the
charged
Higgs, ${\tilde h}^\pm$, and so depends on the two unknown parameters
in the mass matrix, $\mu$ and $M_2$.  (See Eq. \ref{charmass}).
The calculation of the cross section is presented in the Appendix.

The search proceeds by looking for the decay
${\tilde \chi}^\pm_1 \rightarrow {\tilde\chi}^0_1 
{\overline f}^\prime f$.
The assumption is made that the ${\tilde \chi}^0_1$ is
stable and escapes the detector unseen.   Using this 
technique, ALEPH obtains a limit,\cite{aleph}
\beqn
M_{\tilde \chi^\pm}&>& 85.5 ~GeV\quad {\hbox {for}}~  \mu=-500
	~GeV,~~\mid \mu\mid >> M_2
\nonumber \\
M_{\tilde \chi^\pm}&>& 85.0~GeV\quad {\hbox {for}}~ M_2=500
	~GeV,~~~M_2 >> \mid \mu\mid
\eeqn
based on a total of $21.5~pb^{-1}$ of data at energies between
$\sqrt{s}=161.3~GeV$ and $172.3~GeV$ and assuming
$M_{\tilde{\nu}}>200~GeV$.
This limit is not very sensitive to $\tan\beta$.
In the gaugino region, $\mid \mu\mid >>M_2$,
 there is a strong sensitivity to
the mass of the sneutrino as the sneutrino mass is lowered.

It is interesting to compare the search for charginos and 
neutralinos at LEP with the search at the Tevatron
and the LHC. At
the LHC one clear signature will be,\cite{chargino} 
\beq
pp\rightarrow {\tilde \chi}^\pm_1 {\tilde \chi}^0_2
\nonumber 
\eeq
with,
\beqn 
{\tilde \chi}^\pm_1&\rightarrow &l^{\prime\pm} \nu {\tilde \chi}_1^0
\nonumber \\  
{\tilde \chi}^0_2&\rightarrow& l {\overline l} {\tilde \chi}_1^0
\quad . 
\eeqn
The cross section for this process is $\sigma \sim 1-100~pb$  
for masses  below  $1~TeV$.
This gives a ``tri-lepton signature" with three hard, isolated leptons,
significant $E_T$ and little jet activity.  The dominant Standard 
Model backgrounds are from $t {\overline t}$ production
(which  can be eliminated by requiring that the 2 fastest
leptons have the same sign) and $W^\pm Z$ production (which
is eliminated by requiring that $M_{ll}\ne M_Z$).

At the LHC the largest background
to chargino and neutralino production is from other
SUSY particles, such as squark and gluino production,
which also give events with leptons, multi-jets, and missing $E_T$.
\begin{itemize}
\item The biggest background to SUSY is SUSY itself.
\end{itemize}
  Since the
squarks and gluinos are strongly interacting, they will generate
more jets and a harder missing $E_T$ spectrum than the charginos
and neutralinos.
This    
 can  be used to separate 
squark and gluino production from the chargino and neutralino
production process of interest.

\begin{itemize}
\item
The tri-lepton signal offers the possibility of untangling the
${\tilde \chi}^+{\tilde \chi}^0$ signal from the gluino
and squark background.  
\end{itemize}

CDF has searched for this decay chain and obtains a limit
which is sensitive to $\tan \beta$ and $\mu$.  For
$\tan\beta=4$ and $\mu=-200~GeV$, they obtain a limit
$M_{{\tilde \chi}^+} > 70~GeV$.\cite{cdflim}
This is somewhat weaker than
the limit found at LEPII.

Aside from observing the process and verifying
the existence of charginos and neutralinos,
 we would also like to obtain a 
handle on the masses of the SUSY particles.
  The kinematics are such that,
\beq
0 < M_{ll} < M_{{\tilde \chi}^0_2}-M_{{\tilde \chi}^0_1}
\quad ,
\eeq
and hence the distribution $d\sigma/dM_{ll}$ has a sharp cut-off
at the kinematic boundary which can be used to
obtain information on the masses.  Recently, significant
progress has been made in our understanding of the capabilities
of a hadron collider for extracting values of the SUSY particle
masses from different event distributions.\cite{paigetasi} 

\subsection{Squarks, Gluinos, and Sleptons}
 
Squarks and sleptons,(${\tilde f}_i$),  can be
 produced at both $e^+e^-$ and
hadron colliders.  At LEP and LEPII,
 they would be  pair produced via
\beq
e^+e^-\rightarrow \gamma,Z
\rightarrow {\tilde f}_i {\tilde f}^*_i.
\eeq 
If there were a scalar with mass  less than half the
$Z$ mass,  it would 
increase the total width of the $Z$, $\Gamma_Z$.
Since $\Gamma_Z$ agrees quite precisely with the
Standard Model prediction,  
the measurement of the $Z$ lineshape gives
\beq
M_{\tilde q}> 35-40~GeV\eeq  
 for squarks and sleptons.\cite{pdg}  
The limit from the $Z$ width is  
particularly important  because it is independent
of the squark or slepton decay mode and so applies for
any model with low energy supersymmetry.
(Remember that the coupling of the sfermions to $\gamma,~Z$ is
fixed by gauge invariance.)

\begin{figure}[tb] 
\vspace*{.3in}  
\centerline{\epsfig{file=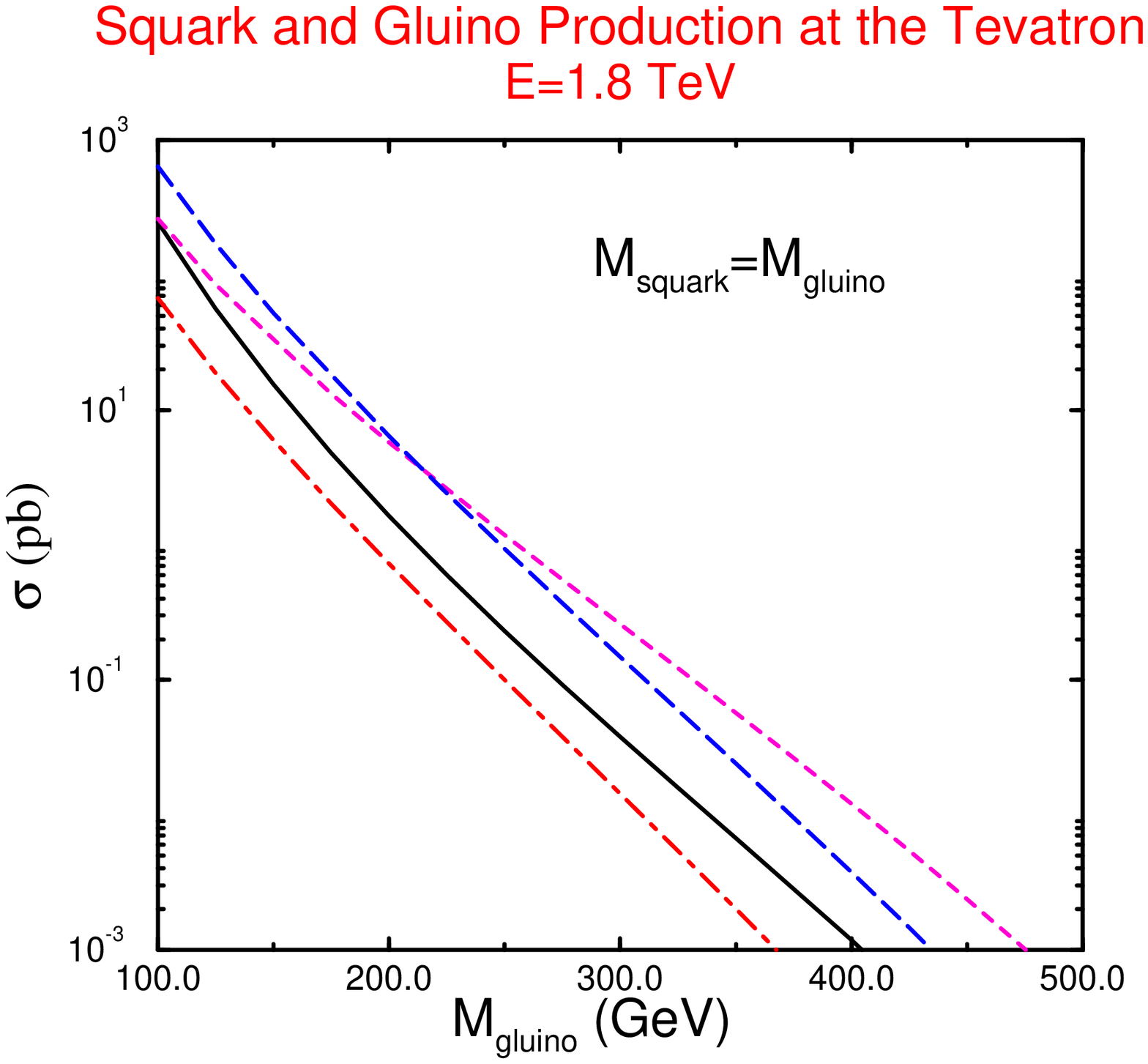,height=3.5in}}
\caption{Squark and gluino production
at the Tevatron assuming $M_{\tilde q}=M_{\tilde g}$.
  The solid line is $p {\overline p}
\rightarrow {\tilde g}{\tilde g}$, the dot-dashed
${\tilde q}{\tilde q}$,
 the dotted ${\tilde q}{\tilde q}^*$,
 and the dashed is ${\tilde q}{\tilde g}$. This figure includes only
the Born result and assumes $10$ degenerate squarks.}
\end{figure}
  
\begin{figure}[tb]
\vspace*{.3in} 
\centerline{\epsfig{file=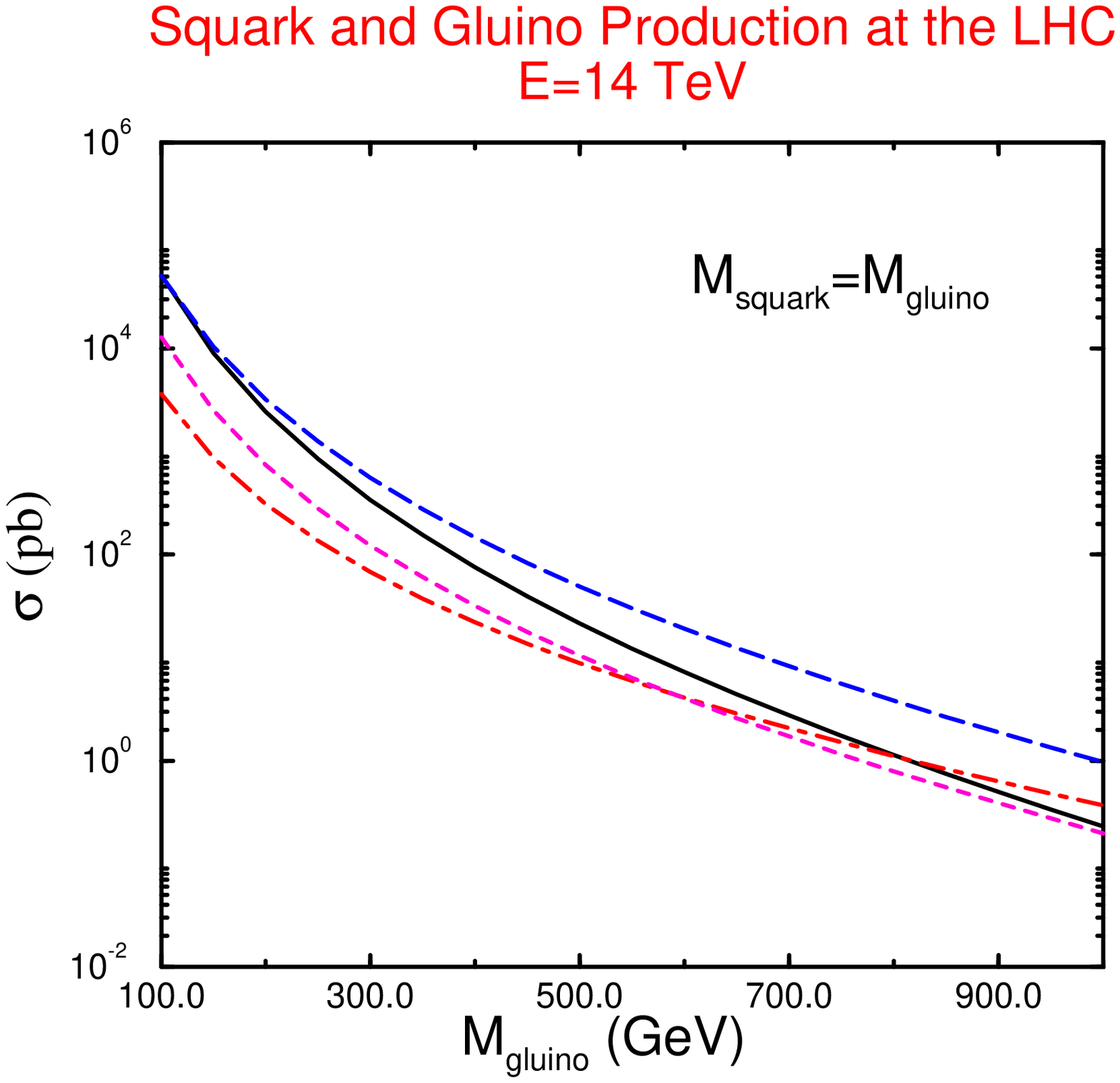,height=3.5in}}
\caption{Squark and gluino production at the LHC 
assuming $M_{\tilde q}=M_{\tilde g}$.
The solid line is $ p p
\rightarrow {\tilde g}{\tilde g}$, the dot-dashed 
${\tilde q}{\tilde q}$, the dotted ${\tilde q}{\tilde q}^*$,
and the dashed is ${\tilde q}{\tilde g}$.  This figure
includes only the Born result and assumes $10$ degenerate 
squarks.}
\vspace*{.5in} 
\end{figure}

There are also limits on the direct production of
 squarks and gluinos from
the Tevatron.  The rates for squark and
gluino production at both the Tevatron and the
  LHC are shown in Figs. 17 and 18 and analytic expressions for
the Born cross sections can be found in Ref.~\citenum{sigsusy}.
The QCD radiative corrections to these process are
large and increase the cross sections 
by up to a factor of two.\cite{squglu} 
 We neglect the mixing
effects in the squark mass matrix and assume that there are $10$
degenerate squarks associated with the light quarks.  
(The top squarks are assumed to be different since here 
mixing effects can be relevant.) 
   The cross sections are significant, around
$1~pb$ for squarks and gluinos in the few hundred GeV range
at the Tevatron.

The cleanest signatures for squark and gluino production
are jets plus missing $E_T$ from
the undetected  LSP, 
assumed to be ${\tilde \chi}_1^0$,
 and jets plus multi- leptons
plus missing $E_T$.\cite{squark}
  It will clearly be exceedingly difficult
to separate the effects of squarks and gluino production,
since they both contribute to the same experimental signature.
Limits from the Tevatron require (at the $95\%$ confidence
level) that \cite{tevlimits}
\beqn
M_{\tilde g}&>& 175~GeV\nonumber \\
M_{\tilde q}&>& 175~GeV\quad {\hbox{for~}} M_{\tilde g}< 300~GeV
\quad .
\eeqn
Details about squark and gluino searches at the Tevatron can be
found in the lectures of Lammel
at this school, Ref.~\citenum{lamhoo}, and 
about searches at the LHC in Ref.~\citenum{paigetasi}.

Limits on the stop squark are particularly interesting since in many models 
the stop
is the lightest squark.  There are two types of stop squark decays which 
are relevant.  The first is,
\beq
{\tilde t}\rightarrow b {\tilde \chi}^+_1 
\rightarrow b  f {\overline f}^\prime {\tilde \chi}_i^0
\quad .
\eeq 
The signal for this decay channel  is jets plus missing 
energy.  This signal shares many features with the
dominant
top quark decay, $t\rightarrow b W^+$.
Another possible decay
chain for the stop squark
is 
\beq
{\tilde t}\rightarrow c {{\tilde \chi}_1^0},
\eeq
which also leads to jets plus missing energy.  The two cases
must be analyzed separately. 
Limits from LEPII require that the lightest stop squark mass
be greater than $67~GeV$.  This limit is independent of the 
mixing in the stop mass sector, but is sensitive to the lightest
neutralino mass.\cite{lepstop} A slightly higher bound is found
at the Tevatron, 
\beq
M_{\tilde t}>93~GeV
,
\eeq
again depending on the mass of the lightest neutralino.\cite{d0stop}

 From the examples we have given, it is clear that searching for
SUSY at a hadron collider is  particularly
challenging since there will
typically be many SUSY particles which are kinematically
accessible.  Hadron colliders  thus have a large discovery
potential, but it is difficult to separate the various processes.
To a large extent, one must trust the generic signatures of
supersymmetry:  $E_T^{\rm miss}$, plus multi-jet and multi-lepton
signatures.  One will need to observe a signal in many channels in
order to verify the consistency of the model.
 
\begin{figure}[tb]
\vspace{.2in} 
\centerline{\epsfig{file=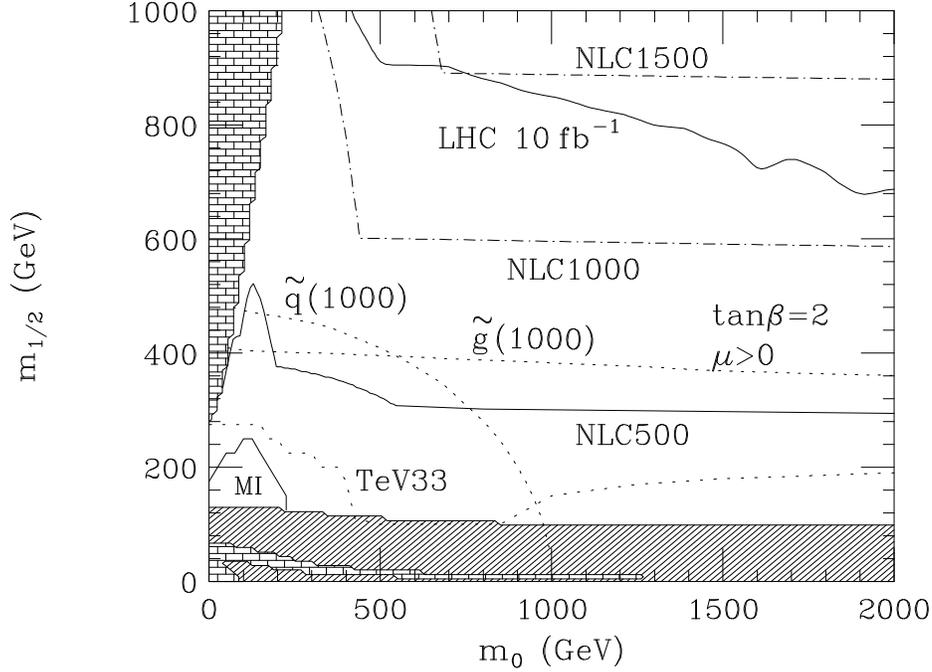,height=3.5in}}
\caption{Discovery reach in the CMSSM at various possible
future accelerators.}
\label{snowfig}
\end{figure}

\section{CONCLUSION}

A preliminary investigation of the differing capabilities
for observing supersymmetry at various colliders was made
at the Snowmass 1996 meeting.
This study considered the
CMSSM and mapped out the region in $m_0-m_{1/2}$ space which
would be accessible at the different machines.
 For each machine,
a number of different discovery channels were
considered.   By combining the various channels,
the curves of Fig. \ref{snowfig},
were obtained.   The study came
to the rough conclusion that a high energy $e^+ e^-$ collider
with $\sqrt{s}\sim 1.2-1.5~TeV$ had a similar SUSY discovery
reach to that of the LHC with $10~fb^{-1}$, as can be seen in 
Fig. \ref{snowfig}.\cite{snow2}  Such a strong conclusion is
only possible because the CMSSM relates the masses of the 
various particles to each other.  The Snowmass study only required
the discovery of the existence of  supersymmetry in a single
channel and did not consider how to
elucidate the characteristics of a SUSY model.

Weak scale supersymmetry is a theory in
need of experimental confirmation.  The 
theoretical framework has evolved to a point where
predictions for cross sections, branching ratios, and decay
signatures can be reliably made.  In many cases,
calculations exist beyond the leading order in perturbation
theory.   However, without
experimental observation of a SUSY particle or a
precision measurement which disagrees with the Standard
Model (which could be explained by
SUSY particles in loops) there is no way of
choosing between the many possible manifestations of low
energy SUSY and thereby  fixing the parameters in the soft SUSY
breaking Lagrangian. 
With the coming of LEPII, the Fermilab Main Injector, and the
LHC, large regions of SUSY parameter space will be 
explored and we can only hope that some evidence for
supersymmetry will be uncovered.

\section*{Appendix:  $e^+e^-\rightarrow {\tilde \chi^+}_i {\tilde \chi^-}_j$}

In this appendix we compute the cross section for 
$e^+e^-\rightarrow {\tilde \chi^+}_i {\tilde \chi^-}_j$ using
the Lagrangian of Eq.\ref{charglag}.
This result can be found in many
standard references \cite{eech1,eech2} and our derivation
follows that of Ref.~\citenum{eech2}.  There are $3$ diagrams which contribute
to this process; $s$-channel $\gamma$ and $Z$ exchange
and $t-$ channel sneutrino exchange.
The amplitudes in terms of $4-$ component Dirac spinors are:
\beqn
{\cal A}_\gamma&=&
{e^2\over s} {\overline v}(e^+) \gamma^\mu u(e)
	{\overline u}({\tilde \chi}^+)\gamma_\mu u({\tilde \chi}^-)
\nonumber \\
{\cal A}_Z&=&
{g^2\over 2 \cos^2\theta_W}
{1\over s-M_Z^2}
 {\overline v}(e^+) \gamma^\mu\biggl( R_e P_++
L_e P_-\biggr) 
\nonumber \\ && \qquad \cdot u(e)
	{\overline u}({\tilde \chi}^+)\gamma_\mu 
\biggl(C_{ij}^+P_++C_{ij}^-P_-\biggr)
u({\tilde \chi}^-)
\nonumber \\
{\cal A}_{\tilde \nu}&=&
- {g^2\over t-{\tilde m}^2_\nu}V_{i1}V_{j1}^* {\overline u}({\tilde \chi}^-)
P_- u(e^-) {\overline v}(e^+) P_+ u({\tilde \chi}^+)
~,
\eeqn
with $P_{\pm} ={1\over 2} (1\pm\gamma_5),~ R_e=2\sin^2\theta_W,~
L_e=2\sin^2\theta_W$, and $C_{ij}^\pm$ defined in Eq. \ref{cchargdef}. 
Since the charginos are Majorana particles, there is no distinction
between spinor and anti-spinor for them.  (In obtaining ${\cal A}_Z$
from Eq. \ref{charglag},
 we have used Eq. \ref{switch}.)  We can use the Fiertz rearrangement,
\beq
({\overline \psi}_1P_- \psi_2)({\overline \psi}_3 P_+ \psi_4)=
{1\over 2}({\overline \psi}_3 \gamma^\mu P_- \psi_2)
({\overline \psi}_1 \gamma_\mu P_+ \psi_4)
\eeq
to rewrite the sneutrino contribution so that it has the
same form as the $\gamma$ and $Z$ contributions,
\beq
{\cal A}_{\tilde \nu}=
{1\over 2}
 {g^2\over (t-{\tilde m}^2_\nu)}V_{i1}V_{j1}^* 
{\overline v}(e^+) \gamma^\mu P_- u(e^-)
{\overline u}({\tilde \chi}^+)
\gamma_\mu
P_-  u({\tilde \chi}^-) \quad .
\eeq
In terms of helicity eigenstates the total amplitude is then
\beq
{\cal A}(e^+e^-\rightarrow {\tilde \chi}^+_i
{\tilde \chi}^-_j)={e^2\over s}\sum_{m,n=\pm}
X_{mn}
{\overline v}(e^+) \gamma^\mu P_m u(e^-)
{\overline u}({\tilde \chi}^+)
\gamma_\mu
P_n  u({\tilde \chi}^-)
\eeq
where
\beqn
X_{++}&=& 1+{L_e C_{ij}^-\over 2\sin^2\theta_W\cos^2\theta_W}
{1\over 1-M_Z^2/s} +{ V_{i1}V^*_{j1}\over 2 \sin^2\theta_W}
{s\over t - {\tilde m}_\nu^2}
\nonumber \\
X_{--}&=& 1+{R_e C_{ij}^+\over 2\sin^2\theta_W\cos^2\theta_W}
{1\over 1- M_Z^2/s}
\nonumber \\
X_{+-}&=& 1+{R_e C_{ij}^-\over 2\sin^2\theta_W\cos^2\theta_W}
{1\over 1- M_Z^2/s}
\nonumber \\
X_{-+}&=& 1+{L_e C_{ij}^+\over 2\sin^2\theta_W\cos^2\theta_W}
{1\over 1- M_Z^2/s}
\quad .
\eeqn
This result makes clear the importance of the relative sign between the
sneutrino contribution and the $s-$ channel diagrams.
The sneutrino exchange interferes destructively with
the $\gamma$ and $Z$ diagrams for  small sneutrino
mass.  In Fig. 20, we show the cross section for pair 
production of the lightest chargino.  In the Higgsino-like
region, $\mid \mu\mid << M_2$,
the cross section is relatively insensitive to
the sneutrino mass, while in the gaugino-like region,
$\mid \mu\mid >> M_2$,  there is 
a strong suppression of the cross section for ${\tilde m}_\nu < 200~
GeV$.  

It is straightforward to find the differential cross section,
\beqn
{d \sigma\over d\cos \theta}&=& {\alpha^2 \beta \pi \over 8 s}
\biggl\{
(X_{++}^2+X_{--}^2) (1-\beta\cos\theta)^2
\nonumber \\
&&+(X_{+-}^2+X_{-+}^2) (1+\beta\cos\theta)^2
\nonumber \\  && 
+{8 M_{\tilde \chi^+}^2\over s} (X_{++}X_{-+}+X_{--}X_{+-})
\biggr\}\quad , 
\eeqn
where $\beta^2=1-4 M_{\tilde \chi^+}^2/s$.    

\begin{figure}[tb]
\vspace{.3in} 
{\centerline{\epsfig{file=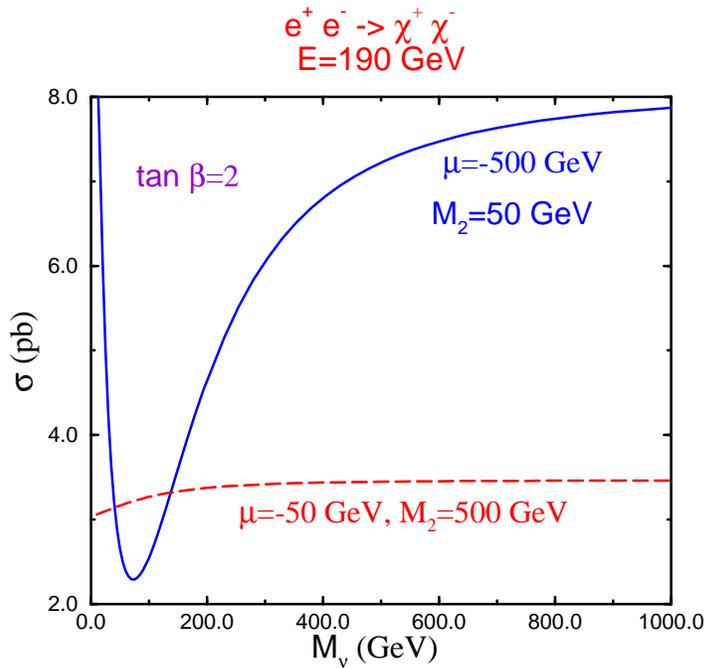,height=3.5in}}}
\caption{Cross section for pair production of
the lightest chargino as a function of the sneutrino mass,
$M_{\tilde \nu}$.}
\end{figure}

\section*{Acknowledgments}

I thank all the students at this school who asked such
wonderful questions and really made me think about 
supersymmetry. 
Helpful discussions with Frank Paige are   gratefully
acknowledged.
This work has been supported by the DOE under contract
number DE-AC02-76-CH-00016.


\section*{References}
\begin{thebibliography}{99}

\bibitem{lqt}{B.~Lee, C.~Quigg, and H.~Thacker, {\it Phys.
Rev.} {\bf D16} (1977) 1519;
D.~Dicus and V.~Mathur, {\it Phys. Rev.} {\bf D7} (1973) 3111.}  
 

\bibitem{hkrep} {H.~Haber and G.~Kane, {\it Phys. Rep.}
{\bf 117C} (1985) 75.}

\bibitem{bagtasi}{J.~Bagger, Lectures presented at 
the 1991 Theoretical Advanced Study Institute, Boulder,
CO, June, 1991;
Lectures presented at the 1995 Theoretical Advanced
Study Institute, Boulder, CO, June, 1995, hep-ph/9604232;
  H.P.~Nilles, {\it Phys. Rep.}
{\bf 110} (1984) 1;
H.~Haber, Lectures presented at the 1986 
Theoretical Advanced Study Institute, Santa Cruz, 
CA, June, 1986;  R.~Arnowitt, A.~Chamseddine and P.~Nath,
{\it Applied N=1 Supergravity}, (World Scientific, 1984);
V.~Barger and R.~Phillips, {\it Recent Advances
in the Superworld}, J.~Lopez and D.~Nanopoulos, Ed.
(World Scientific, 1994).} 

\bibitem{martin} {S.~Martin, Contribution to {\it Perspectives
	in Supersymmetry}, ed. G. ~Kane, (World Scientific,
Singaport, 1997), hep-ph/9709356.}


\bibitem{xerxes} {X. Tata, Lectures presented at the 1995
Theoretical Advanced Study Institute, {\it QCD~and~Beyond},
Boulder, CO, June, 1995, hep-ph/9510287.}  

\bibitem{peskin}{H.~Murayama and M.~Peskin, {\it Ann. Rev. Nucl. Part.
Sci.} {\bf 46} (1996) 533.}
    

\bibitem{drees} {M.~Drees, Contribution to Inauguration
Conference of the Asia Pacific Center for Theoretical
Physics, Seoul, Korea, June, 1996, hep-ph/9611409.} 

\bibitem{bd}{J.~Bjorken and S.~Drell, {\it Relativistic
	Quantum Mechanics}, (McGraw Hill, New York, 1964).}

\bibitem{wess}{J.~Wess and J.~Bagger, {\it Supersymmetry
and Supergravity}, (Princeton University Press,
Princeton, N.J. 1983);
P.~Fayet and S.~Ferrara, {\it Phys. Rep.}
{\bf 32} (1977) 249.}  

\bibitem{west} {See the lectures of P.~West in this volume for
an introduction to the formal aspects of supersymmetry.}

\bibitem{gates} {See the lectures of  J.~Gates in this volume
for an introduction to superfields.}  


\bibitem{anoms}{D.~Gross and R.~Jackiw, {\it Phys.Rev.}
{\bf D6} (1972) 477;
C.~Bouchiat, J.~Iliopoulos and P. Meyer, {\it Phys. Lett.}
{\bf B38} (1972) 519;
H.~Georgi and S.~Glashow, {\it Phys. Rev.} {\bf D6} (1972) 429;
L.~Alvarez-Gaume and E.~Witten, {\it Nucl. Phys. } 
{\bf B234} (1983) 269.} 
  
\bibitem{quaddiv}{E.~Witten, {\it Nucl. Phys.} {\bf B185} (1981) 513;
M.~Dine, W.~Fischler, and M.~Srednicki, {\it Nucl. Phys.}
{\bf B189} (1981) 575;  S.~Dimopoulos and S.~Raby, {\it Nucl.
Phys. } {\bf B192} (1981) 353;
J.~Polchinski and L.~Susskind, {\it Phys. Rev.} {\bf D26} (1982) 3661;
L.~Ibanez and G.~Ross, {\it Phys. Lett.} {\bf B105} (1981) 439.}

 
\bibitem{early}{S.~Dimopoulos and H.~Georgi, {\it Nucl.
Phys. } {\bf B193} (1981) 150;
N.~Sakai, {\it Z. Phys. }{\bf C11} (1981) 153;
P.~Fayet, {\it Phys. Lett.} {\bf B69} (1977) 489;
{\bf B84} (1979) 416.}



\bibitem{hz} {L. Hall~ and M.~Suzuki, {\it Nucl Phys.}
{\bf B231} (1984) 419; T. Banks, Y. Grossman, E. Nardi, and
Y. Nir, {\it Phys. Rev. } {\bf D} (1995) 5319;
B. de Carlos and  P. White, {\it Phys. Rev.} {\bf D54}
(1996) 3427;
E. Nardi, {\it Phys. Rev. } {\bf D55} (1977) 5772.}

\bibitem{proton}{S.~Weinberg, {\it Phys. Rev.}{\bf D26}
(1982) 287; N.~Sakai and T.~Yanagida,
{\it Nucl. Phys. } {\bf B197} (1982) 533;
S.~Dimopoulos, S.~Raby, and F.~Wilczek,
{\it Phys. Lett.} {\bf B112} (1982) 133;
J.~Ellis, D.~Nanopoulos, and S.~Rudaz, {\it Nucl. Phys.}
{\bf B202} (1982) 43.} 


\bibitem{sher}{ C.~Carlson, P.~Roy, and M.~Sher,
{\it Phys. Lett} {\bf B357} (1995) 99;
 G.~Bhattacharyya, hep-ph/9709395.}

\bibitem{goity} {J.~Goity and M.~Sher, {\it Phys. Lett.}
{\bf B346} (1995) 69.}

\bibitem{dreiner} {H.~Dreiner, to be published in
{\it Perspectives on Supersymmetry},
ed. G. Kane, (World Scientific, Singapore, 1998), hep-ph/9707435;
	G.~Altarelli, {\it et. al.}, hep-ph/9703276.}

\bibitem{rp}{G.~Farrar and P.~Fayet, {\it Phys. Lett.}
{\bf B76} (1978) 575;
F.~Zwirner, {\it Phys. Lett.} {\bf 132B}
(1983) 103;
J.~Ellis, G.~Gelmini, C.~Jarlskog, G.~Ross, and
J.~Valle, {\it Phys. Lett.} {\bf B150}
(1985) 142;
G.~Ross and J.~Valle, {\it Phys. Lett.} {B151} (1985) 375;
S.~Dawson, {\it Nucl. Phys.} {\bf B261}(1985) 297;
S.~Dimopoulos and L.~Hall, {\it Phys. Lett.}{\bf B207}
(1988) 210.}

\bibitem{pdg}{Particle Data Group, {\it Phys. Rev.}
{\bf  D 54}, (1996) 1.} 


\bibitem{farr} G.~Farrar, Contribution to SUSY 97, University
of Pennsylvania, May 1997, hep-ph/9710277.

\bibitem{baer1}{H.~Baer, C. Kao, and X.~Tata,
{\it Phys. Rev. } {\bf D51} (1995) 2180;
H.~Baer, C.~Chen, and X. Tata,
{\it Phys. Rev. } {\bf D55} (1997) 1466.}

\bibitem{snow}{ J. Amundson {\it et. al.},
Contribution to 1996 Snowmass Proceedings, hep-ph/9609374.}


\bibitem{hera}{J. Breitweg {\it et. al.}, ZEUS collaboration,
DESY 97-25, hep-ex/9702015; C. Adloff {\it et. al.} H1 collaboration,
DESY 97-24, hep-ex/9702012, B.~Straub, Proceedings of the 1997
Lepton-Photon Conference, Hamburg, Germany, 1997.}


\bibitem{wein}{L.~Hall, J.~Lykken, and S.~Weinberg,
{\it Phys. Rev.} {\bf D 27} (1973) 2359.}  

\bibitem{soft}{L.~Giradello and M.~Grisaru,
{\it Nucl. Phys.} {\bf B194} (1982) 65;
K.~Harada and N.~Sakai, {\it Prog. Theor. Phys.}
{\bf 67} (1982) 67.}

\bibitem{hnew}{H.~Haber, Contribution to SUSY 97,
University of Pennsylvania, May 1997, hep-ph/9709450.}  

\bibitem{hhg}{J.~Gunion, H.~Haber, G.~Kane, and S.~Dawson,
{\it The Higgs Hunter's Guide} (Addison Wesley, Menlo Park,
CA) 1990.}  

\bibitem{massloop}{P.~Chankowski, S.~Pokorski, and J.~Rosiek,
{\it Phys. Lett.} {\bf B274} (1992) 191; {\bf B281} (1992) 100;
Y.~Okada, M.~Yamaguchi, and T.~Yanagida, {\it
Prog. Theor. Phys.} {\bf 85} (1991) ;
{\it Phys. Lett.} {\bf B262} (1991) 54;
J.~Espinosa and M.~Quiros, {\it Phys. Lett.}
{\bf B267} (1991) 27;
{\it Phys. Lett.} {\bf B266} (1991) 389;  
H.~Haber and R. Hempfling, {\it Phys. Rev.}
{\bf D48} (1993)4280;
{\it Phys. Rev. Lett.} {\bf 66} (1991) 1815;
J.~Gunion and A. Turski, {\it Phys. Rev.}
{\bf D39} (1989) 2701; 
{\bf D40} (1990) 2333;
M.~Berger, {\it Phys. Rev. } {\bf D41} (1990) 225; 
K.~Sasaki, M.~Carena and C.~Wagner, {\it Nucl.
Phys.} {\bf B381} (1992) 66; R.~Barbieri and M.~Frigeni,
{\it Phys. Lett.} {\bf B258} (1991) 395; 
J.~Ellis, G.~Ridolfi and F.~Zwirner, {\it Phys.
Lett.} {\bf B257} (1991) 83; {\bf B262} (1991) 477;
R.~Hempfling and A.~Hoang, {\it Phys. Lett.} {\bf B331}
(1994) 99; R.~Barbieri, F. Caravaglios, and M.~Frigeni,
{\it Phys. Lett.} {\bf B258} (1991)167; 
H.Haber, R.~Hempfling, and H.~Hoang, 
{\it Z. Phys.} {\bf C75} (1997) 539;
M.Carena, M.~Quiros, and C.~Wagner, {\it Nucl. Phys.}
{\bf B461} (1996) 407;
M.~Carena, J.~Espinosa, M.~Quiros, and C.~Wagner,
{\it Phys. Lett.} {\bf B355} (1995) 209.}


\bibitem{quiros}{M. Quiros, {\it XXIV International Meeting
on Fundamental Physics: From Tevatron to LHC},
Gandia, Spain, 1996, hep-ph/9609392; T.~Elliot, S.~King, and
P.~White, {\it Phys. Lett.} {\bf B305} (1993) 71;
G.~Kane, C.~Kolda, and J.~Wells, {\it Phys. Rev. Lett.}
{\bf 70} (1993) 2686.}

\bibitem{singlet}{U.~Ellwanger, M. Rausch de Traubenberg, and
C. A. Savoy, {\it Nucl. Phys. } {\bf B492} (1997) 21; {\it
Phys. Lett.} {\bf B315} (1993) 331; S.F. King, P.White,
{\it Phys. Rev. } {\bf D53} (1996) 4049;
J.Ellis, J. Gunion, H. Haber, L. Roszkowski, and F. Zwirner,
{\it Phys. Rev.} {\bf D39} (1989) 844.}  

\bibitem{habergun}{H.~Haber and J.~Gunion, {\it Nucl. Phys.}
{\bf B272} (1986) 1; {\it Nucl. Phys. } {\bf B278} (1986) 449;
erratum, {\bf B402} (1993) 567.}

\bibitem{squrad}{The FORTRAN program HDECAY is documented
in M.~Spira, CERN-TH-95-285, hep-ph/9610350 along with
references to the original calculations.}


\bibitem{cvetic} G.~Cleaver {\it et.al.}, hep-ph/9705391.

\bibitem{lephiggs}{M.~Carena, P.~Zerwas, {\it et.al.},
{\it Higgs Physics at LEPII}, hep-ph/9602250, 1996.}  

\bibitem{eech1} {A.~Bartl, H.~Fraas, W.~Majerotto, and B.~Mosslacher,
{\it Z. Phys.} {\bf C55} (1992) 257;  A.~Bartl, H.~Fraas,
and W.~Majerotto, {\it Z. Phys.} {\bf C30} (1986) 441;
J.~M.~Frere and G.~Kane, {\it Nucl. Phys.} {\bf B223}
(1983) 331;  S.~Dawson, E.~Eichten and C.~Quigg,
{\it Phys. Rev.} {\bf D31} (1985) 495;
D.~Dicus, S.~Nandi, W.~Repko, and X.~Tata, 
{\it Phys. Rev. Lett.} {\bf 51} (1983) 1030;
J.~Ellis and G.~Ross, {\it Phys. Lett.} {\bf 117B}
(1982) 397;  V.~Barger, R.~Robinett, and W.~Y.~Keung, and 
R.~J.~Phillips, {\it Phys. Lett.} {\bf 131B}  (1983) 372.}

\bibitem{eech2} {J.~Feng and M.~Strassler, {\it Phys. Rev.}
{\bf D51} (1995) 4661; {\bf D55} (1997) 1326.}


\bibitem{betafuns} {J.~Bagger, S.~Dimopoulos, and E.~Masso,
{\it Phys. Lett.} {\bf B156} (1985) 357;
{\it Phys. Rev. Lett.} {\bf 55} (1985) 920;
M.~Einhorn and D.~Jones, {\it Nucl. Phys.} {\bf B196} (1982) 475.}



\bibitem{unif}{S.~Dimopoulos, S.~Raby, and F.~Wilczek, {\it Phys. Rev.}
{\bf D24} (1981) 1681; U.~Amaldi {\it et.al.}, {\it Phys.
Rev.} {\bf D36} 1987 1385; P.~Langacker and M.~Luo,
{\it Phys. Rev.} {\bf D44} (1991) 514; J.~Ellis, S.~Kelley,
and D.~Nanopoulos, {\it Phys. Lett.} {\bf B260} (1991) 447;
U.~Amaldi, W.~deBoer, and H.~Furstenau, {\it Phys. Lett.}
{\bf B260} (1991) 447; N.~Sakai, {\it Z. Phys} {\bf C11} (1982) 153.}

\bibitem{ccunif}{J.~Ellis, S.~Kelley, and D.~Nanopoulos,
{\it Phys. Lett.} {\bf B260} (1991)131;
	P.~Langacker and M.~Luo, {\it Phys. Rev.}
	{\bf D44} (1991) 817; U.~Amaldi,
	W.~deBoer, and H.~Furstenau, {\it Phys. Lett.}
	{\bf B260}(1991) 131;
	M.~Carena, S.~Pokorski, and C.~Wagner, {\it Nucl. Phys.}
	{\bf B406} (1993) 59;
	P.~Langacker and N.~Polonsky,
	{\it Phys. Rev.} {\bf D47} (1993) 4028.}

\bibitem{coups2}
	{	J.~Bagger, K.~Matchev, and D.~Pierce,
	{\it Phys. Lett.} {\bf B348} (1995) 443;
	D.~Pierce, J.~Bagger, K.~Matchev, and R. Zhang,
	{\it Phys. Rev. Lett.} {\bf 78} (1997) 1002, erratum,
	{\bf 78} (1997) 2497; {\it Nucl. Phys. } {\bf B491}
	(1997) 3.}

\bibitem{pierce}{D.~Pierce, Contribution to the {\it
Proceedings of the 1996 SLAC Summer Institute}, hep-ph/9701344.} 

\bibitem{fp}{H.~Baer, C.~Chen, F.~Paige, and X.~Tata,
{\it Phys. Rev.} {\bf D54} (1996) 5866; 
{\it op. cit.} {\bf D53} (1996) 6241;
{\it op.cit.}
{\bf D52} (1995) 1565;
{ \bf D52} (1995) 2746.}
 
                        
\bibitem{mess}{M.~Dine, A.~Nelson, Y.~Shirman,
{\it Phys. Rev. }{\bf D51} (1995) 1362;
M.~Dine, A.~Nelson, Y.~Nir, and Y.~Shirman, 
{\it Phys. Rev.} {\bf D53} (1996) 2658.}  


\bibitem{kolda}{For a recent review of GMSB, see C. Kolda,
 contribution to SUSY 97,
University of Pennsylvania, May 1977, hep-ph/9707450.}  



\bibitem{yukren}{M.~Machacek and M.~Vaughn, {\it Nucl. Phys.}
{\bf B222} (1983) 83; C.~Ford, D.~Jones, P.~Stephenson, and M.~
Einhorn, {\it Nucl. Phys. } {\bf B395} (1993) 17.}
  
\bibitem{ewsbtop}{L.~Ibanez, {\it Nucl. Phys.}
{\bf B218} (1983) 514;
{\it Phys. Lett.} {\bf B118} (1982) 73;
L.~Ibanez and G.~Ross, {\it Phys. Lett.} {\bf B110} (1982) 215;
J.~Ellis, D.~Nanopoulos, and K.~Tamvakis, {\it Phys. Lett.}
{\bf B121} (1983) 123;
L.~Alvarez-Gaume, J.~Polchinski, and M.~Wise,
{\it Nucl. Phys.} {\bf B221} (1983) 495;
B.~Ananthanarayan, G.~Lazarides, and Q.~Shafi,
{\it Nucl. Phys.}{\bf D44} (1991) 1613.}

\bibitem{masssamp}{V.~Barger, M.~Berger, and P.~Ohmann,
{\it Phys. Rev.} {\bf D49} (1994) 4908.}  

\bibitem{btau}{B.~Pendleton and G.~Ross, {\it Phys. Lett.}
{\bf B98} (1981)291;
V.~Barger, M.~Berger, P.~Ohmann, and R.~Phillips,
{\it Phys. Lett.} {\bf B314} (1993) 351;
S.~Kelley, J.~Lopez, and D.~Nanopoulos, {\it Phys. Lett.}
{\bf B274} (1992) 387; M.~Carena, M.~Olechowski, S.~Pokorski,
and C.~Wagner, {\it Nucl. Phys.} {\bf B426} (1994) 269;
N.~Polonsky, {\it Phys. Rev.} {\bf D54} (1996)4537;
N.~Polonsky, {\it Phys. Rev.} {\bf D54}
(1996) 4537.}

\bibitem{largeb}  H.~Baer {\it et.al.}, hep-ph/9712305;
T.~Blazek and S.~Raby, hep-ph/9712257; H. Baer {\it et.al},
{\it Phys. Rev. Lett.} {\bf79} (1997) 986; A.~Bartl, W.~Majerotto,
and W.~Porod, {\it Z. Phys.} {\bf C64} (1994) 499, erratum,
{\bf C68} (1995) 518. 



\bibitem{deb}{W.deBoer {\it et.al.}, hep-ph/9712376;
{\it Z.~Phys.} {\bf C75} (1997) 627.}

\bibitem{ewsusy}{P.~Chankowski and S.~Pokorski,
{\it Acta. Phys. Polon.} {\bf 27} (1996) 1719;
 G.~Kane, R.~Stuart, and J.~Wells,
{\it Phys. Lett.} {\bf B354} (1995) 350;
T.~Blazek, M.~Carena, S.~Raby, and C.~Wagner,
{\it Phys. Rev.} {\bf D56} (1997) 6919; T. Blazek and S. Raby,
Presented at {\it International Workshop on Quantum Effects
in the Minimal Supersymmetric Standard Model}, Barcelona,
Spain, 1997, hep-ph/9712255.}

\bibitem{rhosusy}{ W. Beenakker, R. Hopker, and P. Zerwas,
{\it Phys. Lett.} {\bf B349} (1995) 463;  A.~Djouadi
{\it et. al.}, hep-ph/9710438.} 

  
\bibitem{cleo}{M.~Alam {\it et.al.}, 
(CLEO Collaboration), {\it Phys. Rev. Lett} {\bf 74} (1995) 2885.} 
 
\bibitem{bsg}{ S. Bertolini, F.~Borzumati, A.~Masiero and
G.~Ridolfi, {\it Nucl. Phys.} {\bf B353} (1991) 591;
R.~Barbieri and G.~Giudice, {\it Phys. Lett.} {\bf 309} (1993) 86;
P.~Nath and R.~Arnowitt, {\it Phys. Lett.} {\bf B336} (1994) 395;
G.~Kane, C.~Kolda,
L.~Roszkowsi, and J.~Wells, {\it Phys. Rev.}
{\bf D49} (1994) 6173;
V.~Barger, M.~Berger, P. Ohmann, and R. Phillips, {\it Phys.
Rev. } {\bf D51} (1995) 2438;
B.~deCarlos and J.~A.~Casas, {\it Phys. Lett.} {\bf B349} (1995) 300,
{\it ibid} {\bf B351} (1995) 604.} 

\bibitem{bsg2}{W.~deBoer {\it et.al.}, 
{\it Z.~Phys.} {\bf C71} (1996)415.}

\bibitem{bb}{H.~Baer and M. Brhlik, {\it Phys. Rev.}
{\bf D55} (1997) 3201.}
 
 
\bibitem{fcnc}{F.~Gabbiani, E.~Gabrielli, A.~Masiero,
 {\it Nucl. Phys.}
{\bf B477} (1996) 321;
A.~Masieiro and L.~Silvestrin, Contribution to
{\it 35th Course of International School on Subnuclear
Physics}, Erice, 1997, hep-ph/9711401;
L.~Hall, A.~Kostelecky, and S.~Raby,
{\it Nucl. Phys. } {\bf B267} (1986) 415;
L.~Hall and L.~Randall,
{\it Phys. Rev. Lett.} {\bf 65} (1990) 2939;
M.~Dine, R.~Leigh, and A.~Kagan,
{\it Phys. Rev.} {\bf D48} (1993) 4269;
Y.~Nir and N.~Seiberg, {\it Phys. Rev. Lett.} {\bf B309}
(1993) 337 ;  J.~Bagger, K. Matchev, R.-J. Zhang,
hep-ph/9707225.} 

\bibitem{halltasi}{See the lectures of L.~Hall at this school
for a discussion of flavor physics in SUSY models.}
  
\bibitem{lamhoo}{See the lectures of R. van Kooten and 
S. Lammel for discussions of experimental limits on SUSY
particles at LEPII and the Tevatron.
 See also, M.~Carena {\it et.al.}, hep-ex/9712022 for
experimental limits from the Tevatron. }


\bibitem{sopc} A.~Sopczak, Proceedings of {\it The First International
Workshop on Non-Accelerator Physics},
Dubna, July, 1997, hep-ph/9712283.



\bibitem{bargermm}{V.~Barger, M.~Berger, J.~Gunion, and T.~Han,
	{\it Phys. Rev. } {\bf D55} (1997) 142;
{\it Phys. Rev. Lett.}
{\bf 78} (1997) 3991; {\it Nucl Phys. Proc. Suppl.}
{\bf 51A} (1996) 13.}


\bibitem{nlchiggs}{A.~Djouadi {\it et. al.}, {\it
Proceedings of the Workshop Physics with $e^+e^-$ Linear
Colliders}, (Annecy-Gran Sasso-Hamburg, 1995), Ed. P.~
Zerwas, hep-ph/9605437.} 



\bibitem{spira}{M.~Spira, A.~Djouadi, D.~Graudenz, and
P.~Zerwas, {\it Nucl. Phys.} {\bf B453} (1995) 17;
{\it Phys. Lett.} {\bf B318} (1993) 347;
S.~Dawson, A.~Djouadi, and M.~Spira, {\it Phys. 
Rev. Lett.} {\bf 77} (1996) 16.}   
 

\bibitem{lhchiggs}{J.~Gunion, A. ~Stange, and S.~Willenbrock,
{\it Electroweak Symmetry Breaking and Physics at the TeV
Scale}, Ed. T.~Barklow, S.~Dawson, H.~Haber, and J.~Siegrist,
(World Scientific, 1996), hep-ph/9602238.}  


\bibitem{lhcprop}{ATLAS Collaboration, Technical Proposal,
LHCC/P2 (1994); CMS Collaboration, Technical Proposal,
LHCC/P1 (1994).}    


\bibitem{wm}{A.~Stange, W.~Marciano, and S.~Willenbrock,
	{\it Phys. Rev.} {\bf D50} (1994) 4491; {\bf D49}
	(1994) 1354.}  

\bibitem{dfr}{D.~Froidevaux {\it et.al.} ATLAS internal note,
PHYS-No-74 (1995).}       

\bibitem{paigetasi}{See the lectures  of F.~Paige for a
discussion of the prospects for observing SUSY at the LHC.}


\bibitem{aleph}{
ALEPH Collaboration, Ref. 614 (1997), submitted to the 1997
EPS-HEP Conference, Jerusalem.}

\bibitem{chargino}{H.~Baer, C.~Kao, and X.~Tata, {\it Phys. 
Rev.} {\bf D48} (1993) 5175;
H.~Baer, C.~Chen, F.~Paige, and X.~Tata, {\it Phys.
Rev.} {\bf D50} (1994) 4516; {\bf D55} (1996) 4508.}


\bibitem{cdflim}{J. Conway, Contribution
to EPS 97, Jerusalem, 1997.}

\bibitem{snow2}{J.~Bagger, U.~Nauenberg, X.~Tata, and A.~White,
Proceedings of Snowmass, 1996, hep-ph/9612359.}

\bibitem{sigsusy}{S.~Dawson, E.~Eichten, and C.~Quigg,
{\it Phys. Rev.} {\bf D31} (1985) 1581; H.~Baer, A.~Bartl, D.~
Karatas, W.~Majerotto, and X.~Tata, {\it Int. Jour. Mod.
Phys.} {\bf A4} (1989) 4111.}

\bibitem{squglu}{W.~Beenakker, R.~Hoper, M.~Spira, and P.~Zerwas,
{\it Z. Phys.} {\bf C69} (1995) 163.}    

\bibitem{squark}{H.~Baer, J.~Ellis, G.~Gelmini, D.~Nanopoulos,
and X.~Tata, {\it Phys. Lett.} {\bf B} (1985) 175; H.~Baer, V.~Barger,
D.~Karatas, and X.~Tata, {\it Phys. Rev.} {\bf D36} (1987) 96;
R.~Barnett, J.~Gunion, and H.~Haber, {\it Phys. Rev.} {\bf D37} (1988)
1892; {\it Phys. Lett.} {\bf B315} (1993) 349; H.~Baer, C.~Kao, and
X.~Tata, {\it Phys. Rev.} {\bf D48} (1993) 2978.}  


\bibitem{tevlimits} F.~Abe {\it et.al.}, CDF Collaboration,
 {\it Phys. Rev. Lett.}
{\bf 75} (1995) 613;  S.~Abachi, D0 collaboration, 
 {\it Phys. Rev. Lett.}
{\bf 75} (1995) 618. 

\bibitem{lepstop}{ R. Barate {\it et.al.}, ALEPH Collaboration,
CERN-PPE/97-084, hep-ex/970812;
K.~Ackerstaff {\it et. al.},
{\it Z. Phys} {\bf C75} (1997) 409.}  


\bibitem{d0stop} S.~Abachi, D0 collaboration, 
 {\it Phys. Rev. Lett.}
{\bf 76} (1996) 2222.




\end{thebibliography}
\end{document}